\documentclass[12pt]{emulateapj}

\usepackage{amsmath,amssymb,natbib,graphicx,hyperref}

 \newcommand{\ie}{{i.e.}}

 \newcommand{\msun}{\mbox{$M_{\odot}$}}

 \newcommand{\ergsec}{\mbox{erg s$^{-1}$}}
 \newcommand{\rvir}{\mbox{$R_{200\rm{c}}$}}
 \newcommand{\mvir}{\mbox{$M_{200\rm{c}}$}}
\newcommand{\zp}{\mbox{$z_{\rm p}$}}
 \newcommand{\zs}{\mbox{$z_{\rm s}$}}
\newcommand{\zG}{\mbox{$z_{\rm G}$}}
\newcommand{\pz}{\mbox{$\mathcal{P}(z)$}}
\newcommand{\flux}{erg cm$^{-2}$ s$^{-1}$ }

\shorttitle{Galaxies in X-ray Groups}
\shortauthors{George et al.}
 
\begin{document}
  

 \title{Galaxies in X-ray Groups I: Robust Membership Assignment and
   the Impact of Group Environments on Quenching}



\author{
Matthew~R.~George\altaffilmark{1,2},
Alexie~Leauthaud\altaffilmark{2,3},
Kevin~Bundy\altaffilmark{1},
Alexis~Finoguenov\altaffilmark{4,5},
Jeremy~Tinker\altaffilmark{6}, 
Yen-Ting~Lin\altaffilmark{7,8}, 
Simona~Mei\altaffilmark{9,10},  
Jean-Paul~Kneib\altaffilmark{11}, 
Herv\'{e}~Aussel\altaffilmark{12},
Peter~S.~Behroozi\altaffilmark{13},
Michael~T.~Busha\altaffilmark{13,14},
Peter~Capak\altaffilmark{15},  
Lodovico~Coccato\altaffilmark{16},
Giovanni~Covone\altaffilmark{17},
Cecile~Faure\altaffilmark{18},
Stephanie~L.~Fiorenza\altaffilmark{19},
Olivier~Ilbert\altaffilmark{11},
Emeric~Le~Floc'h\altaffilmark{12},
Anton~M.~Koekemoer\altaffilmark{20},
Masayuki~Tanaka\altaffilmark{7},
Risa~H.~Wechsler\altaffilmark{13},
Melody~Wolk\altaffilmark{21}
}

\altaffiltext{1}{Department of Astronomy, University of California,
  Berkeley, CA 94720, USA}

\altaffiltext{2}{Lawrence Berkeley National Laboratory, 1 Cyclotron
  Road, Berkeley CA 94720, USA}

\altaffiltext{3}{Berkeley Center for Cosmological Physics, University
  of California, Berkeley, CA 94720, USA}

\altaffiltext{4}{Max-Planck-Institut f\"{u}r Extraterrestrische
  Physik, Giessenbachstra\ss{}e, 85748 Garching, Germany}

\altaffiltext{5}{University of Maryland Baltimore County, 1000 Hilltop
  Circle, Baltimore, MD 21250, USA}

\altaffiltext{6}{Center for Cosmology and Particle Physics, Department
of Physics, New York University, 4 Washington Place, New York, NY
10003, USA}

\altaffiltext{7}{Institute for the Physics and Mathematics of the
  Universe, The University of Tokyo, Kashiwa, Chiba 277-8568, Japan}

\altaffiltext{8}{Institute of Astronomy \& Astrophysics, Academia
Sinica, Taipei, Taiwan}

\altaffiltext{9}{University of Paris Denis Diderot, 75205 Paris CEDEX 13, France}

\altaffiltext{10}{GEPI, Observatoire de Paris, Section de Meudon, 92195 Meudon CEDEX, France}

\altaffiltext{11}{Laboratoire d'Astrophysique de Marseille, CNRS
  Universit\'{e} de Provence, 38 rue F. Joliot-Curie, 13388 Marseille
  Cedex 13, France}

\altaffiltext{12}{Service d'Astrophysique, CEA-Saclay, Orme de
  Merisiers, Bat. 709, 91191 Gif-sur-Yvette, France}

\altaffiltext{13}{Kavli Institute for Particle Astrophysics and
  Cosmology; Physics Department, Stanford University; and SLAC
  National Accelerator Laboratory, Stanford CA 94305, USA}

\altaffiltext{14}{Institute for Theoretical Physics, University of
  Zurich, 8057 Zurich, Switzerland}

\altaffiltext{15}{Spitzer Science Center, 314-6 Caltech, 1201
  East California Boulevard, Pasadena, CA, 91125, USA}

\altaffiltext{16}{European Southern Observatory, Karl-Schwarzchild-str. 2, 85748 Garching, Germany}

\altaffiltext{17}{Universit\`{a} di Napolia “Federico II”, Dipartimento
 di Sciennze Fisiche and INAF – Observatorio Astronomico di
 Capodimonte, v. Moiariello 16, 80131 Napoli, Italy}

\altaffiltext{18}{Laboratoire d'Astrophysique, Ecole Polytechnique F\'{e}d\'{e}rale de
Lausanne (EPFL), Observatoire de Sauverny, 1290 Versoix, Switzerland}

\altaffiltext{19}{Astrophysical Observatory, City University of New York, College of
Staten Island, 2800 Victory Blvd., Staten Island, NY 10314, USA}

\altaffiltext{20}{Space Telescope Science Institute, 3700 San Martin
  Drive, Baltimore, MD 21218, USA}

\altaffiltext{21}{Institut d'Astrophysique de Paris, UMR 7095, 98 bis Boulevard Arago, 75014 Paris, France}


\email{mgeorge@astro.berkeley.edu}

\begin{abstract}
Understanding the mechanisms that lead dense environments to host
galaxies with redder colors, more spheroidal morphologies, and lower
star formation rates than field populations remains an important
problem. As most candidate processes ultimately depend on host halo
mass, accurate characterizations of the local environment, ideally
tied to halo mass estimates and spanning a range in halo mass and
redshift are needed.  In this work, we present and test a rigorous,
probabalistic method for assigning galaxies to groups based on precise
photometric redshifts and X-ray selected groups drawn from the COSMOS
field.  The groups have masses in the range $10^{13} \lesssim
M_{200\rm{c}}/M_{\odot} \lesssim 10^{14}$ and span redshifts
$0<z<1$. We characterize our selection algorithm via tests on
spectroscopic subsamples, including new data obtained at the VLT, and
by applying our method to detailed mock catalogs.  We find 
that our group member galaxy sample has a purity of $84\%$ and
completeness of $92\%$ within $0.5\rvir$. We measure the impact of
uncertainties in redshifts and group centering on the quality of the
member selection with simulations based on current data as well as
future imaging and spectroscopic surveys. As a first application of
our new group member catalog which will be made publicly available, we
show that member galaxies exhibit a higher quenched fraction compared
to the field at fixed stellar mass out to $z \sim 1$, indicating a
significant relationship between star formation and environment at
group scales.  We also address the suggestion that dusty star forming
galaxies in such groups may impact the high-$\ell$ power spectrum of
the cosmic microwave background and find that such a population cannot
explain the low power seen in recent SZ measurements. 
\end{abstract}
 

 
 


\section{Introduction}
\setcounter{footnote}{0}

Galaxies in dense cluster regions have long been known to have
different characteristics than counterparts in the field, with redder colors,
a greater tendency for spheroidal morphologies, and suppressed star
formation rates. Dense clusters are also the sites of the most massive
and luminous galaxies. Much effort has been made to find the redshift, halo
mass, and cluster-centric distance at which these distinctions between
galaxy populations are imprinted and the process by which these
transformations occur \citep[e.g.,][]{Oemler1974, Dressler1980, Butcher1984, Dressler1997,
Poggianti1999, Lewis2002, Goto2003, Balogh2004, DePropris2004,
Kauffmann2004, Lin2004, Blanton2005, Cucciati2006, Cooper2006, Weinmann2006, Capak2007a,
Gerke2007, Blanton2009, Hansen2009, Mei2009, Feruglio2010}.
While massive clusters present clear
examples of galaxy transformations due to gas stripping, merger
activity, and tidal disruption \citep[e.g.,][]{Kenney1995,
  Gavazzi2001, Cortese2007}, the extent to which these
processes  
affect the majority of galaxies which live in less dense environments
is uncertain. Extending cluster samples to groups with lower halo
masses and higher redshifts is challenging because it requires
significant observational expenditures and careful analysis to isolate
such environments from the field. 

Recent analyses at low redshift have confirmed the existence of an
environmental dependence of galactic structure and colors across a
range of environments \citep[e.g.,][]{Kauffmann2004, Baldry2006,
  Bamford2009}. The corresponding picture at $z\sim1$ has been less clear. With
pointed observations around high-redshift galaxy clusters, several
studies have found significant trends in morphology, color, and
star-formation rate with local galaxy density
\citep[e.g.,][]{Postman2005, Smith2005, Tanaka2005, Poggianti2008}.
However, some find that the relations disappear in stellar mass-selected
samples, arguing that environmental trends are due to differences in
the stellar mass distribution between environments rather than
physical processes acting in dense regions
\citep[e.g.,][]{Poggianti2008}. 

In field surveys reaching $z\sim1$, results from
the VIRMOS-VLT Deep Survey \citep[VVDS;][]{Scodeggio2009} and zCOSMOS
\citep{Tasca2009,  Cucciati2010, Iovino2010, Kovac2010} show little or
no environmental influence on morphology and color especially at high
stellar masses ($\log(M_{\star}/M_{\odot})\gtrsim10.7$), while results
from DEEP2 \citep{Cooper2010} and others from zCOSMOS \citep{Peng2010}
show a clear relationship between color and environment. These papers
generally find weakening environmental trends with increasing
redshift, but differ in the redshift at which the trends
disappear. \citet{Cooper2007} and \citet{Cooper2010} discuss the discrepancies in
environmental trends seen in high-redshift field surveys and suggest
that the non-detection by some studies could be due to
the use of less confident spectroscopic redshifts and lower sampling rates, as
well as increased difficulty with determining environmental densities
using optical spectroscopy at high redshift, while \citet{Peng2010}
attributes the differences to the definitions used to characterize
environments.

The aim of this work is to define a clean sample of galaxies in dense
group environments out to redshift $z=1$ to address these issues. We study groups from the
COSMOS survey that have been identified as sources of extended X-ray
emission \citep[][and in prep.]{Finoguenov2007}, which is a strong
indication that they are virialized structures and not chance
associations of galaxies. The groups have halo masses in the range
$10^{13} \lesssim M_{200\rm{c}}/M_{\odot} \lesssim 10^{14}$ as
determined by weak lensing \citep{Leauthaud2010}.  In a companion
paper, we describe weak lensing tests to optimize the identification
of halo centers (Paper II; George et al., in prep.). We select member
galaxies based on photometric redshifts derived from
extensive multi-wavelength imaging, which provides a
much greater sampling density than existing spectroscopic
surveys. Using a spectroscopic subsample and mock catalogs, we
carefully evaluate our member selection for potential biases or
contamination, and account for photometric redshift
uncertainties. This robust sample of group members can be used to
address unsettled questions about the link between galaxies and their
environments.

A key challenge is to disentangle the intrinsic and extrinsic factors
that may play a role in shaping galaxy properties. For instance, galaxies
in dense regions have a higher characteristic stellar mass than in
less dense environments \citep[e.g.,][]{Baldry2006}, so a
morphology-mass relation could be conflated with a morphology-density
relation. Since stellar mass plays an important role in determining
galaxy properties, and mass-to-light ratios are strongly affected by
star formation activity, recent environmental studies have stressed
the use of stellar mass-selected samples rather than
luminosity-selected samples to make a fair comparison across
environments \citep[e.g.,][]{VanDerWel2007, Scodeggio2009,
  Cooper2010}.

In addition to controlling for intrinsic galaxy
properties in these studies, defining and measuring the
``environment'' presents another problem. The distance to the $N$th
nearest neighbor or the mean density of galaxies inside a fixed radius
are commonly used as environmental indicators
\citep[e.g.,][]{Dressler1980}. \citet{Kauffmann2004}
show that galaxy properties correlate most tightly with
local density on scales below $\sim1~\rm{Mpc}$ and are uncorrelated
with the density on larger scales once the small-scale density is
fixed. Several studies have shown evidence
that galaxy properties correlate most tightly with density within
their halo, and have emphasized that the aperture used for comparing
equivalent regions must scale with halo mass to avoid confusion
between local and global densities \citep{Hansen2005, Weinmann2006,
  Blanton2007, Haas2011}. Instead of using the galaxy density field to
define environment, one can define a catalog of galaxy groups and
clusters and study their properties as a function of halo mass and
group-centric distance. In this paper we use the term ``group'' to
denote a set of galaxies with a common dark matter halo and to
emphasize the low mass range studied, making no formal distinction
between groups and clusters.

Catalogs of galaxy groups have been constructed from both optical
surveys identifying galaxy overdensities and X-ray surveys detecting
the hot gas trapped by deep gravitational potentials. Optical catalogs
often employ matched filters \citep[e.g.,][]{Postman1996} and tesselations
\citep[e.g.,][]{Marinoni2002, Gerke2005} to isolate groups from the
background field, and red sequence methods have proven efficient at
identifying groups over large volumes \citep[e.g.,][]{Gladders2005,
  Koester2007}. These catalogs 
typically assign the brightest member galaxy as the center of
each group, and use the richness determined by the number of members
as a proxy for the total mass. X-ray detections can reduce the
likelihood of projection effects, improve mass estimates, help with
the determination of group centers, and shed light on the interplay
between the hot gas and stellar content of groups
\citep[e.g.,][]{Mulchaey2003, Lin2004, Finoguenov2007,
  Sun2009}.

Our understanding of group properties has benefited from small,
well-studied samples at low redshift \citep[e.g.,][]{Mulchaey1998, Zabludoff1998,
  Tran2001, Sun2009} along with larger statistical samples taken over wide
areas or to high redshifts \citep[e.g.,][]{Eke2004, Gerke2005,
  Gladders2005, Miller2005, Yang2005, Berlind2006,
  Hansen2009}. These studies have established many similarities
between groups and more massive clusters, including their extended
dark matter halos and elevated fractions of quenched early type
galaxies relative to the field. Groups show some differences from
clusters including gas mass fractions that are typically lower in less
massive systems, and the differences between physical processes acting on
galaxies in groups and those in clusters are still being explored.
Recent and ongoing surveys are pushing to greater 
sample sizes and higher redshifts with multi-wavelength observations and
spectroscopic campaigns \citep[e.g.,][]{Osmond2004, Driver2009,
  Milkeraitis2010, Adami2010}. Several large imaging surveys in
development plan to study the growth of structure without significant
spectroscopic observations initially (Dark Energy
Survey\footnote{\url{http://www.darkenergysurvey.org}}, Hyper
Suprime-Cam\footnote{\url{http://sumire.ipmu.jp/en}}, Large Synoptic
Survey Telescope\footnote{\url{http://www.lsst.org}},
EUCLID\footnote{\url{http://sci.esa.int/euclid}}), so photometric
redshifts will be important for identifying member galaxies using
techniques such as those outlined in this paper.

We study groups in the COSMOS field where a unique data set has been compiled for studying the
interplay between galaxies, intragroup gas, and dark matter in galaxy
groups out to $z\sim1$. The COSMOS survey has obtained X-ray observations for group detections, deep
imaging data spanning ultraviolet (UV), optical, and infrared (IR)
wavelengths for precise photometric redshifts and stellar masses,
extensive spectroscopic coverage, and high resolution imaging from the {\sl
  Hubble Space Telescope} (HST) for measuring morphologies and weak
lensing. Group catalogs have been  
constructed in this field from X-ray data \citep[][and in prep.]{Finoguenov2007},
zCOSMOS spectroscopy \citep{Knobel2009}, photometric redshifts
\citep{Gillis2011}, CFHTLS-Deep photometry
\citep{Olsen2007,Grove2009}, and with a combination of weak lensing
and matched filters \citep{Bellagamba2011}. Additionally,
\citet{Scoville2007b} studied large-scale structures in this field
using photometric redshifts, and \citet{Kovac2010a} measured the
galaxy density field using zCOSMOS redshifts to probe a large dynamic
range of environments. Here we focus on the X-ray selected group
catalog to ensure a pure sample of virialized structures whose masses
have been characterized with weak lensing \citep{Leauthaud2010}.
\citet{Giodini2009} have studied the stellar
mass content of these X-ray groups; we expand upon their work with a
thorough characterization of a new member selection algorithm and
develop a group member catalog for a variety of applications.

The data used in making our group catalog are described in \S~\ref{s:data}. 
In \S~\ref{s:sample}, we present the sample selection and sensitivity
limits, along with tests of the quality of the
photometric redshifts constructed from the imaging data. 
Our selection algorithm for the member catalog is described in
\S~\ref{s:membership}, where we associate member galaxies with groups based
on their proximity to the X-ray center and their photometric redshifts.
In \S~\ref{s:tests}, we characterize the reliability of our selection
with mock catalogs from simulations and by comparing our photometric
redshift selection to the subsample of sources with spectroscopic redshifts.
We make the catalog of group membership assignments and galaxy
properties publicly available, describing the format and release in \S~\ref{s:catalog}. We
discuss in \S~\ref{s:discussion} some of our initial findings from the
catalog, including the influence of the group environment on galaxy colors
out to $z\sim 1$. We find evidence of suppressed star formation in
galaxies in group environments over the entire redshift range studied, and
briefly discuss how the low incidence of star forming galaxies in
groups cannot play a significant role in explaining recent
observations of a deficit of power from the Sunyaev-Zel'dovich effect
\citep[SZ;][]{Sunyaev1972} in the angular spectrum of the cosmic
microwave background \citep[CMB; e.g.,][]{Lueker2010, Fowler2010}.

We adopt a WMAP5 $\Lambda$CDM cosmology to determine distances and
halo masses with $\Omega_{\rm m}=0.258$, $\Omega_\Lambda=0.742$,
$H_0=72$~\rm{km~s}$^{-1}$~\rm{Mpc}$^{-1}$
\citep{Dunkley2009}, the same values used by \citet{Leauthaud2010} to
calibrate the masses of this group sample. Distances are expressed in physical units of
\rm{Mpc}, magnitudes are given on the AB system, X-ray
luminosities are expressed in the rest-frame 0.1-2.4 keV band, and
logarithmic quantities use base 10. Group
masses are estimated from their X-ray luminosity using the $L_X-M$
relation derived in \citet{Leauthaud2010} and concentrations are then derived
from the mass-concentration relation of \citet{Zhao2009} assuming an
NFW density profile \citep*{Navarro1996}. We estimate the virial radius
of groups as \rvir, the radius within which 
the mean density is 200 times the critical density of the Universe at
the redshift of the group, $\rho_{\rm{c}}(\zG)$, and use the
corresponding halo masses defined as $\mvir \equiv
(200\rho_{\rm{c}}(\zG))(4\pi/3)\rvir^3$. We also make use of the NFW
scale radius, defined as $R_{\rm s}=\rvir/c_{200\rm{c}}$, where
$c_{200\rm{c}}$ is the concentration parameter.


\section{COSMOS Data}
\label{s:data}

The COSMOS field has been observed in a broad range of wavelengths,
with imaging data from X-ray to radio and a large spectroscopic
follow-up program (zCOSMOS) with the Very Large Telescope (VLT)
\citep{Scoville2007a,Koekemoer2007,Lilly2007}. We have added to the
spectroscopic sample in groups with a recent campaign using the
Focal Reducer and low dispersion Spectrograph 2 (FORS2) at the VLT
(Program ID 084.B-0523; PI: Mei). 
X-ray imaging has been taken with the {\sl XMM-Newton} \citep[1.5~Ms covering
2.13~deg$^2$;][]{Hasinger2007,Cappelluti2009} and {\sl Chandra} 
observatories \citep[1.8~Ms covering 0.9~deg$^2$;][]{Elvis2009}.  Imaging
obtained through the F814W filter of the Wide Field Channel (WFC) of
the Advanced Camera for Surveys (ACS) on HST adds accurate shape measurements for
morphologies and weak lensing \citep{Scarlata2007,Leauthaud2007}.
Observations of over thirty photometric bands covering the ultraviolet,
optical, and infrared ranges have enabled the determination 
of precise photometric redshifts \citep{Capak2007b,Ilbert2009}, with
typical redshift uncertainty $\sigma_{\mathcal{P}}\lesssim0.01$ for galaxies
with F814W~$<22.5$, and $\sigma_{\mathcal{P}}=0.03$ for F814W=24, at
$z<1.2$ (see \S\S~\ref{s:photoz} and \ref{s:photoz_tests} for details
and tests of the photometric redshifts used in this paper).

\subsection{X-ray Catalog}
\label{s:xray}

The entire COSMOS region has been mapped through 54 overlapping {\sl
  XMM-Newton} pointings and additional {\sl Chandra} observations
covering the central region (0.9 deg$^2$) with higher spatial
resolution. A mosaic combining these two data sets has been used to
find and measure the fluxes of groups using a wavelet transform method
described in \citet{Vikhlinin1998}. The data reduction process
including the combination of X-ray data sets and identification of
optical counterparts follows that of \citet{Finoguenov2009,
  Finoguenov2010}. An initial group catalog from the COSMOS field is
presented in \citet{Finoguenov2007}.

Briefly, extended objects
are detected in the mosaic when the sum of the flux on scales of
$32\arcsec$ and $64\arcsec$ is greater than the flux on small scales
by a given threshold. Detections on smaller scales tend to be
contaminated by point sources, which are cleaned from {\sl XMM} and {\sl Chandra}
data separately to allow for variability. The flux is calculated using a
scaling relation that incorporates a $\beta$-model fit to the surface
brightness within $32\arcsec$, resulting in a $4\sigma$ detection
limit of the group sample of $1.0\times 10^{-15}$ \flux over $96\%$ of
the ACS field. Once extended X-ray sources are detected, a red
sequence finder is employed on galaxies with a projected distance less
than $0.5$~Mpc from the centers to identify an optical counterpart and
determine the redshift of the group, which is then refined with
spectroscopic redshifts when available. The red sequence finder only
requires an overdensity of red galaxies and not a deficiency of blue
galaxies, meaning that it does not specifically require an enhanced
red fraction to identify groups.

A quality flag (hereafter \textsc{xflag}) is assigned to the
reliability of the optical counterpart, with 
flags 1 and 2 indicating a secure association, and higher flags
indicating potential problems due to projections with other sources or
bad photometry due to bright stars in the foreground. We
run our membership algorithm on all detections in the catalog with
$\zG<1$ but in later analyses limit the sample to groups with
\textsc{xflag}~$=1$ and 2, which have reliable spectroscopically confirmed optical
counterparts. The difference in these two flags 
reflects the uncertainty in the X-ray position; for \textsc{xflag}~$=2$ the
uncertainty in each coordinate is assumed equal to the wavelet scale
of $32\arcsec$, and for \textsc{xflag}~$=1$ sources, which have more certain centers,
the uncertainty is $32\arcsec$ divided by the significance of the flux
measurement. The mean uncertainty in right ascension and declination
for sources with flags 1 and 2 is $23\arcsec$ ($120~\rm{kpc}$ at
$z=0.4$ and $170~\rm{kpc}$ at $z=0.8$ for our adopted cosmology).

The main changes from the catalog described in \citet{Finoguenov2007}
are the detection of fainter groups thanks to deeper {\sl XMM}
coverage, a more conservative point-source removal procedure,
increased redshift accuracy due to the availability of more
spectroscopic data and improved photometric redshifts, and some
changes in quality flags after visual inspection of optical
counterparts. In total, the catalog used in this paper contains 211
extended X-ray sources over 1.64 deg$^2$, spanning the
redshift range $0<z<1$ and with a rest-frame 0.1--2.4 keV luminosity
range of $41.3<\log(L_X/\ergsec)<44.1$, and 165 of these groups and
clusters have secure optical counterparts with \textsc{xflag}~$=1$ or
2. X-ray detections without clear optical counterparts are likely a
mix of unresolved active galactic nuclei (AGN), projections of
multiple systems, and background fluctuations; tests of the
identification method using a larger spectroscopic sample will be
presented with the updated X-ray catalog in a separate paper
(Finoguenov et al., in prep.).

\subsection{Spectroscopic Data}
\label{s:spectroscopy}

The COSMOS field has been targeted by a number of spectroscopic
campaigns. We use spectra from the zCOSMOS
``20K sample'' (Lilly et al., in prep.) which targeted galaxies in the ACS area to a
magnitude limit of $i^{+}=22.5$, along with other spectroscopic data
sets from Keck, MMT, SDSS, and VLT \citep{Prescott2006, Capak2010}. We
include in this paper a new sample from our 
recent program with FORS2/VLT (see \S~\ref{s:fors2}). Each redshift has
an associated confidence flag; we use only those of class 3 or 4 meaning that the
redshift is secure or very secure. In repeat observations, zCOSMOS
targets with these confidence classes have a verification rate of
$>99\%$ \citep{Lilly2007}. For galaxies in the ACS field with F814W~$<24.2$ and any
redshift passing this quality cut, the spectroscopic sample has 529
galaxies from FORS2, 11619 from zCOSMOS, and 1527 from 
other sources. These spectra are distributed throughout the redshift
range used in this paper, with 1931 at $0.05<z\le0.25$, 4257 at
$0.25<z\le0.50$, 4184 at $0.50<z\le0.75$, and 1980 at
$0.75<z\le1.00$.

A spectroscopically-selected group catalog has been constructed by
\citet{Knobel2009}. We defer a detailed comparison between the
galaxy content of spectroscopically-selected groups and X-ray selected
groups to future work (Finoguenov et al., in prep.). \citet{Kovac2010a}
showed that there is good general correspondence between the overdense
regions in the galaxy density field, the spectroscopically-selected
groups with $N_{\rm mem}\ge4$, and the X-ray detected groups.

Our primary use of the spectra is to obtain precise group redshifts and to
verify the accuracy of photometric redshifts, which are critical for
both the membership selection and the weak lensing analysis. Roughly
$20\%$ of group members have spectroscopic redshifts in addition to
the photometric redshifts used for member selection. 

\subsubsection{FORS2 Spectra}
\label{s:fors2}
We have recently obtained additional spectra of galaxies in the COSMOS
field with the FORS2 spectrograph at the VLT. Targets were selected
for a number of scientific goals including velocity dispersion
measurements of massive central galaxies and a comparison sample of field ellipticals, a
study of merger rates within groups based on the abundance of close
pairs, and refined redshift determinations for 
group members. Data were taken on four clear nights with excellent
conditions and 0.8\arcsec\ typical seeing from 2010 February 14-18. The
FORS2 instrument was used in MXU mode with the 600RI grism and GG435
order separation filter, providing a wavelength range of roughly
$4500-9000$\AA. There were 27 masks each observed in 4 exposures of
650 seconds. Each mask had roughly 50 slits of width $0.6\arcsec$ for
bright targets and $1\arcsec$ for fainter ones, and a typical slit
length of $8\arcsec$.

These data have been reduced using the standard {\sc esorex} reduction
pipeline\footnote{\url{http://www.eso.org/cpl/esorex.html}}. In short, for
each mask and detector we performed bias subtraction and overscan
removal, determined a wavelength solution from He, HgCd, Ar, and Ne arc
lamps, and found slit extraction regions using a pattern recognition
algorithm on the arc and flat lamp exposures. Science exposures were
bias subtracted and flat fielded before a median combination. The flat
fields were first normalized by dividing out a smooth component
calculated using a $10\times10$ pixel median filter to account for the
intrinsic shape of the flat lamp spectrum. A local sky subtraction was
performed on each CCD column prior to rectification, then cosmic rays
were removed and object spectra were optimally extracted. 

For the extracted objects, we measured redshifts with a modified
version of the {\sc zspec} software used for the DEEP2 survey (Cooper
et al., in prep.). Each one-dimensional spectrum was fit by a linear
combination of galaxy eigenspectra and also compared with stellar and
quasar templates over a range of redshifts to find possible redshift
values. Spectral features important for fitting in this range of
wavelengths and redshifts include [\ion{O}{2}], CaK, CaH, G-Band,
H$\beta$, [\ion{O}{3}], Mgb, and NaD. Each spectrum was visually
inspected by at least two co-authors alongside the two-dimensional spectral image to choose the
best redshift and assign a quality flag according to the zCOSMOS
system \citep{Lilly2007}. In cases where the first two inspectors
disagreed on the redshift or quality flag, a third person viewed the
spectrum independently to reconcile differences. We
include only those objects with a secure redshift (quality flag = 3,4)
in our sample, which amounts to $529$ galaxies. Our redshift success
rate will improve with continuing reduction efforts to handle cases
where slits were tilted to cover close pairs or to measure
velocity dispersions along the major axis of a galaxy. There are $8$
objects in this sample that have been observed by SDSS, with a median
and scatter between redshift measurements of $16$ and $43
\rm{km~s^{-1}}$, respectively. For $126$ objects observed by both
FORS2 and zCOSMOS, the scatter in redshifts is $160~{\rm km~s^{-1}}$
after removing $2$ outliers with $|\Delta z|>0.002$. We have detected
a median offset of $\sim100~{\rm km~s^{-1}}$ between zCOSMOS redshifts
and those measured by FORS2 and SDSS which is still under
investigation, but since the magnitude of this offset is a factor of 3
smaller than the typical group velocity dispersion and several times
smaller than photometric redshift errors it should not impact our results.

\subsection{Photometric Redshifts}
\label{s:photoz}

Despite the extensive spectroscopic data available, coverage of group
members is incomplete. We instead use photometric redshifts (hereafter
photo-$z$s) to determine distances to galaxies. \citet{Ilbert2009} constructed spectral energy
distributions (SEDs) from over 30 bands of UV, optical,
and IR data described in \citet{Capak2007b}, and compared these SEDs
with templates from galaxies at known redshifts supplemented with
stellar population synthesis models. They computed photo-$z$s
from SEDs using a $\chi^2$ template fitting method which included a
treatment of emission lines. The derived $\chi^2(z)$ function was used
to compute a probability density function (PDF), \pz, which is the
likelihood that a galaxy lives at a redshift $z$ given the photometric
data and the spectral templates used. Rather than collapsing this
function to a single value at the mean, median, or peak and assuming
Gaussian uncertainty as is often done, we make use of the full PDF to
determine group membership, described in \S~\ref{s:membership}. 
In this paper we use an updated version (pdzBay\_v1.7\_010809) of the
photo-$z$ catalog presented in \citet{Ilbert2009} with additional deep
$H$ band data and small improvements in the template fitting
techniques.

\citet{Ilbert2009} demonstrated that these photo-$z$s are precise and
accurate thanks to the broad wavelength range covered by the
photometric data and the many bands into which it is divided.
Those authors discussed the quality of the photo-$z$s in comparison with
a number of samples of spectroscopic redshifts using the
normalized median absolute deviation \citep[NMAD $=1.48 \times$
median$(|\zs-\zp|/(1+\zs))$; ][]{Hoaglin1983}, which is an estimator for
$\sigma_{\Delta z/(1+\zs)}$ that is robust to
outliers. \citet{Ilbert2009} showed that the distribution of offsets
between the photometric and spectroscopic redshifts is well-fit by a
Gaussian with a standard deviation equal to the NMAD. Applying this
estimator to galaxies considered for group membership, i.e., those
with F814W~$<24.2$, $z_{p}<1.2$, and an available stellar mass estimate (see
\S~\ref{s:stellarmass}), there are over 12000 spectroscopic redshifts
and the overall agreement with photo-$z$s is $\sigma_{\Delta z/(1+\zs)} =
0.008$. However, the spectroscopic sample is dominated by the
zCOSMOS survey, which has a magnitude limit of
$i^+=22.5$. The other spectroscopic samples have a variety of
selection functions, so we cannot assume that the sample is
representative of the full photometric sample of galaxies.

A second measure of uncertainty of a photometric redshift comes
directly from the width of the PDF. \citet{Ilbert2009} have shown
that the shape of \pz\ is broadly consistent with the distribution of
offsets between photometric and spectroscopic redshifts. For example,
$65\%$ of objects have a redshift offset within the $68\%$ uncertainty
on the PDF, $\sigma_{\mathcal{P}}$. In \S~\ref{s:photoz_tests}, we
discuss further tests on the agreement between these two estimates of
redshift uncertainty, $\sigma_{\mathcal{P}}$ and $\sigma_{\Delta z}$ (we
henceforth drop the conventional factor of $1+\zs$ to make direct
comparisons between the two quantities), and we study variations in
photo-$z$ quality that could bias our selection against different
galaxy populations.

\subsection{Stellar Mass Estimates}
\label{s:stellarmass}

Stellar masses are used in the identification of group centers (see
\S~\ref{s:centers}) and are estimated using the Bayesian code
described in \citet{Bundy2006a}, with good agreement to the
masses determined by \citet{Drory2009}. For each galaxy, the SED and
photo-$z$ described above are referenced to a grid of stellar population
synthesis models constructed using the \citet{Bruzual2003} code and
assuming an initial mass function from \citet{Chabrier2003}.  The
grid includes models that vary in age, star formation history, dust
content, and metallicity.  At each grid point, the probability that
the observed SED fits the model is calculated, and the corresponding
stellar mass is stored. 
By marginalizing over all parameters in the grid, the stellar mass
probability distribution is obtained.  The median and width of this
distribution are taken respectively as the stellar mass estimate and
the uncertainty due to degeneracies and the model parameter space.
The final stellar mass error estimate also includes uncertainties from
the K-band photometry and the expected error on the luminosity
distance that results from the photo-$z$ uncertainty, producing a typical
final uncertainty of 0.2-0.3 dex. Stellar mass estimates in this paper
require a $3\sigma$ detection in $K_{\rm s}$-band, which
is complete to a typical depth of $K_{\rm s}=24$ \citep{McCracken2010}.


\section{Sample Limits and Quality of Photometric Redshifts}
\label{s:sample}

\subsection{Mass Limits and Quality Flags}

In this section we present the sample selection and sensitivity limits
for galaxies and groups used in our analysis. Since one of our
limiting factors is the decline in photo-$z$ precision at faint magnitudes,
we also discuss tests of the accuracy and precision of
photo-$z$s for different galaxy populations to show that our group member
sample is not biased by variations in the quality of photo-$z$s.

As mentioned in \S~\ref{s:xray}, we consider X-ray detected groups at
redshifts $0<\zG<1$. Identifying optical associations with X-ray
groups becomes more challenging at $\zG>1$ and typically requires
dedicated spectroscopic followup, so we omit high redshift candidates
from this work. In addition to the flags from the X-ray catalog
describing the quality of the optical identification and centroid
uncertainty, we record three additional flags for each group:
\begin{itemize}
\item{\textsc{mask}: more than $10\%$ of the area within \rvir\ or within
    $R_{\rm{s}}$ of the X-ray center is masked in optical images or
    falls outside the edges of the ACS field}
\item{\textsc{poor}: 3 or fewer member galaxies are associated with the group
    (using $P_{\rm mem}>0.5$, see \S~\ref{s:membership})}
\item{\textsc{merger}: the projected radius (\rvir) drawn from the X-ray center of one
    group overlaps with that of another by more than $25\%$ and the
    group redshifts are consistent ($|\Delta z|<0.01$).}
\end{itemize}
We flag groups in masked regions because their membership may not be
adequately represented and central galaxies may not be
properly identified. Poor groups are flagged as possibly questionable
optical associations or redshift determinations, and merging groups
are flagged because the algorithm may confuse membership
assignments. Of the $165$ X-ray groups with a clear optical
counterpart (\textsc{xflag}~$=1$ or 2), 10, 12, and 15 groups are
assigned the \textsc{mask}, \textsc{poor}, and \textsc{merger} flags,
respectively. Our rationale for assigning these flags is to attain a
group catalog that is as pure as possible, though not necessarily complete.

\begin{figure*}[htb]
\epsscale{1.17}
\plottwo{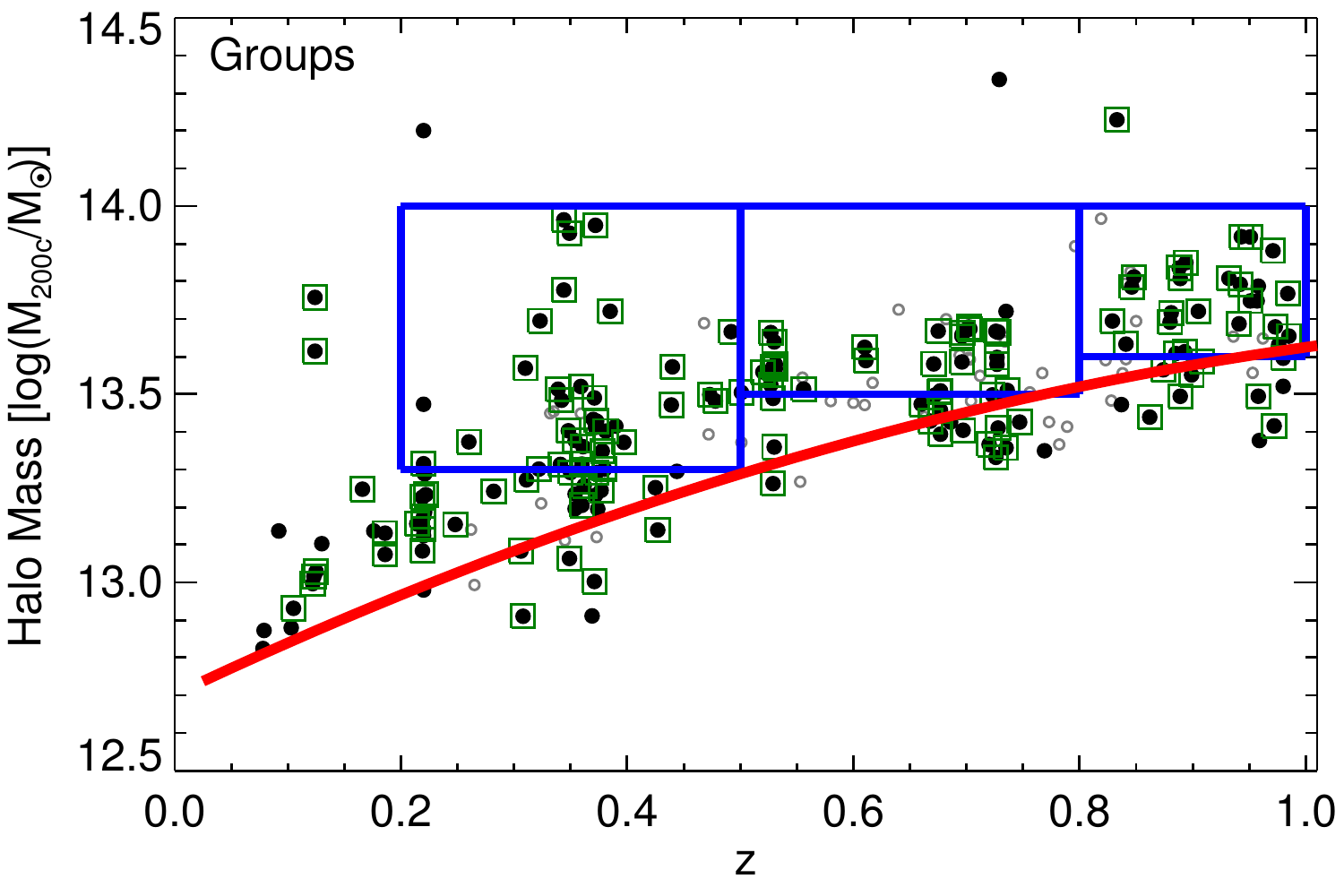}{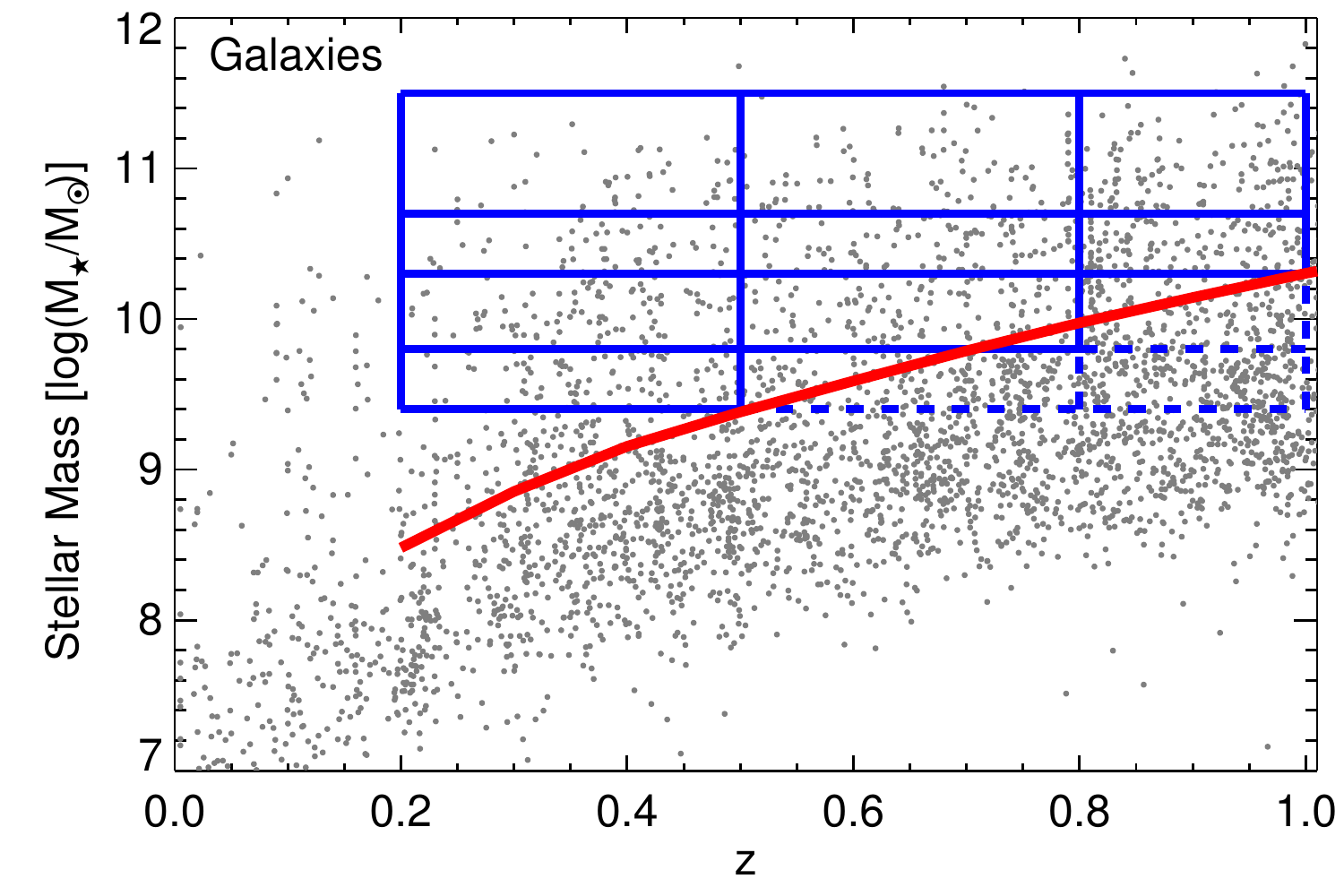}
\caption{Mass limits for groups (left) and galaxies (right) over the
  redshift range $0<z<1$. Symbols for groups denote their quality
  flags; green squares have \textsc{xflag}=1,2 and
  \textsc{mask}=\textsc{poor}=\textsc{merger}=0, black filled circles have
  \textsc{xflag}=1,2, and gray open circles are the rest. Red curves
  denote the mass sensitivity corresponding with the X-ray flux limit
  (left) and the stellar mass limits for a passive galaxy
  (right). Blue boxes show the mass and redshift bins used for analysis in
  \S~\ref{s:discussion}, with dashed boxes denoting stellar mass bins with
  significant incompleteness. Only every third galaxy is plotted for visual clarity.}
\label{fig:mass_limits}
\end{figure*}

The left panel of Figure~\ref{fig:mass_limits} shows the halo masses
and redshifts for the group sample. Green squares represent the
cleanest sample of $129$ groups with \textsc{xflag} $=1$ or 2, and
none of the other flags set, black points relax the restrictions on
the \textsc{mask}, \textsc{poor}, and \textsc{merger} flags, and gray
dots represent the remaining sources in the catalog with higher values
of \textsc{xflag}. The red curve shows the $4\sigma$ X-ray flux limit
reached in $96\%$ of the field of $1.0 \times 10^{-15}$ \flux converted to a limiting group
mass. Coverage is non-uniform, so some groups are detected below this
threshold in areas with deeper coverage. Blue lines show the mass and
redshift bins used for later analysis.

To be considered for group membership and to derive stellar mass
estimates, galaxies must be brighter than F814W~$<24.2$ and have a
photo-$z$ in the range $0<\zp<1.2$. Galaxies must also have a
$3\sigma$ $K_{\rm s}$-band detection, 
for which the typical limiting depth is $K_{\rm s}=24$. Though the
photometry in COSMOS is complete to $i^+=26.2$ and has similar depths
in other optical filters \citep{Capak2007b}, the $K_{\rm s}$-band
detection requirement causes detections in the ACS imaging to
become incomplete near F814W=24.2, which is also in the magnitude
range where photo-$z$ quality deteriorates rapidly (see
\S~\ref{s:photoz_tests}). The F814W filter magnitude correlates more
strongly with photo-$z$ precision than longer wavelength filters (the
$4000$\AA\ break enters the filter range at $z\sim 0.75$ and remains
in that range beyond our redshift limit), so we use it to apply the
formal magnitude cut at F814W=24.2. Taking this as our primary
magnitude cut, we find that only $5\%$ of the sample with F814W~$<24.2$
is excluded due to a nondetection in $K_{\rm s}$ or a failure to find
an acceptable stellar mass fit, with $2\%$ of bright objects
(F814W~$<22.5$) and $8\%$ of faint obects ($23.5<$~F814W~$<24.2$) being cut.
Because of photo-$z$ uncertainties, we allow galaxies to have a
higher redshift limit than the groups in which they reside, giving the
$\zp<1.2$ cut.

The right panel of Figure~\ref{fig:mass_limits} shows the stellar
masses and photometric redshifts for galaxies meeting the selection criteria. We
plot only every third galaxy for clarity. The red curve shows the
$85\%$ stellar mass completeness limit calculated for the oldest
allowable stellar template at each redshift for the combined
requirements of $K_{\rm s}<24$ and F814W~$<24.2$. This passive limit
is conservative, as younger stellar populations have lower
mass-to-light ratios. At $z=1$, our stellar mass limit is roughly 
$\log(M_{\star}/M_{\odot})=10.3$, or $0.25 M^{*}$ (\citealt{Drory2009} found
$\log(M^{*}/M_{\odot}) \approx 10.9$ for the massive end of a
double-Schechter function fit to the stellar mass function at $z \sim
1$, with little redshift evolution). Solid 
blue lines in the figure show the mass and redshift bins for later
analyses, and dashed lines are drawn for stellar mass bins that extend
significantly below
our completeness limits. Our sample has fewer galaxies and groups at
$z<0.2$ than at higher redshifts because of the smaller volume probed,
but the increasing volume at higher redshifts allows for good
statistical samples.

\subsection{Tests of Photometric Redshifts}
\label{s:photoz_tests}

We have described how the increase in photo-$z$ errors at faint magnitudes
partially motivates our selection cut on galaxies brighter than
F814W~$=24.2$, with the implicit concern that poorer photo-$z$s degrade
our ability to assign galaxies to groups. Here we test how photo-$z$
quality varies with other galaxy properties to ensure that our
selection is not biased by systematic errors for certain galaxy populations.

In principle, photo-$z$ quality can depend on any property of an SED or the
templates used in the fitting process. Redshifts of red galaxies with strong
$4000$\AA\ breaks have traditionally been easier to constrain than their
bluer counterparts. Fainter galaxies have larger photometric
uncertainties which propagate into their photo-$z$s. Galaxy mass and
environment may play a role if, for example, the photo-$z$ templates are
not representative of evolutionary histories unique to dense group
regions. Morphology can also have a subtle effect since the
inclination of disks alters the extinction along the line of sight
\citep{Yip2011}. 

Motivated by these possible sources of variation in
photo-$z$ quality, we divide our sample into different populations and
quantify the precision and accuracy of their redshift estimates.
We also compare two estimates of photo-$z$ uncertainty, the $68\%$ width
of the PDF ($\sigma_{\mathcal{P}}$), and the deviation between
photometric and spectroscopic redshifts ($\sigma_{\Delta z/(1+\zs)}$).

In order to test the reliability of the redshift uncertainty
for different galaxy populations, we
slice the galaxy sample into bins based on their 
brightness, redshift, color, morphology, stellar mass, and
environment. Here we use unextincted rest frame
colors derived from the best-fitting templates using the
difference between absolute magnitudes in near-ultraviolet (NUV) and R
bands ($C \equiv M(NUV)-M(R)$) as described by \citet{Ilbert2010}. In
that paper, spectral classes were identified with the following cuts
on $C$ from blue to red:
\begin{eqnarray} 
C < 1.2 & &\textrm{``high activity"} \nonumber \\
1.2 < C < 3.5 & &\textrm{``intermediate activity"} \nonumber \\
C > 3.5 & &\textrm{``quiescent."} \nonumber
\end{eqnarray}
These classes were found to correlate with visually classified
morphologies as expected. For these tests, we use
morphologies determined using the Zurich Estimator of Structural Types
\citep[ZEST; ][]{Scarlata2007} on the ACS images. The results are compiled in
Table~\ref{t:photoz_tests}, in which we present the size and average
magnitude of each population, along with the two measures of photo-$z$
uncertainty and the fraction of sources for which the photo-$z$ deviates
significantly from the spectroscopic redshift.

The two independent measures of photo-$z$
uncertainty are in good agreement, suggesting that we can safely use
PDF widths to quantify the precision of a given photo-$z$. Furthermore,
we do not see strong trends in photo-$z$ quality with galaxy type or
environment, and the outlier fraction is typically no larger than a
few percent. In particular, the photometric depth in many bands and 
the treatment of emission lines in fitting SEDs by
\citet{Ilbert2009} appears to balance the weakening $4000$\AA\ break
for bluer galaxies, so that photo-$z$ quality does not significantly
depend on color. The lack of strong variations in photo-$z$ uncertainties, and 
the agreement between the two measures of photo-$z$ uncertainties across
galaxy types and environments demonstrates the robustness of these
redshifts for different populations. We have not included the photo-$z$s
from \citet{Salvato2009} for AGN due to their rarity and the reasonable
accuracy of the photo-$z$s of \citet{Ilbert2009} for these sources, but
future work focusing on AGN may benefit from the improved redshift
accuracy.

\begin{deluxetable*}{lrcccccc}
\tablecaption{Photo-$z$ quality for galaxies with spectroscopic redshifts}
\tablehead{\colhead{Sample} & \colhead{$N_{obj}$} &
  \colhead{$\langle z\rangle $} & \colhead{$\langle \rm{F814W} \rangle$} &
  \colhead{med$(\zs-\zp)$} &
  \colhead{$\sigma_{\Delta z}$\tablenotemark{a}} &
  \colhead{$\sigma_{\mathcal{P}}$\tablenotemark{b}} & \colhead{$\eta(\%)$\tablenotemark{c}} }
\startdata
                 All &  12370 & 0.52 & 21.3 &  0.003 &  0.012 &  0.013 & 1.1 \\
\hline
       Bright; low z &   5768 & 0.31 & 20.7 &  0.001 &  0.009 &  0.011 & 1.0 \\
      Bright; high z &   5613 & 0.71 & 21.6 &  0.006 &  0.015 &  0.014 & 1.0 \\
        Faint; low z &    280 & 0.27 & 23.0 & -0.002 &  0.014 &  0.016 & 3.9 \\
       Faint; high z &    709 & 0.86 & 23.0 &  0.007 &  0.027 &  0.024 & 2.7 \\
\hline
        Bright; blue &   6840 & 0.52 & 21.3 &  0.002 &  0.011 &  0.013 & 1.1 \\
       Bright; green &   2548 & 0.44 & 21.0 &  0.006 &  0.014 &  0.013 & 1.2 \\
         Bright; red &   1993 & 0.55 & 20.6 &  0.003 &  0.010 &  0.011 & 0.4 \\
         Faint; blue &    731 & 0.70 & 23.1 &  0.000 &  0.021 &  0.021 & 3.8 \\
        Faint; green &    259 & 0.64 & 23.2 &  0.011 &  0.028 &  0.027 & 2.7 \\
          Faint; red &     67 & 0.92 & 23.1 &  0.013 &  0.031 &  0.020 & 7.5 \\
\hline
  Bright; early type &   2450 & 0.53 & 20.4 &  0.004 &  0.011 &  0.011 & 0.8 \\
   Bright; late type &   7248 & 0.49 & 21.3 &  0.003 &  0.012 &  0.013 & 0.7 \\
   Bright; irregular &   1444 & 0.59 & 21.3 &  0.002 &  0.012 &  0.012 & 2.1 \\
\hline
   High stellar mass &   4216 & 0.62 & 20.8 &  0.005 &  0.013 &  0.012 & 1.1 \\
    Low stellar mass &   8154 & 0.47 & 21.5 &  0.002 &  0.012 &  0.013 & 1.2 \\
\hline
         Near groups &    961 & 0.45 & 20.6 &  0.002 &  0.011 &  0.012 & 0.9 \\
      Outside groups &   9691 & 0.54 & 21.4 &  0.003 &  0.012 &  0.013 & 1.2 \\
\hline
       Clean regions &  10987 & 0.53 & 21.3 &  0.003 &  0.012 &  0.013 & 1.2 \\
      Masked regions &   1439 & 0.51 & 21.2 &  0.003 &  0.013 &  0.012 & 2.5 \\
\hline
  MMGG$_{\rm scale}$ &    126 & 0.49 & 19.3 &  0.000 &  0.009 &  0.010 & 0.0 \\
                 AGN &    229 & 0.56 & 20.3 &  0.003 &  0.015 &  0.011 & 1.3
\enddata
\tablenotetext{a}{NMAD $=1.48\times$~median$[|\zs-\zp|]$}
\tablenotetext{b}{$1.48\times$~median$[|68\%$ uncertainty on photo-$z$ PDF$|]$}
\tablenotetext{c}{Fraction of objects with $|\zs-\zp|/(1+\zs) > 0.1$}

\tablecomments{Brightness bins are
  divided at F814W=22.5 which is the limiting magnitude for
  zCOSMOS; redshift bins are split at $z=0.5$; color bins are
  $M(NUV)-M(R) < 1.2$ (blue), $1.2 < M(NUV)-M(R) < 3.5$ (green), and
  $M(NUV)-M(R) > 3.5$ (red); 
  morphologies are categorized by ZEST; stellar masses are separated at
  $\log(M_{\star}/\msun)=10.5$; group environments are classified as
  ``near" within \rvir\ of an X-ray group center and where
  $|z_s-\zG|/(1+\zG) < 0.005$, and ``outside"
  beyond $3 \rvir$ and where $|z_s-\zG|/(1+\zG) 
  > 0.01$. Masked regions are areas in the optical images with bright
  foreground stars, satellite trails, or image defects. MMGG$_{\rm
    scale}$ are the most massive group galaxies within 
  an NFW scale radius of the X-ray center (see
  \S~\ref{s:centers}). AGN have been identified in {\sl
    Chandra} X-ray data \citep{Elvis2009}.\label{t:photoz_tests}}
\end{deluxetable*}

\begin{figure*}[htb]
\epsscale{1.0}
\plotone{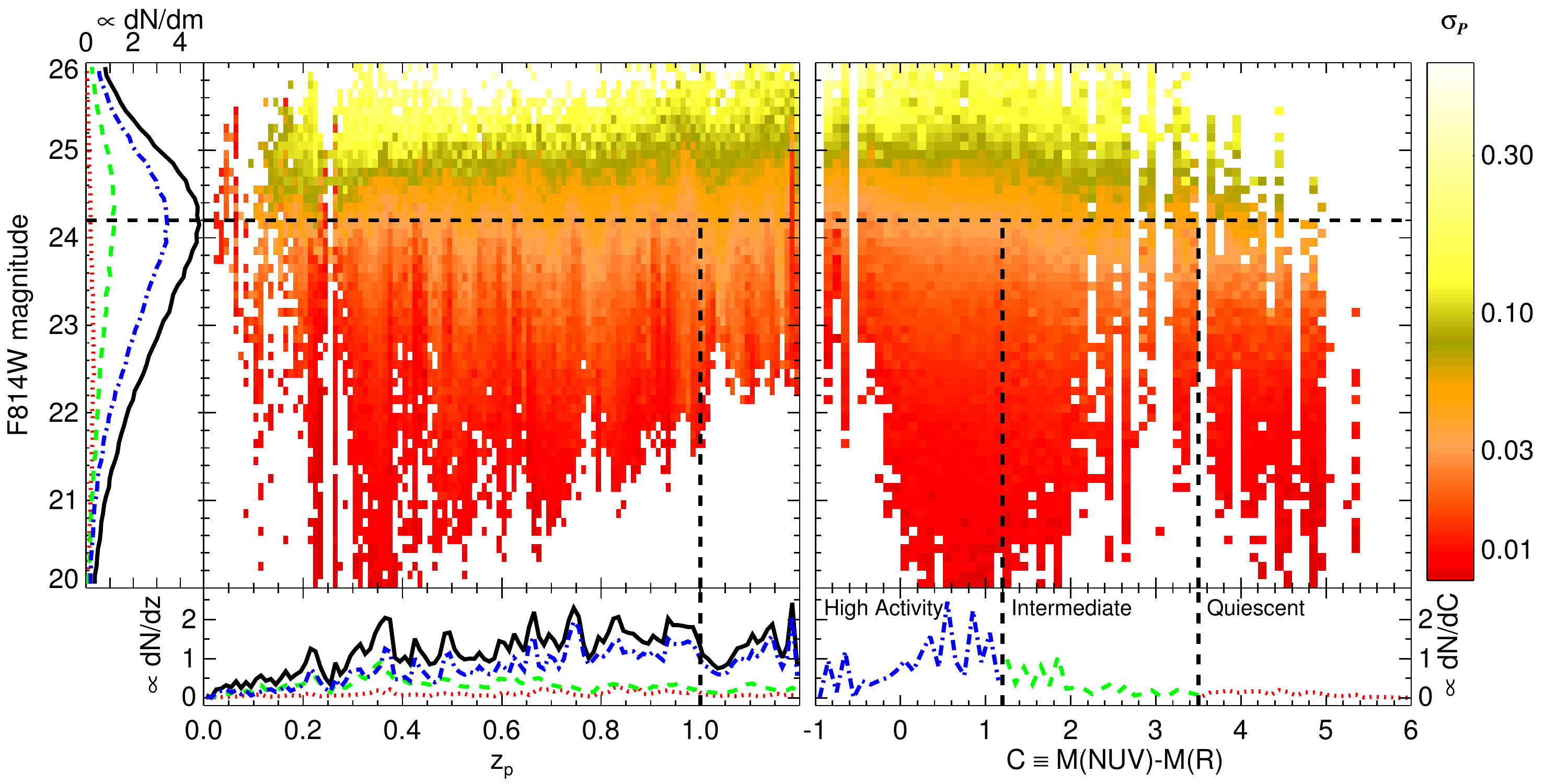}
\caption{Photo-$z$ uncertainties ($\sigma_{\mathcal{P}}$) as a function
  of magnitude and redshift (left), and color (right). Main panels
  show photo-$z$ uncertainty in bins colored according to the scale at
  right. Margin plots on the left side, bottom left, and bottom right show the magnitude,
  redshift, and color distributions, respectively. Curves are
  separated by color classification showing all galaxies considered
  (solid black), ``high activity'' galaxies (blue dot-dashed),
  ``intermediate activity'' galaxies (green dashed), and 
``quiescent'' galaxies (red dotted). Dashed black lines show the
galaxy magnitude and group redshift cuts at F814W=24.2 and $\zG=1$ for the
sample, as well as the divisions between color types. Ordinate axes on margin
plots should be multiplied by $10^4$ (left side) and $10^5$ (bottom)
for normalization. We do not see strong variations in photo-$z$ precision
with redshift or color, but there is a significant magnitude-dependence.}
\label{fig:zerr}
\end{figure*}

While the photo-$z$ accuracy is good across the sample, the quality does
decrease at fainter magnitudes. We account for this effect when
selecting member galaxies by allowing larger tolerances in redshift
space for fainter sources. There is also some degradation at higher redshift,
but since our sample is not as heavily weighted toward high redshifts
as it is toward faint magnitudes, we do not currently account for the
redshift dependence of photo-$z$ accuracy when selecting group members.
We note that Table~\ref{t:photoz_tests} shows mean magnitudes and
photo-$z$ errors for objects that also have spectroscopic redshifts;
these errors are representative of the PDF uncertainties for the full
galaxy sample in the bright bins, but in the faint bins the
spectroscopic sample is brighter than the full population and
photo-$z$ errors are smaller than average.

Figure~\ref{fig:zerr} illustrates how the photo-$z$ PDF uncertainty varies with
magnitude, redshift, and color for the full galaxy sample. The parameter space is divided into
bins of $\Delta\zp=0.01$, $\Delta \rm{F814W}=0.1$, and $\Delta C=0.1$. For
bins containing at least 10 galaxies, the half-width of the median
$68\%$ uncertainty on \pz\ is computed and plotted according to the
color scale shown. Additionally, we plot the redshift, magnitude, and
color distributions of galaxies to characterize the catalog. Clearly
the strongest trend in photo-$z$ precision is the decrease in quality at
faint magnitudes, and there is only a weak dependence on redshift and galaxy
color. Where the group member selection
algorithm requires an estimate of redshift uncertainty, we consider
only the magnitude dependence of the photo-$z$ uncertainties, ignoring
the smaller variations due to color and redshift.


\section{Group Membership Selection}
\label{s:membership}
\subsection{Overview}

This is not a paper about \textit{finding} galaxy groups; instead our aim is to
associate galaxies with groups that have already been identified as
extended X-ray sources. Our basic strategy is to take the locations of
groups from the X-ray catalog described in \S~\ref{s:xray} and
\citet[][and in prep.]{Finoguenov2007} and assign galaxies
to groups based on their positions and redshifts. Previous work on
finding group and cluster members has often included assumptions about
properties such as their red sequence content, luminosity function, and radial
distribution. Because galaxy group populations have not been
well-characterized in the mass and redshift range probed by this data 
set, we do not apply such filters to select members, with the hope
that we can then measure these properties in an unbiased manner.

Effectively, we are selecting galaxies in a cylinder oriented along
the line of sight around the X-ray position and redshift for each
group. The radius chosen for this cylinder is the estimated \rvir\ of
each group based on the total mass derived from the X-ray luminosity
versus \mvir\ relation for the group sample as determined by weak
lensing \citep{Leauthaud2010}. The depth of the cylinder
in redshift space is allowed to vary for each candidate member galaxy
according to the typical photo-$z$ uncertainty for its apparent
magnitude (see Fig.~\ref{fig:zerr}).

Photometric redshift uncertainties are larger than the typical intrinsic span
of a galaxy group in redshift space. A typical photo-$z$ error of
$\sigma_{\mathcal{P}} = 0.01$ in redshift-space corresponds to an uncertainty
of roughly $40~\rm{Mpc}$ in distance along the line of sight, while a halo with
$\log(\mvir/M_{\odot})=13.5$ has a velocity dispersion of $\Delta\zG
\approx 0.001$ \citep{Evrard2008} or a line of sight distance
uncertainty of roughly $4~\rm{Mpc}$ at $z=0$. As a result, we must 
account for contamination of the member sample by galaxies at a similar
redshift and position that do not belong to the group. One option is
to subtract a mean background density from the number of galaxies
found near the group. This statistical background subtraction can be
extended to other quantities of interest, such as the total stellar
mass in a group, by measuring those quantities averaged over regions
away from the group and subtracting them from the values measured
at the position of the group. One is left with the measured aggregate quantities
for each group, but not a clear list of members and non-members. 
Another approach is to assign each galaxy a membership probability
reflecting the likelihood that it belongs in a group, given some
information about the relative number of field galaxies and group
members. One can then determine properties of the group by selecting
members above a given probability threshold, or by weighting members
according to their probability of being a member.

We adopt this Bayesian approach to produce a group member
catalog, which can in turn be used to measure a variety of properties
about each group without requiring a new statistical background
subtraction for each quantity. The selection algorithm thus assigns a
probability of membership in a particular group to each galaxy given a
number of observables: the projected separation of the group and
galaxy in units of the group radius, the redshifts of the galaxy and
group along with the typical photo-$z$ uncertainty for the magnitude of
the galaxy, and an estimate of the number density of field galaxies
relative to group members. Additionally, stellar masses are used to
select a central galaxy from the membership list, refining the somewhat
uncertain X-ray positions (see \S~\ref{s:centers}).

\subsection{Algorithm}
 
In this section, we explain in detail how our selection algorithm
works. We reiterate that our task is to identify galaxies that belong
to groups rather than to find groups themselves. Our use of photo-$z$
PDFs to associate galaxies to known groups and clusters is similar to
the method outlined by \citet{Brunner2000}; we extend this method to
incorporate varying photo-$z$ errors and a prior on the relative
fractions of galaxies in groups and the field. The approach presented
here was designed with COSMOS data in mind, but may be applicable to
other multi-wavelength group and cluster studies, such as optical
imaging surveys in fields with SZ or X-ray data. In
\S~\ref{s:tests}, we consider the quality of our resulting member
catalog and how it could be modified by these different data sets. We
attempt to keep the discussion here general while inserting details
specific to the COSMOS data when necessary. To find the center of a group, we
start with the X-ray centroid and then refine this position using
the most massive member galaxy near the X-ray position, and finally we
update the member list around the new central galaxy (more details on centering are
presented in \S~\ref{s:centers} and Paper II.)

\begin{figure*}[htb]
\epsscale{0.73}
\plotone{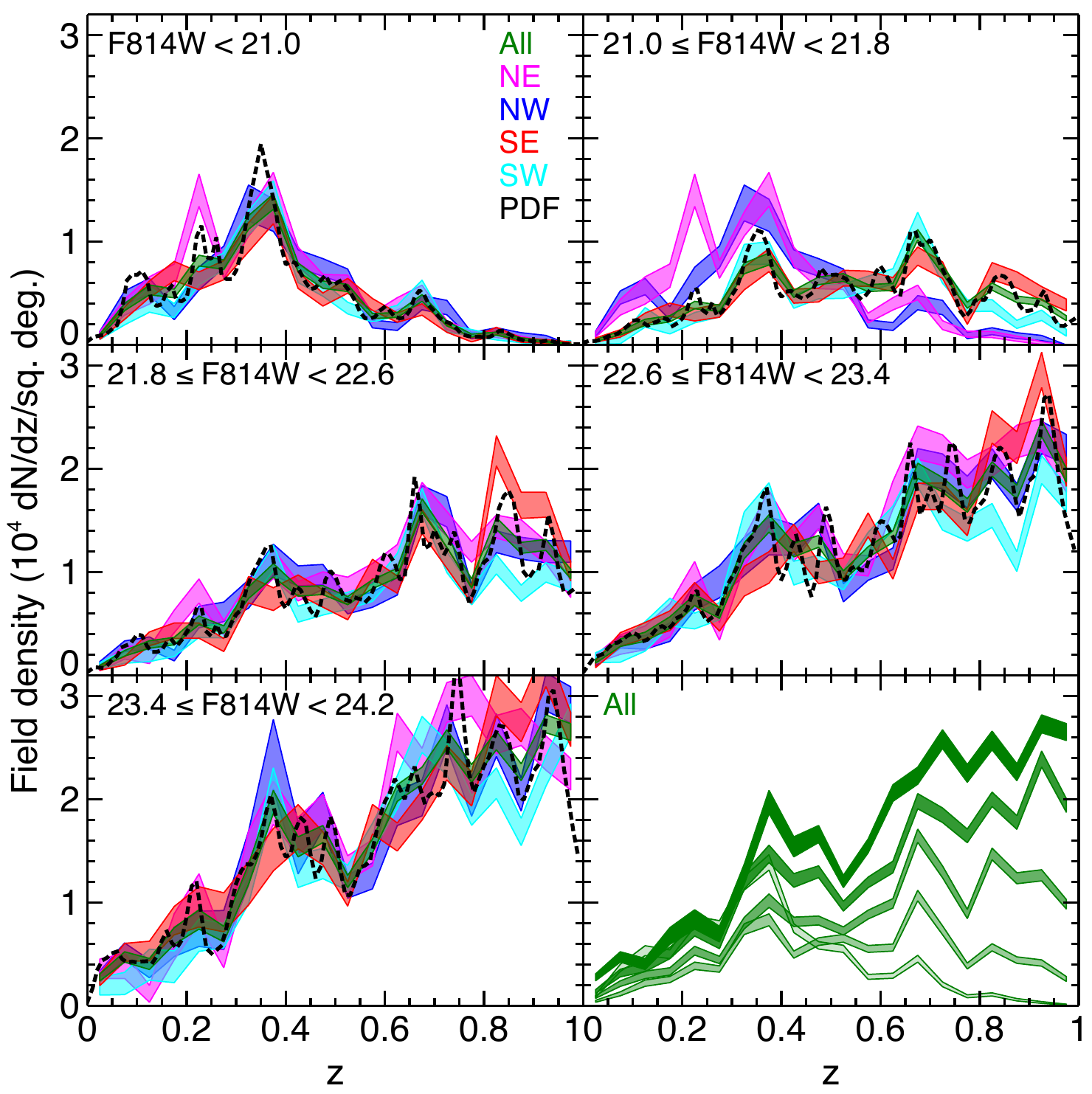}
\caption{Field density as a function of position, redshift, and
  magnitude. Different colored curves correspond to quadrants of the COSMOS
  field. The thickness of each curve corresponds to the Poisson
  uncertainty in each measurement. The dashed black line gives the
  stacked $\mathcal{P}(z)$ for all galaxies sampled at redshift
  intervals of 0.01, agreeing nicely with the measurement from
  counting galaxies in photo-$z$ intervals of 0.05. The variation between
  quadrants is not large, so we use the mean density across the whole
  field, shown by the green curve in each of panel and repeated in the
  bottom right panel for each magnitude bin.}
\label{fig:fielddensity}
\end{figure*}

We first consider the field galaxies that can contaminate our selection.
The background density of galaxies varies with position, redshift, and
magnitude. We measure the number of galaxies in redshift bins
($\Delta z=0.05$) and magnitude bins (starting at F814W~$<21.0$, then
using a width of 0.8 mag, and ending at $23.4<\rm{F814W}<24.2$). This count excludes
the volume within $3\rvir{}$ and $z_{\rm{G}}\pm 5\sigma_{\mathcal{P}}(\bar{m})$
around all groups in the catalog regardless of flags,
where $\bar{m}$ is the mean magnitude of galaxies in the bin. The
final results are not strongly sensitive to the choice of volume removed around
groups. Figure~\ref{fig:fielddensity} shows this field density
$n_F(\rm{F814W},z)=dN_F/dz/d\Omega$, which is similar to the quantity
shown in the bottom left panel of Figure~\ref{fig:zerr} but split
into magnitude bins. Figure~\ref{fig:fielddensity} also shows the
field density as computed by summing the redshift probability
distribution functions of galaxies for comparison to the approach of
directly counting galaxies in photo-$z$ bins. Despite different sampling
intervals ($\Delta z=0.01$ for the PDFs and 0.05 for bin counting),
the methods show excellent agreement.

One can measure the background density locally around groups to
account for correlated structure or globally across the field to
increase the statistical sample with a larger volume, reducing
noise. We have divided the COSMOS field into four separate quadrants
to look for variations in $n_F$ with position and find that
the values are in reasonable agreement across the field, with
the density in individual quandrants deviating from the mean by typically
no more than the Poisson errors. When smaller volumes are chosen to estimate
the field density surrounding groups, the increased Poisson
uncertainty swamps the constraint on locally correlated structure. We
thus opt to use the entire area to estimate the mean density of
background galaxies as a function of magnitude and redshift. We
discuss further the choice of this method of estimating the field density
in \S~\ref{s:error}.

We next consider candidate member galaxies, constructing a list of
those objects in a cylinder with a projected distance from the group center less
than \rvir\ and a redshift within $3\sigma_{\mathcal{P}}(m_{\rm max})$ of the
group redshift \zG, where $m_{\rm max}$ is the limiting magnitude
F814W=24.2 and $\sigma_{\mathcal{P}}=0.035$. The number of member
candidates can be compared with the field density in
Figure~\ref{fig:fielddensity} to estimate the fraction of galaxies
that are group members.

For each candidate, we compare the photo-$z$ PDF to the expected
redshift distributions of group members and field galaxies.
We assume that each galaxy is either a group member (G) or part of the
field (F), and assign a Bayesian membership probability using the
relative sizes of the group and field populations as a prior to
normalize the distributions. While the intial $3\sigma_{\mathcal{P}}$
cut uses the photo-$z$ value \zp\ to make a rough selection, here we
use the full distribution \pz\ for each galaxy to account for
secondary peaks or other unusual features in the redshift PDF. The
probability that a galaxy belongs to a group given \pz\ can be written as 
\begin{equation}
\label{eq:bayes}
P(g \in G|\pz) = \frac{P(\pz|g \in G) P(g \in G)}{P(\pz)}.
\end{equation}
The term $P(\pz|g \in G)$ is the likelihood of measuring the particular
photo-$z$ PDF for a known group member. The prior $P(g \in G) =
N_{G}/(N_{G}+N_{F}) = 1-P(g \in F)$ is based on the relative number of
group and field galaxies in the cylinder, and 
\begin{eqnarray}
P(\pz) &=& P(\pz|g \in G) P(g \in G) + \nonumber \\
          &  & P(\pz|g \in F) P(g \in F)
\end{eqnarray}
is the probability of measuring \pz\ for any galaxy in the group or
field. Each factor in Equation~\ref{eq:bayes} has an implicit
dependence on magnitude which we omit here and in following equations for notational
simplicity, but we do account for magnitude-dependent variations in
\pz\ and in the field and group densities.

In order to compare the observed \pz\ with that expected for a group
or field galaxy, we must assume a distribution of redshifts for each
population. Since the intrinsic velocity dispersion of groups is
smaller than the uncertainty in \zp\, we model the true group redshift
distribution as a $\delta$-function at \zG, which is then convolved
with a Gaussian of width $\sigma_{\mathcal{P}}(m)$ to account for
photo-$z$ measurement uncertainty. We have tested the effects of
modifying the true group redshift distribution to be  broader than a
$\delta$-function to account for intrinsic velocity dispersion but
found this correction to be negligible. The redshift distribution
of field galaxies is assumed to be uniform near \zG\ and remains
unchanged after accounting for photo-$z$ measurement uncertainty. Each of
these redshift distributions is convolved with the photo-$z$ PDF \pz\
(note that $\int{\mathcal{P}(z)dz}=1$), giving
\begin{eqnarray}
P(\pz|g \in G) &=& \int{\mathcal{P}(z) \mathcal{N}(\zG,\sigma_{\mathcal{P}}) dz}  \\
P(\pz|g \in F) &=& \int{\frac{\mathcal{P}(z)}{w(\sigma_{\mathcal{P}})}dz}
\end{eqnarray}
where $\mathcal{N}(\zG,\sigma_{\mathcal{P}})$ is a Gaussian centered on the
group redshift with width equal to the typical \pz\ uncertainty for the
magnitude of the galaxy considered. The field density distribution is
normalized so that the integral over the redshift range $\zG \pm
3\sigma_{\mathcal{P}}$ is unity, so the width normalization parameter is
$w(\sigma_{\mathcal{P}}(m))=6\sigma_{\mathcal{P}}(m)$. We have written 
these convolutions as indefinite integrals, but in reality they are
discrete sums sampled at the redshift intervals $\Delta z=0.01$ and range $0
\le z \le 6$ for which \pz\ has been calculated. Because \pz\ is
sampled at intervals close to the typical 
photo-$z$ uncertainty, the distribution can effectively become a
$\delta$-function, underestimating the true redshift error which
has contributions from template uncertainties as well as photometric
uncertainties. So we first convolve \pz\ with a Gaussian of width
d$z=0.01$ to account for these uncertainties and avoid sharply peaked
PDFs.

To estimate the prior, $P(g \in G)$, we begin by counting the number
of galaxies in the range $\zG \pm 3\sigma_{\mathcal{P}}(m)$, measuring
$N_{tot}=N_{G}+N_{F}$. The measurement of the field density shown in
Figure~\ref{fig:fielddensity} provides an independent estimate of
$n_{F}$, which allows us to calculate an expected number of field
galaxies in the cylinder, $\hat{N}_F=\int{n_F dz d\Omega}$. For each
galaxy we linearly interpolate the curve in the relevant magnitude bin
to the group redshift, and multiply $n_F$ by the volume searched
around the group, $6\pi\rvir^2\sigma_{\mathcal{P}}(m)$, to determine
$\hat{N}_F$. This value is subtracted from the measured $N_{\rm tot}$ to
determine the expected number of group galaxies in the cylinder,
$\hat{N}_{G}$. We use the estimated values, $\hat{N}_{F}$ and
$\hat{N}_{G}$, to determine $P(g \in G)$ and $P(g \in F)$, and
Equation~\ref{eq:bayes} assigns each galaxy a membership probability
between zero and one. In cases where a group is not well-detected in a given
magnitude bin $(N_{\rm tot}<\hat{N}_{F}$, \ie{} $\hat{N}_{G}<0)$, galaxies
in the bin are flagged and excluded from membership
analysis. Tests in \S~\ref{s:tests} show that excluding these galaxies
does not cause significant incompleteness in the member selection.

It is possible for the search cylinders of different groups to
overlap, either because they reside in neighboring positions at the
same redshift, or because of projections along the line of sight
within the redshift uncertainties. In cases where a galaxy is a
candidate member of multiple groups, each probability is
recorded. A total of 4631 galaxies are assigned high 
probabilities of membership $(P_{\rm mem} \equiv P(g \in G|\pz) > P(g
\in F|\pz)$, i.e. $P_{\rm mem} > 0.5)$ in a group, and of these members only 163 or $3.5\%$ are also
assigned to a second group. For most applications we can restrict our 
analysis to the highest group membership probability for each galaxy
without any significant change in results, but recording each
probability assignment will aid in the study of merging groups.

\subsection{Group Centers}
\label{s:centers}

The robust identification of central galaxies is a challenging task,
and relevant for a range of applications from satellite kinematics to
stacked weak lensing to studying the most massive galaxies
\citep[e.g.,][]{Skibba2011}. Miscentering is a significant source of
systematic uncertainty in measuring the richness and weak lensing
signal in optical groups \citep[e.g.,][]{Johnston2007, Rozo2011,
  Rykoff2011}. X-ray data and weak lensing offer additional
information about the centers of mass of halos, which we use along
with the galaxy content to guide our selection. We outline our approach to
determining the optimal tracer of the center of mass here, and present
our results in further detail in Paper II.

We use the X-ray position as an initial approximation of a group's
center, but for these faint detections the position can be uncertain by
up to the wavelet detection scale of 32\arcsec  ($\sim200~\rm{kpc}$ at
$z=0.5$), so we consider other data to improve upon these constraints
on the centers. Briefly, we have defined multiple candidate 
centers based on luminosity, stellar mass, and proximity to the X-ray
center. By measuring the weak gravitational lensing signal stacked
around each of these positions we can find the optimal center which
maximizes the lensing signal at small radii. Our results indicate that
this optimum center is the member galaxy (i.e., $P_{\rm mem} > 0.5$)
with the highest stellar mass within 
the scale radius plus the X-ray positional uncertainty of the X-ray
center. We refer to this object as the MMGG$_{\rm scale}$, for Most
Massive Group Galaxy within the scale radius. We assign this
galaxy to be the group center, and rerun the algorithm above to find
members within \rvir\ of this galaxy for the final catalog. 

Traditional visual selection of group and cluster centers includes
looking for a bright, usually early type galaxy near the center of the
X-ray or optical distribution, perhaps with an extended stellar
envelope. Visual inspections of the Subaru, ACS, and XMM data
support our objective selection, with broad agreement between the MMGG$_{\rm
  scale}$ and the objects one would traditionally identify as central
galaxies. Visual selection becomes more ambiguous at high
redshift and for groups lacking dominant galaxies, while our selection
algorithm makes an objective choice. In a few percent of
cases the MMGG$_{\rm scale}$ disagrees with a visually identified
central galaxy due to photo-$z$ error or because of a significant
offset from the X-ray position putting it outside the scale radius. We
do not amend these cases, sacrificing a small degree of accuracy for a
uniform and objective selection. 

The selection of group centers used here is different than in
\citet{Leauthaud2010}, which employed a weighting based on stellar 
mass and distance to the X-ray position. Of the groups that have a
confident central galaxy assignment from \citet{Leauthaud2010} and also
satisfy the quality cuts for clean groups in \S~\ref{s:sample},
$80\%$ are assigned the same central galaxy by the two methods, $9\%$
of the centrals identified by \citet{Leauthaud2010} are too distant
from the X-ray center for our method to select, and $4\%$ are not
identified as members with the current algorithm. In these cases of
disagreement, the selection of \citet{Leauthaud2010} tends to favor
more massive galaxies that are farther from the X-ray centroid than the
selection used here, with average differences of 0.2 dex in stellar
mass and $55~\rm{kpc}$ in distance to the X-ray centroid.


\section{Purity and Completeness}
\label{s:tests}

Any selection of group members will have some fraction of false
positives, interlopers selected as members that do not belong to a
group, and false negatives, true member galaxies missed by the
selection. To measure properties of member galaxies, we can weight
each galaxy by its membership probability to account for these
uncertainties. But we must test the reliability of those membership
probabilities, and furthermore, for some applications we wish to
define a set of galaxies exceeding a membership probability threshold
with a reasonable degree of purity and completeness.

Purity and completeness are measures of overlap between the sample of
selected members and the population of true members. We define the purity
of the sample, $\mathsf{p}$, to be the fraction of selected members
which are also true members. The completeness of the sample,
$\mathsf{c}$, is the fraction of true members which are
selected. Interlopers are objects which are selected but are not true
members and missed galaxies are objects which are not selected but are
true members. Formally, 
\begin{eqnarray}
\label{eq:purity} 
\mathsf{p} &=& \frac{N_{\rm selected}-N_{\rm interlopers}}{N_{\rm selected}} \\
\mathsf{c} &=& \frac{N_{\rm true}-N_{\rm missed}}{N_{\rm true}}.
\label{eq:completeness}
\end{eqnarray}
We can use the values of $\mathsf{p}$ and $\mathsf{c}$ to estimate
$N_{\rm true}$ using $N_{\rm true}=N_{\rm selected}+N_{\rm
  missed}-N_{\rm interlopers}$ which can be rearranged into 
\begin{equation}
\frac{N_{\rm true}}{N_{\rm selected}}=\mathsf{\frac{p}{c}}
\label{eq:correction}
\end{equation}
using Equations~\ref{eq:purity} and~\ref{eq:completeness}. This
correction factor, $\mathsf{p}/\mathsf{c}$, can be used to remove bias in the estimate
of the intrinsic number of group members, $N_{\rm true}$, if we understand the purity and
completeness of the selection algorithm.

To measure the purity and completeness of our member selection, we
must have some way of telling which galaxies truly belong to
groups. For our application, we use the subsample of objects with
spectroscopic redshifts as well as mock catalogs to obtain knowledge
of group membership that is independent of our photo-$z$ selection. 
The galaxies with spectroscopic redshifts allow us to test the photo-$z$
selection method on the same catalog, directly probing the effect of
photo-$z$ uncertainties. But constraints on $\mathsf{p}$ and $\mathsf{c}$
are limited by the sparseness of spectroscopic coverage, and biases
could be introduced since spectroscopic coverage is not representative
of the full range of galaxies in the group sample. Furthermore, even
spectroscopic selection of group members can have contamination and
incompleteness \citep[e.g.,][]{Gerke2005}.

We perform further diagnostic tests using mock catalogs from N-body
simulations described in \S~\ref{s:mocks}. Mock galaxies are prescribed to occupy halos according to
a halo occupation distribution (HOD) model constrained by measurements
of clustering, lensing, and stellar mass functions in COSMOS
\citep{Leauthaud2011a, Leauthaud2011b}. After running 
the selection algorithm on a mock catalog, we can estimate its purity
and completeness by comparing the results with the input list of group
members. The mocks allow us to study
greater volumes than the observed region, increasing statistical
precision and providing estimates of the effects of sample variance for
the volume probed. Mocks also give direct knowledge of galaxy group membership in real
space without the redshift space distortions that mar spectroscopic
selection, so we can study how the selection algorithm would perform
on data sets with different errors in redshifts or positions. However,
caution must be taken to ensure that the mock 
galaxies adequately represent the reality of correlated structure for
all relevant properties, particularly in their distribution of
positions, masses, and halo occupation.

In the following sections we describe in more detail our diagnostic
tests on the selection algorithm using spectroscopic redshifts and
mock catalogs. We begin by testing the tradeoff between purity and
completeness for different membership probability thresholds, and
proceed to study the principle sources of contamination and
incompleteness in our selection.

\subsection{Spectroscopic Tests}

Here we consider the subset of galaxies with spectroscopic redshifts
to measure the purity and completeness of the selection and study the
effect of photo-$z$ errors. A ``true'' member in this case is defined to
be a galaxy within \rvir\ of the X-ray center and with $c|\zs-\zG| <
2\sigma_{\rm v}(\mvir,z)(1+\zG)$, where $c$ is the speed of light and
$\sigma_{\rm v}(\mvir,z)$ is the velocity dispersion from the
simulations of \citet{Evrard2008}, assuming that the velocity bias between
galaxies and dark matter is unity.

\begin{figure}[htb]
\epsscale{1.25}
\plotone{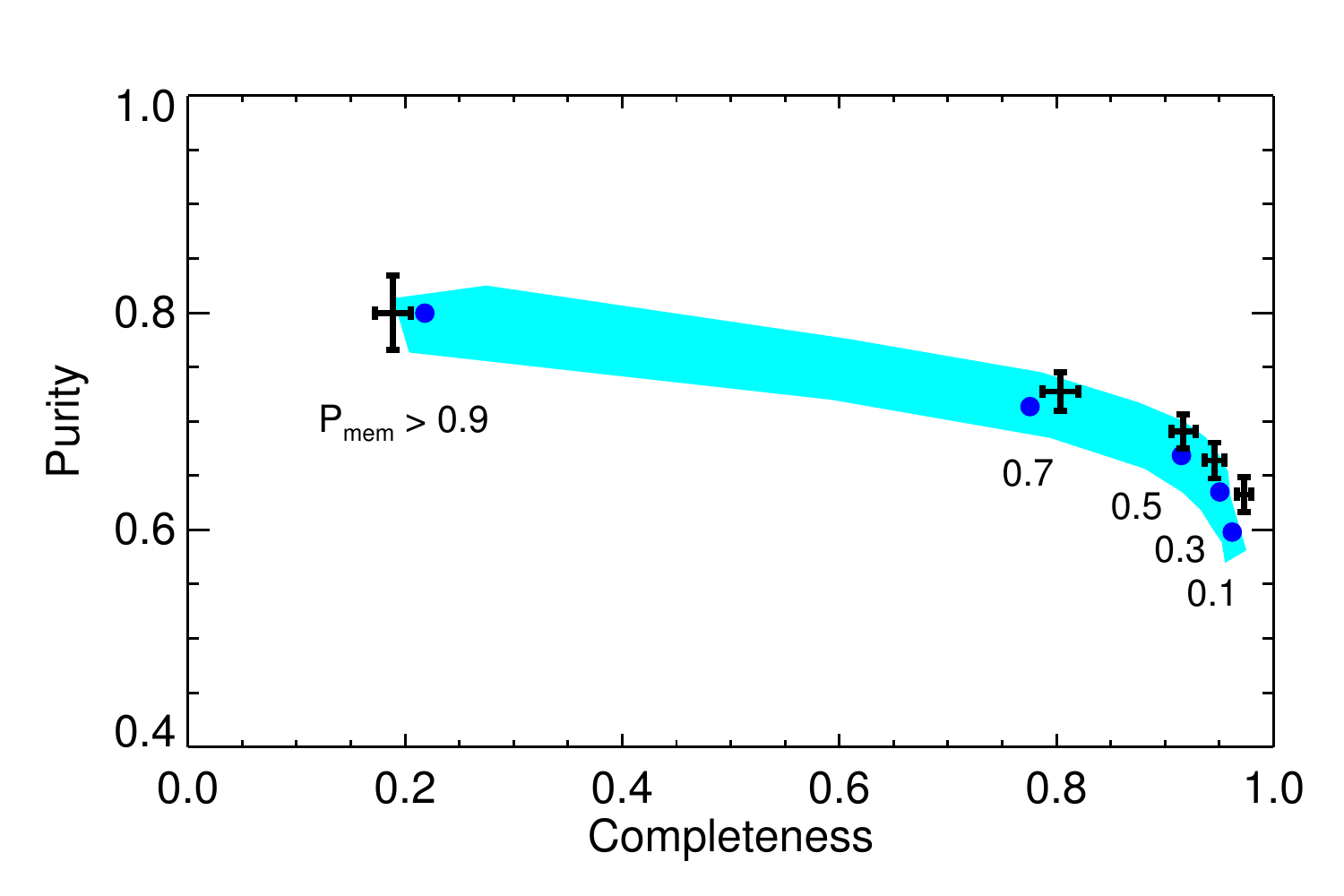}
\caption{Purity and completeness for different membership probability
  thresholds ($P_{\rm mem}>\{0.1,0.3,0.5,0.7,0.9\}$) as measured by the spectroscopic subsample
  (black crosses) and mock catalogs (blue circles, mean of ten lightcones). Error bars
  are the standard deviation from 1000 bootstrap samples of the spectroscopic catalog. The cyan
  shaded band is the region spanned by the ten mock lightcones.} 
\label{fig:memstat_threshold}
\end{figure}

As we vary the membership probability threshold for photo-$z$ selection,
we can see a tradeoff between purity and completeness shown in
Figure~\ref{fig:memstat_threshold} with black points from the
spectroscopic test. Error bars show the standard deviation of 1000
bootstrap samples of the spectroscopic catalog. Restricting the member
list to sources with membership probability $P_{\rm mem} > 0.9$ gives a purity and
completeness of $80\%$ and $19\%$ respectively. Lowering the
membership threshold increases completeness while decreasing
purity. In later sections, we use a threshold of $P_{\rm mem}>0.5$ as
a compromise between these competing factors, which for the
spectroscopic test produces a purity of $69\%$ and a completeness of
$92\%$. 

\begin{figure*}[htb]
\epsscale{1.0}
\plotone{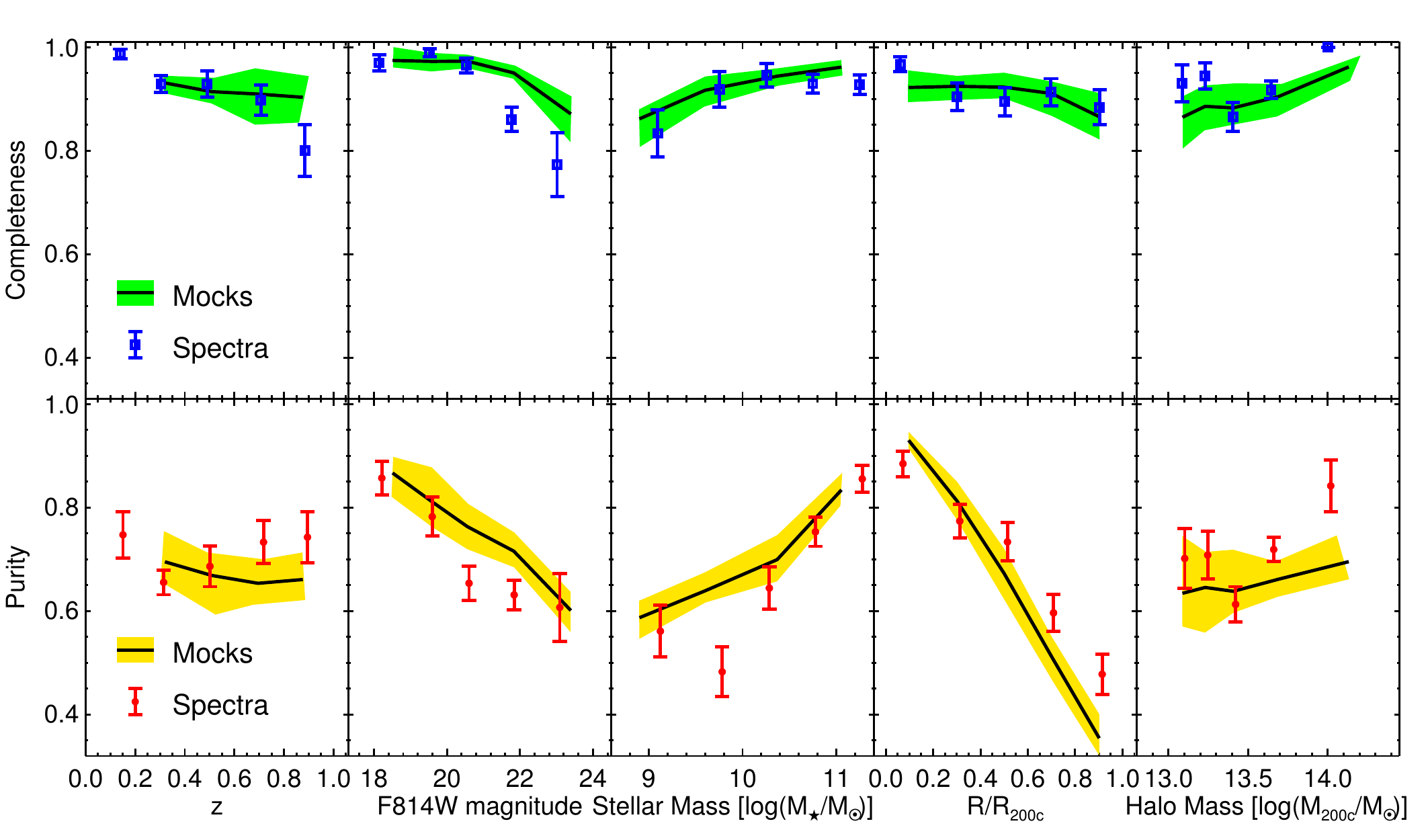}
\caption{Completeness (top row) and purity (bottom row) of the galaxy
  membership selection as measured by the
  spectroscopic subsample (points with error bars) and mock catalogs
  (shaded bands) for galaxies with $P_{\rm mem}>0.5$. Error bars are
  the standard deviation from 1000 bootstrap samples of the
  spectroscopic catalog, and shaded bands show the range spanned by the
  ten mock lightcones, while the solid black curve represents the mock
  mean. Bins were chosen to measure a roughly constant number of galaxies for each 
  property tested while still representing the range of observed properties.}
\label{fig:memstat_all}
\end{figure*}

\begin{figure}[htb]
\epsscale{1.08}
\plotone{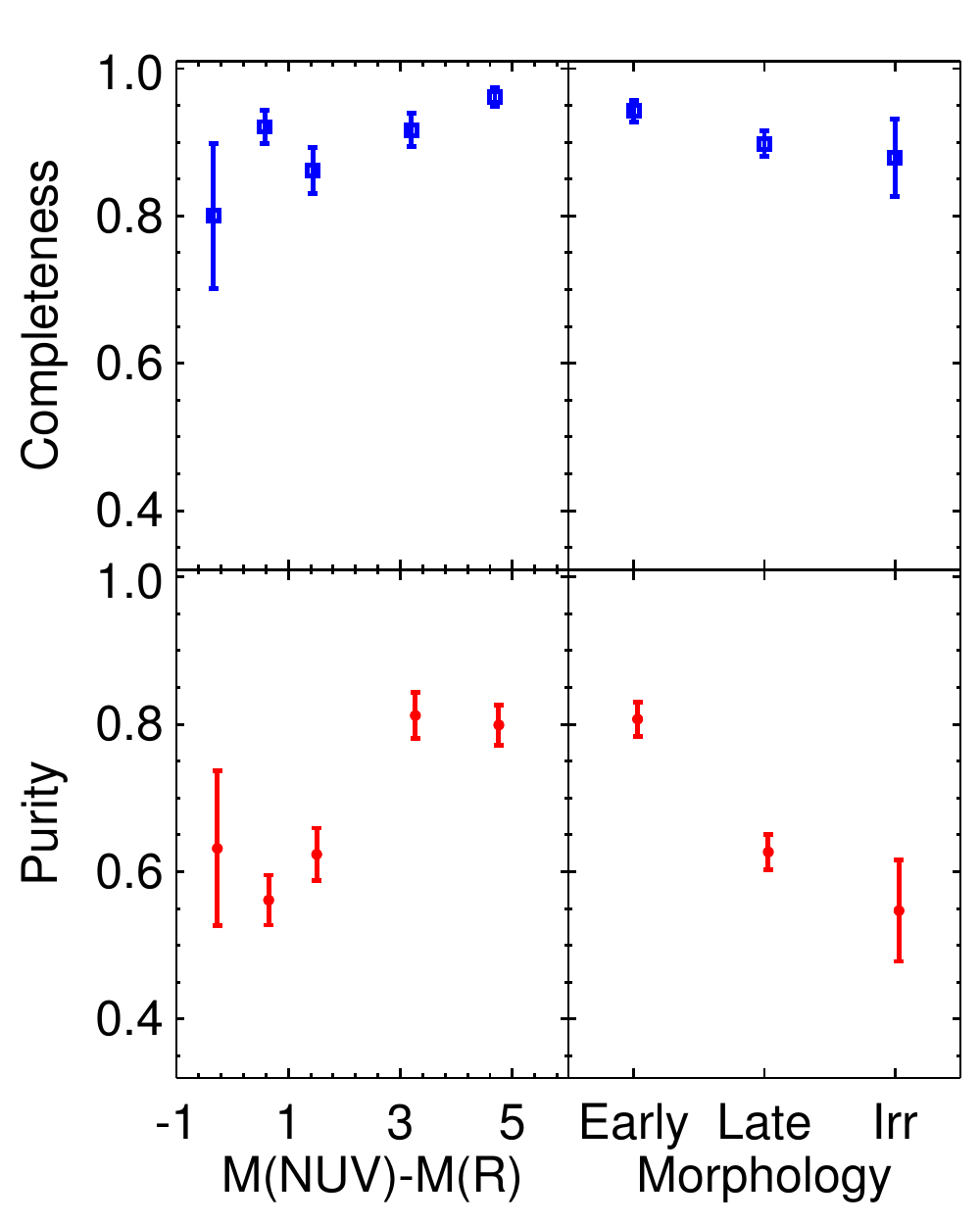}
\caption{Completeness (top row) and purity (bottom row) of the galaxy
  membership selection as measured by the
  spectroscopic subsample for galaxies with $P_{\rm mem}>0.5$. Error bars are
  the standard deviation from 1000 bootstrap samples of the
  spectroscopic catalog. Morphological classes are defined from ZEST
  \citep{Scarlata2007}; the early type category includes ellipticals
  (type=1) and bulge-dominated disks (type=2.0), the late type
  category includes the remaining type=2 sources, and irregulars
  have type=3.} 
\label{fig:memstat_color}
\end{figure}

To further study the quality of the membership selection, we can
measure trends in purity and completeness against other properties,
seen in Figure~\ref{fig:memstat_all}. In this figure, we show how the
selection performs for galaxies of different redshift, magnitude,
stellar mass, group-centric distance, and group halo mass, by
measuring the purity and completeness of objects assigned $P_{\rm
  mem}>0.5$. Figure~\ref{fig:memstat_color} shows
the same tests for color and morphology. The results are discussed in more detail in 
\S~\ref{s:error}, but we can see that the selection quality does not
vary significantly with redshift or group mass, but does degrade in
the outskirts of groups and for faint, low-mass galaxies, which also
tend to have blue colors and late type morphologies. We have tested the
influence of target selection effects on 
these results by restricting the spectroscopic sample to
zCOSMOS galaxies which were uniformly selected at
$i^+<22.5$. The purity and completeness measurements are
consistent within the error bars of the full sample, but have slightly
larger uncertainties due to the smaller sample size.

\subsection{Mock Catalogs}
\label{s:mocks}

We use numerical simulations to construct a series of mock catalogs for a
COSMOS-like survey to test the reliability of our member
selection. Mocks are created from a single simulation (named 
``Consuelo''), part of the Las Damas suite (McBride et al., in prep.)\footnote{Details regarding this simulation can be found at {\url{http://lss.phy.vanderbilt.edu/lasdamas/simulations.html}}}. Consuelo
is a box of $420~h^{-1}~{\rm Mpc}$ on a side with 
$1400^3$ particles of mass $1.87 \times 10^9~h^{-1}~{\rm M}_{\odot}$
and a softening length of $8~h^{-1}~{\rm kpc}$.\footnote{We use
  $H_0=100~h~\rm{km~s^{-1}~Mpc^{-1}}$ in this paragraph only.} 
This simulation can robustly resolve halos with masses above
$\sim10^{11} h^{-1}~{\rm M_{\odot}}$ which corresponds to central
galaxy stellar masses of $\sim10^{8.5}~h^{-1}~{\rm M_{\odot}}$, well-matched
to our completeness limit of F814W=24.2 at $z=0.2$ (see
Figure~\ref{fig:mass_limits}).

We extract ten light cones from the Consuelo simulation that have the
same area as COSMOS and individually non-overlapping volumes. Halos
within the simulation are identified with a friends-of-friends (FOF) halo
finder \citep{Davis1985} with a linking length of $b=0.2$. For typical
halos in the mass range we consider, FOF masses and
spherical overdensity masses (defined within a radius where the mean
density is 200 times the background) typically agree within $\sim10-20\%$
\citep{Tinker2008}; we thus only convert from background to critical
overdensity to obtain \mvir{}.
Halos are populated with galaxies using the HOD model of
\citet{Leauthaud2011a, Leauthaud2011b} 
that simultaneously fits the stellar mass functions, galaxy clustering,
and galaxy-galaxy lensing signals of COSMOS. We adopt the $z \sim 0.6$
HOD model of \citet{Leauthaud2011b} with the following parameters from
Table 5 of that paper:
$\log(M_{1})=12.725$, $\log(M_{*,0})=11.038$, $\beta=0.466$,
$\delta=0.61$, $\gamma=1.95$, $\sigma_{\log \rm  M_{*}}=0.249$,
$B_{\rm cut}=1.65$, $B_{\rm sat}=9.04$, $\beta_{\rm cut}=0.59$, 
$\beta_{\rm sat}=0.740$, $\alpha_{\rm sat}=1$. Details regarding the
parameters in this HOD model can be found in
\citet{Leauthaud2011a}. As shown in \citet{Leauthaud2011b}, there is a
small amount of redshift evolution in this parameter set from $z \sim
0.2$ to $z \sim 1$. However, the redshift evolution should not have a
large impact on our assessment of the completeness and purity of the
group membership selection and so we neglect the redshift evolution of
the HOD in this work.

Galaxies are assigned cosmological redshifts as well as mock
spectroscopic redshifts which include the effect of peculiar
velocities from the velocity dispersion within halos. Photometric
redshifts are drawn from a Gaussian distribution centered around the
spectroscopic redshift with width equal to the photo-$z$ uncertainty for
that magnitude. A Gaussian \pz\ is then centered at \zp\ with the same
width, and sampled at the same redshift interval as the PDF for real
galaxies. We do not include catastrophic photo-$z$ errors which are
shown in Table~\ref{t:photoz_tests} to be a small fraction of the sample. 

The HOD model of \citet{Leauthaud2011a} assigns stellar
masses to mock galaxies but does not assign magnitudes or colors. In
order to apply a similar magnitude cut to the mock galaxies as used in
the selection algorithm, we assign F814W magnitudes to mock
galaxies. For each mock galaxy, we construct a galaxy sample from the
COSMOS data that is matched in redshift and stellar mass in bins of
$\Delta z=0.02$ and $\Delta\log(M_{\star}/M_{\odot})=0.2$. An F814W magnitude
is assigned to each mock galaxy by randomly drawing a magnitude from
the matched sample. We do not assign colors or morphologies to mock
galaxies since the dependence of these properties on redshift and
environment are not well-constrained. We will rely on our
spectroscopic sample in order to determine the completeness and purity
of the group membership selection as a function of color and
morphology instead of using mock catalogs.

Mock halos are given the redshift of the central galaxy and X-ray
luminosities according to the mean $L_{\rm X}-\mvir$ relation of
\citet{Leauthaud2010}. To mimic the position uncertainties of the
X-ray detections, \textsc{xflag} quality flags 1 or 2 are assigned randomly
in proportion to their appearance in the COSMOS group catalog. The nominal
group center is offset from the central galaxy with a Gaussian scatter of $32
\arcsec$ for \textsc{xflag}~=~2 halos which is reduced by the measured flux
significance for \textsc{xflag}~=~1 halos, and we assume a typical
$5\sigma$ flux measurement. The impact of centroiding errors is investigated in
\S~\ref{s:surveys}.

Next we run the membership algorithm described in
\S~\ref{s:membership} on the mock galaxy and halo
catalogs, associating galaxies with halos. We can perform the same
purity and completeness tests as with the spectroscopic sample above,
but this time we know the halo membership \textit{a priori}. The
results from these mock catalog tests are presented alongside those
for the spectroscopic subsample as colored bands in Figures~\ref{fig:memstat_threshold}
and~\ref{fig:memstat_all}.

\subsection{Sources of Error}
\label{s:error}

Results from the tests on spectroscopic data and mock catalogs above
can differ because the spectroscopic sample is weighted toward bright
objects and because our knowledge of true membership in the
spectroscopic data is limited by redshift-space distortions, while
membership in the mock catalogs is known by design. 
The general agreement seen in Figures~\ref{fig:memstat_threshold} and
\ref{fig:memstat_all} between these tests of membership quality is
encouraging, and it suggests that the biases are modest and that the mock
catalogs accurately represent the properties of real galaxies that we
wish to study. The normalization of the purity and completeness curves for the
spectroscopic test has a degree of freedom in the velocity
width used to determine whether a spectroscopic redshift is consistent
with a group redshift. We used the criterion $c|\zs-\zG| < 2\sigma_{\rm
  v}(M,z)(1+\zG)$ for spectroscopic membership; a broader velocity
range for the spectroscopic test would result in a higher measured
level of purity and lower completeness in the photo-$z$ selection, and
the converse holds for a smaller velocity range, shifting the curves
up or down. Though the \textit{absolute} measure of purity and
completeness in the spectroscopic tests holds some degree of
arbitrariness, the \textit{relative} trends shown in
Figure~\ref{fig:memstat_all} are in general agreement with the mocks,
with some offsets likely due to sampling bias and redshift-space
distortions. We study the effects of redshift-space distortions on
member selection in the limit of a completely spectroscopic survey in
\S~\ref{s:surveys}.

Information from the spectroscopic tests has the advantage that it
can probe member selection effects due to properties that cannot easily be
modeled (e.g., galaxy color and morphology), and these tests directly
measure the effects of 
photo-$z$ errors on our selection of galaxies in the same set of
groups. We see in Figure~\ref{fig:memstat_color} that the trends of
selection quality with color and morphology parallel the trends with
magnitude and stellar mass from Figure~\ref{fig:memstat_all}. 
We have shown in \S~\ref{s:photoz_tests} that photo-$z$ quality is
not strongly affected by color or morphology, and no other inputs
to our selection algorithm explicitly depend on these properties. We infer
that the lower completeness and purity seen for faint, low mass, blue, and
late type galaxies is driven by two effects; fainter galaxies have
larger photo-$z$ uncertainties, and galaxies in this population tend to live
outside of dense groups so that they are more likely to be
contaminants when selected.

Because only a fraction of objects have spectroscopic
redshifts, the uncertainties can be large. Tests with mock catalogs
alleviate this issue and provide an
estimate of the sample variance in our selection due to the finite
size of the COSMOS region. An additional advantage of the mocks is
that the central galaxy of  each halo is known, so we can test the
success rate for identifying these objects. We find that $77\%$ of
central galaxies are correctly identified as the MMGG$_{\rm scale}$
galaxies in the corresponding halos, $12\%$ are misidentified as
satellites because the central galaxy is not the most massive member
near the centroid, $5\%$ are misidentified as satellites because the
assigned centroid error puts the galaxy outside of the search region,
and only $5\%$ are assigned to neighboring groups or the field due to
photo-$z$ errors. While the HOD used to create the mocks allows for
satellite galaxies to be more massive than centrals due to scatter in
the relation between stellar mass and halo mass, the fraction of
groups where this occurs is sensitive to the parametrization of the
HOD model and is not well-constrained. The problem of identifying
group centers will be discussed in more detail in Paper II.

For the full sample of mock galaxies with $P_{\rm mem}>0.5$, we find a
mean purity of $67\%$ and completeness of $92\%$. Looking at
Figure~\ref{fig:memstat_all}, it is clear that the dominant source of
impurity comes from galaxies in projection near the outskirts of
groups. We can attribute this contamination to the fact that the
density of true members falls steeply as a function of distance from
group centers while our membership algorithm selects galaxies
uniformly out to \rvir. Faint galaxies are another source of impurity since their
photo-$z$ errors are larger than average. Galaxies with lower masses and
bluer colors are more common in the field than in dense environments
(see \S~\ref{s:discussion}), so a higher contamination fraction from
these populations is to be expected. There is also a slight dependence on halo 
mass, since the density contrast between the field and groups is
smaller for low mass halos, lowering the assigned membership
probabilities of candidate members and reducing the completeness of
the selection. These factors motivate the use of matched filters in
finding groups and clusters when the properties of their galaxy
populations are well-characterized; we have not employed such filters
to avoid biasing our sample and because galaxy properties in this
range of halo masses and redshifts are not thoroughly constrained.

The covariance between these galaxy properties makes it challenging to
isolate their influence on the contamination fraction. For example,
the correlation between the stellar mass and brightness of a
galaxy means that the corresponding panels of
Figure~\ref{fig:memstat_all} are related and not independent probes of
contamination sources. The simplest
way to increase the purity of the group sample is to consider only
galaxies at smaller distances from the group center than the cut of
\rvir\ used here. Restricting the mock sample to $R<0.5\rvir$ results
in a mean purity and completeness of $84\%$ and $92\%$, respectively.

An alternative way to address the contamination and incompleteness of
the selection would be to apply correction factors to the member
selection as a function of these properties, as in
Equation~\ref{eq:correction}. This would amount to introducing strong 
priors to the membership algorithm based on our HOD model, limiting
the independence of the sample. In testing this approach however, we
have noticed that the correction factor as a function of group-centric
radius is not significantly tied to other properties such as
magnitude, stellar mass, or color, indicating that the contamination
is due more to geometry than distinct populations of galaxies. This
suggests that we can reliably study the relative radial trends of
these properties, though the absolute radial trends are subject to
uncertainties in the correction factor.

We can compare the member selection used here with that of
\citet{Giodini2009}, who used a statistical background subtraction on
the same body of data to determine galaxy membership and estimate the total
stellar mass in groups. Because the statistical background approach
does not individually assign galaxies to groups, we cannot
directly compute the purity and completeness of the selection, but we
can compare the total stellar mass estimates from the two selection
methods to the mock values. \citet{Giodini2009} selected candidate members
within a projected radius $R_{500\rm{c}}$ of X-ray centroids and $0.02\times(1+z)$ of the
group redshift, and estimated a mean foreground/background
contribution in 20 non-overlapping field regions of the same size and
redshift. 

We run both member selection methods on the mocks, applying
to each method the same corrections described by \citet{Giodini2009}
to deproject the cylindrical search volume into a sphere of radius
$R_{500\rm{c}}$ and to account for stellar mass contributions below
our sensitivity limit, adapted to the stellar mass function and limits
of our mocks. The mean stellar mass content in groups recovered using
their method is $3\%$ lower ($3\%$ higher) than the input mock value
in the redshift range $0.2<z<0.5$ $(0.5<z<1.0)$. With the same
corrections, our selection method estimates the mean stellar mass to be $3\%$
higher ($9\%$ higher) than the mock values over the same redshift
intervals. The typical scatter of $35\%$ between the recovered values and the
input values for a given group is much larger than the offsets for
both methods, but with these tests on mock catalogs we could remove
the small biases in future measurements. The mean stellar masses
inferred by the two methods happen to be quite similar because they
are typically dominated by massive galaxies for which membership
assignment is relatively straightforward. However, we note
that the full membership selected can be quite different because our approach
optimizes group centers using the weak lensing signal and handles
magnitude-dependent photometric redshift uncertainties, whereas
\citet{Giodini2009} use the X-ray centers and a fixed redshift window. 

We can also test different methods of estimating the field density to
see how it influences our member selection.
Our selection algorithm estimates $\hat{N}_{F}$ from the mean density
across the whole field, but smaller regions could instead be used to
estimate the local density around individual groups. While the local
density estimate has the advantage that it traces correlated structure
around groups, it does suffer from greater shot noise than the density
estimated over a larger volume. We have tested our approach
by using annuli centered on each group with inner and outer radii of
$2\rvir$ and $5\rvir$, while keeping the rest of the selection
algorithm the same. With this approach, the typical field density is
higher due to clustering around groups and the resulting membership probability
is slightly lower (increasing the field density by a factor of 2 typically
lowers the membership probability by only $\sim20\%$), but the purity and
completeness of the sample are essentially unchanged, and the fraction
of members crossing a threshold of $P_{\rm mem}>0.5$ between samples
is less than $10\%$. We obtain
similar results when substituting the background estimation method
used by \citet{Giodini2009} for our field density prior, so the
selection algorithm is not strongly sensitive to the approach used for
background estimation.

\subsection{Applicability to Other Surveys}
\label{s:surveys}

In view of other surveys which will search for groups and clusters in
multi-wavelength data, and to better characterize the advantages or
shortcomings of the COSMOS data used in this analysis, we test our
selection algorithm on mock catalogs with different levels of
uncertainty in redshift and centroid measurements. We consider five
hypothetical data sets: a full spectroscopic survey where all galaxies
have the typical redshift uncertainty in
zCOSMOS\footnote{\url{http://archive.eso.org/archive/adp/zCOSMOS/VIMOS\_spectroscopy\_v1.0/index.html}}
of $\sigma_z=3.7\times10^{-4}$, a low-resolution spectroscopic survey
like PRIMUS \citep{Coil2010} with redshift uncertainties of
$\sigma_z=0.005$, a photometric survey with fewer bands and larger
photo-$z$ errors like SDSS \citep{Csabai2003} or DES \citep{Banerji2008}
with $\sigma_z=0.05$, a deeper X-ray survey with more precise
centroids of  $3\arcsec$, and a lower resolution X-ray or SZ survey with centroid
uncertainties of $1\arcmin$. In the first three mock surveys we vary
only the redshift uncertainty and apply the same centering uncertainty
as the fiducial COSMOS mocks described in \S~\ref{s:mocks} assuming
similar X-ray detections. In the
final two mock surveys we use the magnitude-dependent redshift
uncertainties of the COSMOS mocks and assign centroiding
uncertainties, $\sigma_{\rm X}$, in each dimension on the sky. We
offset the nominal centroid from the central galaxy in each dimension
by a random value drawn from a Gaussian of width $\sigma_{\rm X}$. In all
cases we keep the same group and galaxy detection limits as in the
COSMOS data.

\begin{figure*}[htb]
\epsscale{1.0}
\plotone{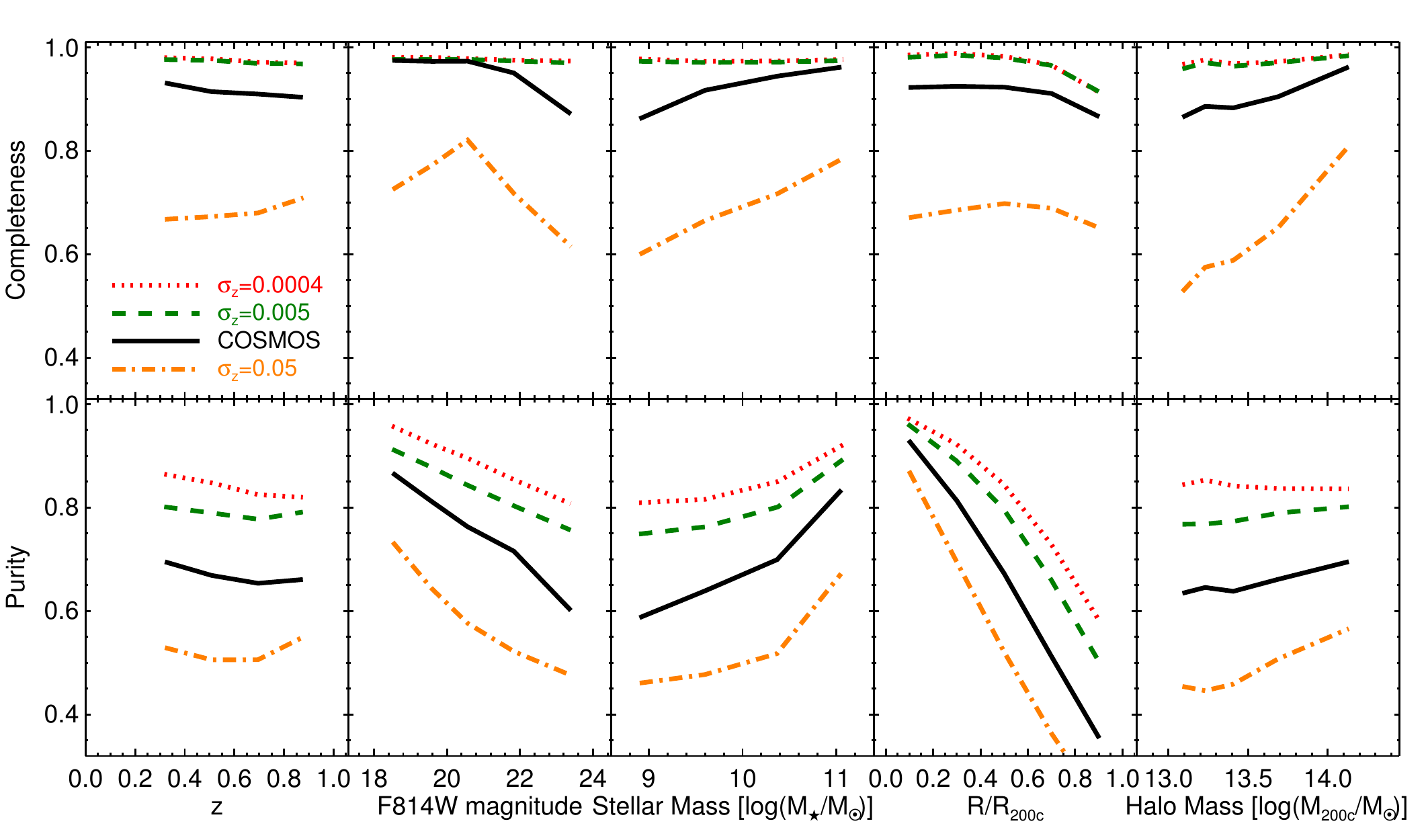}
\caption{Completeness (top row) and purity (bottom row) of the galaxy
  membership selection for hypothetical surveys with different
  redshift uncertainties according to the legend. Each curve
  represents the mean of ten mock lightcones. The fiducial survey is the same as
  that plotted in Figure~\ref{fig:memstat_all} for COSMOS. Note that
  the completeness curves for $\sigma_z=0.0004$ and $0.005$ lie atop
  one another.}
\label{fig:memstat_surveys_zerr}
\end{figure*}

\begin{figure*}[htb]
\epsscale{1.0}
\plotone{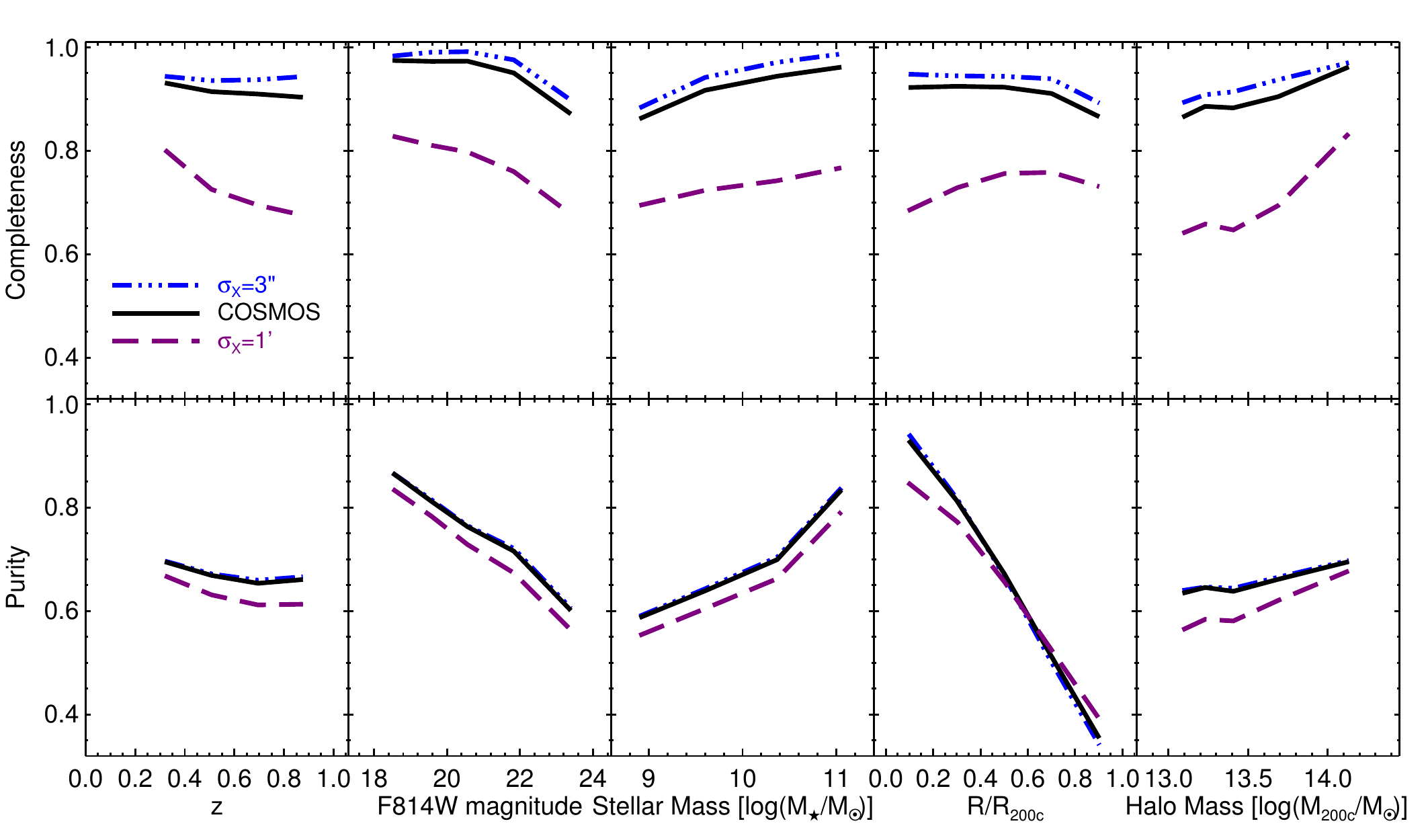}
\caption{Completeness (top row) and purity (bottom row) of the galaxy
  membership selection for hypothetical surveys with different
  centroid uncertainties according to the legend. As in
  Figure~\ref{fig:memstat_surveys_zerr}, each curve represents the
  mean of ten mock lightcones, and the fiducial survey is the same as 
  that plotted in Figure~\ref{fig:memstat_all} for COSMOS.}
\label{fig:memstat_surveys_centroid}
\end{figure*}

Figures~\ref{fig:memstat_surveys_zerr} and
\ref{fig:memstat_surveys_centroid} show the purity and completeness 
obtained when applying our member selection algorithm to these mock
surveys, in a manner similar to Figure~\ref{fig:memstat_all}. We
reiterate that these statistics describe the accuracy of 
the assignment of galaxies to groups, and not the detection of groups
themselves. The figure illustrates that purity and completeness
improve as redshift and centroid uncertainties decline. A number of
other points can be made about these results:

\begin{itemize}
\item{Deeper and more complete spectroscopic coverage would improve our member
    selection, increasing the purity of the sample from $\sim70\%$
    with photo-$z$s to
    $\sim85\%$. Improvements for completeness would mainly be gained
    from faint galaxies near our magnitude limit.}
\item{Among spectroscopic redshifts, high precision is not
    critical. The completeness of the $\sigma_z=3.7\times10^{-4}$ and
    $0.005$ samples are nearly identical, and the higher precision
    spectra provide only a modest improvement in sample purity over
    the low-resolution spectra, from
    $\sim80\%$ to $\sim85\%$. Once the redshift measurement
    uncertainty becomes comparable to the magnitude of intrinsic
    redshift distortions due to peculiar velocities in groups,
    additional spectral resolution does not greatly improve our ability to
    identify members. PRIMUS data in the COSMOS field will improve
    upon the existing sampling of zCOSMOS, but we note that the completeness
    limit for that survey is $i=22.5$ with sparse sampling to
    $i=23.5$, still shallower than our photo-$z$ depth of F814W=24.2.}
\item{The precise photometric redshifts available in the COSMOS field
    are critical for identifying members using our approach. Redshifts
    that are less accurate or precise show significantly reduced purity
    and completeness.}
\item{The precision of X-ray centroids for COSMOS groups is quite
    sufficient for member selection. Improving the positional
    uncertainty by roughly a factor of eight from the mean COSMOS
    value results in only a few percent improvement in completeness
    and a negligible gain in purity. Conversely, less precise centers
    (such as those available from SZ measurements) produce a sample
    with lower purity in the central region ($\sim85\%$ instead of
    $\sim95\%$) and significantly lower completeness ($\sim75\%$
    instead of $\sim90\%$). Though the existing COSMOS centroids are
    adequate for assigning member galaxies to groups, we note that
    several aspects of groups could still be studied with deeper X-ray
    or SZ data including physical offsets between central galaxies and
    hot gas, and the relationships between temperature or entropy with
    other group properties.}
\item{Note that we do not optimize our selection algorithm for these
    hypothetical data sets. Combining catalogs built from different
    observables \citep[e.g.,][]{Cohn2009}, and other techniques such
    as iterative centering or matched filters could improve results.}
\end{itemize}

We can compare the results of our mock spectroscopic selection to
other methods in the literature. We must note that our definitions of purity and
completeness refer to the success rates for assigning members to known
groups, while previous spectroscopic group-finding efforts have typically
quantified the purity and completeness of the identified group catalog in
addition to the galaxy membership assignment. In our mock tests, we
have implicitly assumed that the identification of groups is pure and
complete. It is also difficult to make direct comparisons across surveys
because of differences in data sets, limiting depths, and mock
catalogs. However, looking briefly at the quoted purity and
completeness of spectroscopic group catalogs, we can assess our
algorithm and see the advantage to assigning group membership when the
existence of a group is already know (e.g., from an X-ray detection).

Using a tesselation method to find galaxy groups in DEEP2 with a
limiting galaxy magnitude of R$_{\rm AB}=24.1$,
\citet{Gerke2005} attained a mean interloper fraction (analogous to
our impurity, $1-\mathsf{p}$) of $f_I=0.458\pm0.004$ and a mean galaxy
success rate (analogous to our completeness) of $S_{\rm
  gal}=0.786\pm0.006$ in their mock tests. They reported a one-way group identification
purity of $P_1=0.545\pm0.005$ and completeness of
$C_1=0.782\pm0.006$. \citet{Knobel2009} found that a
friends-of-friends approach performed better than the tesselation
method for identifying groups in zCOSMOS to a limiting magnitude of
I$_{\rm AB}=22.5$, and reported values of
$f_I=0.29$, $S_{\rm gal}=0.84$, $P_1=0.66$, and $C_1=0.81$ from their mocks. These
values are for groups with $N_{\rm mem}\ge2$, and while
\citet{Gerke2005} showed values for group purity and completeness that
were roughly constant with group velocity dispersion,
\citet{Knobel2009} showed that each of these statistics improved when
restricting the sample to higher richness groups, flattening out for
groups with $N_{\rm mem}\gtrsim5$. Disentangling the effects of group
identification from galaxy membership assignment is difficult, but
since the reported $S_{\rm gal}/C_1$ and $(1-f_I)/P_1$ are both
approximately unity, it appears that the main challenge in assigning
galaxy membership with these algorithms is in identifying real
groups. This fact illustrates the advantage of combining group-finding
methods to ensure a reliable sample of groups before assigning members.

\section{Member Catalog}
\label{s:catalog}

In the spirit of public releases of COSMOS data, we make our
membership assignments available as machine-readable files through the
NASA/IPAC Infrared Science Archive
(IRSA)\footnote{Upon publication, see \url{http://irsa.ipac.caltech.edu/Missions/cosmos.html}}.
These data include galaxy positions, redshifts, stellar masses,
colors, and membership probabilities, along with group
identifications. For each group we provide the X-ray position, flux,  
and luminosity, along with the redshift, halo mass, quality flags, and
the position and stellar mass of the central galaxy MMGG$_{\rm scale}$.
For reference, the basic parameters describing the catalog are compiled in
Table~\ref{t:catalog_properties}. For analyses requiring a clean
selection of galaxy groups, we restrict the sample to groups with
\textsc{xflag}~$=1$ or 2 and the \textsc{mask}, \textsc{poor}, and
\textsc{merger} flags blank to ensure that groups
and members have been reliably identified; in the group catalog we
define a new property, \textsc{flag\_include}, to encode this combination of
selection cuts. When a pure and complete
sample of members is needed, we select galaxies with $P_{\rm mem}>0.5$
in the inner regions of groups, $R<0.5\rvir$, with stellar masses
above our sample limit shown in Figure~\ref{fig:mass_limits}.

\begin{deluxetable*}{lc}
\tablecaption{Basic catalog properties\label{t:catalog_properties}}
\tablehead{\colhead{Property} & \colhead{Value}}
\startdata
Field coordinates (J2000) & RA=(149\fdg4, 150\fdg8), Dec=(1\fdg57, 2\fdg90) \\
Group redshift & $0 < \zG\ < 1$ \\
Galaxy magnitude & F814W~$<24.2$ \\
Halo mass & $12.8 < \log(\mvir/M_{\odot}) < 14.3$ \\
X-ray luminosity & $41.3 < \log(L_{\rm X}/\ergsec) < 44.1$ \\
N$_{\rm groups}$ & $211$ \\ 
N$_{\rm groups}$ (\textsc{xflag}~$=1,2$) & $165$ \\  
N$_{\rm groups}$ (clean groups\tablenotemark{a}) & $129$ \\  
N$_{\rm mem} (P_{\rm mem}>0.5)$ & $4631$ \\
N$_{\rm mem} (P_{\rm mem}>0.5$, clean groups) & $3406$ \\ 
N$_{\rm mem} (P_{\rm mem}>0.5, R<0.5\rvir, \log(M_{\star}/M_{\odot})>10.3)$ & $868$ \\
N$_{\rm mem} (P_{\rm mem}>0.5, R<0.5\rvir, \log(M_{\star}/M_{\odot})>10.3$, clean groups) & $657$
\enddata

\tablenotetext{a}{\textsc{xflag}~$=1,2$, \textsc{mask}=\textsc{poor}=\textsc{merger}=0}
\end{deluxetable*}

In addition to the catalog described above using photometric
redshifts, we have also produced a catalog replacing photo-$z$s with
spectroscopic redshifts when available. We use the same selection
algorithm and replace \pz\ from the photo-$z$ with a Gaussian of width
equal to the typical uncertainty in zCOSMOS, $\sigma_z=3.7\times10^{-4}$, and
sample each distribution at intervals of $10^{-5}$ in redshift. This
catalog has better purity and completeness than the photo-$z$ catalog
because of the improved redshift accuracy, but the selection is less
homogeneous because spectroscopic sampling is not representative or
complete.


\section{Analysis and Discussion}
\label{s:discussion}

With the catalog of group members identified and the purity and
completeness of the sample characterized, we provide a first look at
the properties of galaxies in these groups. Here we present
an analysis of the colors of group members
relative to the field. Future papers will study member properties in
more detail, including galaxy morphologies, star formation rates, and
AGN activity with respect to group properties like redshift, halo
mass, and group-centric distance.

\begin{figure}[htb]
\epsscale{1.2}
\plotone{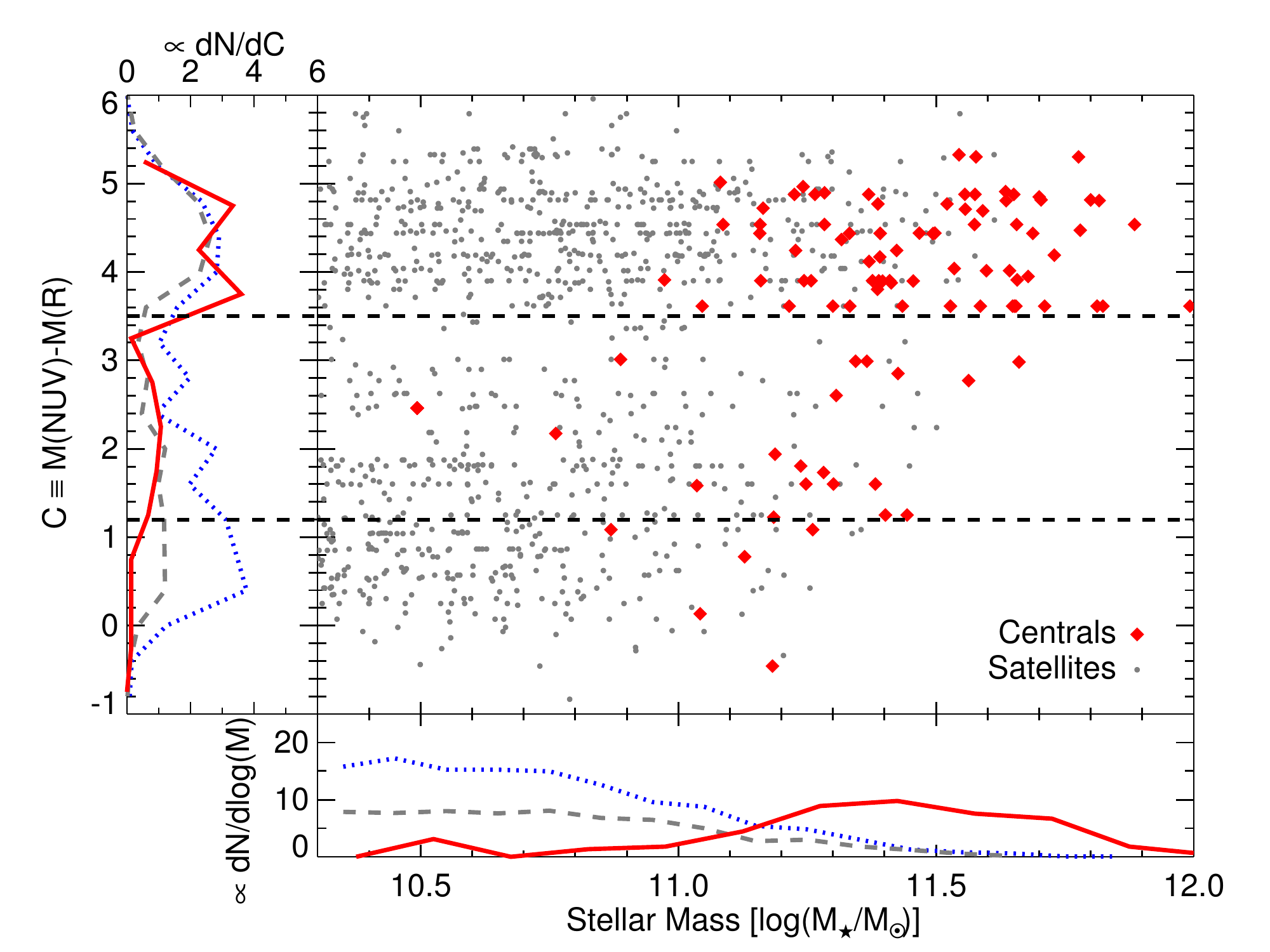}
\caption{Stellar masses and unextincted rest-frame template colors for
  central galaxies (red diamonds) and satellites (gray dots). We plot
  only the unambigous centrals, \ie those 
  designated as the MMGG$_{\rm scale}$ in groups which do not have a more
  massive galaxy in the outskirts or a discrepancy between
  identification with spectroscopic and photometric redshifts. Horizontal 
  dashed black lines show the galaxy spectral classes of
  \citet{Ilbert2010}. Margin plots show the distribution of colors and
stellar masses for centrals (red solid), satellites (gray dashed), and
a redshift-matched sample of field galaxies (blue dotted), rescaled
for comparison. Axis labels should be multiplied by 15 (centrals), 150
(satellites), and 500 (red-shift matched field sample)
to obtain the normalized distributions. The lower end of the stellar
mass range plotted corresponds to our completeness limit at $z=1$.}
\label{fig:group_cm}
\end{figure}

Figure~\ref{fig:group_cm} shows the unextincted rest-frame NUV-R
colors from \citet{Ilbert2010} for group members. We use the clean
sample of groups, selecting members with $P_{\rm mem}>0.5$ within
\rvir\ of the group center. The apparent banding in colors is due to
the finite number of templates used. We call
galaxies identified as the MMGG$_{\rm scale}$ ``centrals'' with the
other members as ``satellites.'' We see a
bimodal distribution in color space for both central and satellite
galaxies, though redder colors are more common than blue for both
types of members. In the margin plots, we show the distributions of
colors and stellar masses for centrals and satellites, as well as
field galaxies not assigned to any X-ray detected group ($P_{\rm
  mem}=0$). The field sample has been selected to match the redshift
distribution of group members in bins of $dz=0.1$. Although the color
distribution of field galaxies is also bimodal, it is clear that bluer
galaxies are more common in the field than in groups, a well-known
result in clusters and dense environments over a range of mass scales
and redshifts \citep[e.g.,][]{Gerke2007}.

We also see that there exist a number of blue centrals, which are of
interest because they suggest that star formation can persist or be
reactivated in the centers of dense groups, or that AGN exist there. The set of red
points in Figure~\ref{fig:group_cm} omits ambiguous cases where there
is a more massive galaxy in the outskirts of a group or where the
MMGG$_{\rm scale}$ differs between the photo-$z$-only catalog and
the one supplemented with available spectroscopic redshifts; $79\%$ of
groups satisfying the quality cuts of \S~\ref{s:catalog} have an
unambiguous central according to these criteria. While the majority of
this sample of centrals are red (77 out of 102), there are 5 centrals
with blue colors indicative of active star formation or AGN, and 20 centrals
with colors indicating intermediate activity. This population warrants
further study to verify that they are accurate centers and to
determine what environmental factors could contribute to the star
formation or AGN activity.

\subsection{Environmental Dependence from $z=0.2$ to $1$}

The higher fraction of blue galaxies in the field compared to groups
shown in Figure~\ref{fig:group_cm}
indicates that star formation is less common in dense
environments. As discussed in the introduction, much work has been
carried out to determine whether this well-known effect is due to a physical
process acting on galaxies in dense enviroments to suppress their star
formation rates, or due to the intrinsic properties of galaxies that
exist in these regions. 

To distinguish between the possibilities of environmental influence
and innate differences, we compare galaxy colors in group and field
environments within fixed stellar mass bins. We measure the fraction
of galaxies of the quiescent type defined by \citet{Ilbert2010}, \ie\
those with $M(NUV)-M(R)>3.5$. These are unextincted rest-frame colors
from the spectral template that best fits each galaxy's SED, allowing
us to study intrinsic colors that are related to specific star
formation rates without the obscuring effects of dust. The correction
is important because galaxies in low density environments have a higher dust
content, even among massive galaxies \citep{Kauffmann2004}. 

\begin{figure*}[htb]
\epsscale{1.2}
\plotone{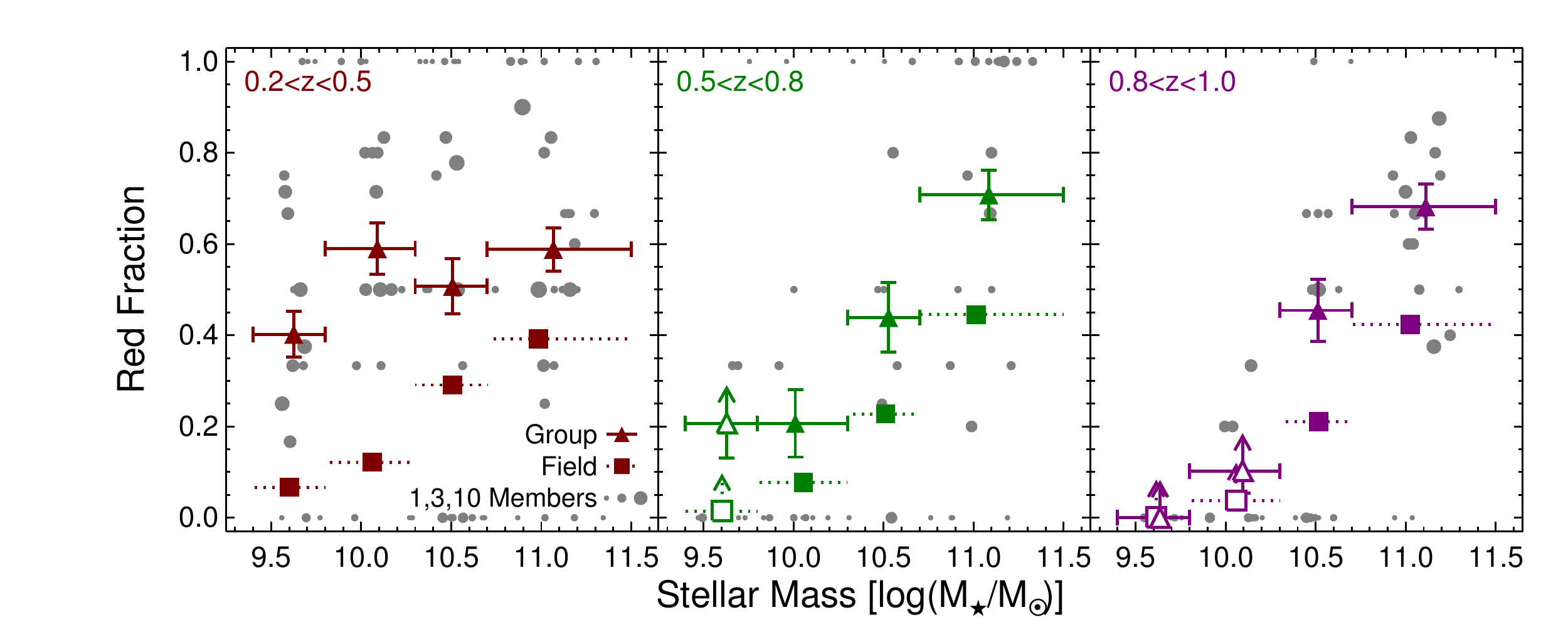}
\caption{Fraction of quenched galaxies as a function of stellar mass
  in different redshift bins (separate panels). Triangles show the
  quenched fraction of members with $P_{\rm mem}>0.5$ and
  projected group-centric distance within $0.5$\rvir\ of groups in the
  mass bins from Figure~\ref{fig:mass_limits}. Squares show the
  quenched fraction of field galaxies with $P_{\rm mem}=0$. Open
  symbols show bins with stellar mass incompleteness; the arrows above
  these symbols indicate that they are likely to be biased lower than
  the true values. Vertical error bars show the standard deviation of 1000
  bootstrap samples and horizontal bars represent the stellar mass bin widths. Gray circles
  show the quenched fraction in a given stellar mass bin for
  individual groups, with size proportional to the number of members
  in the bin. Note that the range of halo masses for group members
  varies between redshift bins.}
\label{fig:quenchz}
\end{figure*}

The fraction of red galaxies in fixed stellar mass bins is plotted
for three redshift ranges in Figure~\ref{fig:quenchz}. For group
members, we consider only galaxies with membership probability $P_{\rm
mem}>0.5$ within a projected distance of $0.5\rvir$ of the center of
groups in the mass completeness-limited bins of
Figure~\ref{fig:mass_limits}. The radial cut on the group sample is to avoid contamination
in the outskirts discussed in \S~\ref{s:error}. The plot includes both centrals and
satellites; excluding centrals leaves the results essentially
unchanged because the sample is dominated by satellites even in the
highest stellar mass bin plotted. The fraction of red galaxies in
this population is plotted against the mean stellar mass for each
bin. We also plot the red fraction in individual groups for galaxies
in the same stellar mass bins to show the
variation between groups. At all redshifts, higher mass galaxies tend
to have higher red fractions than lower mass galaxies. In addition to
this stellar mass dependence, we see a clear separation
between the group and field populations at all redshifts for the
stellar masses probed. The field sample plotted matches the
the redshift bins used for group members; matching the redshift
distribution on finer scales results in quenched fractions
that are at most a few percent different from those plotted.

We also see evidence for an increase in the red fraction with
decreasing redshift among low mass group galaxies. Redshift 
trends are somewhat difficult to interpret because the
range of group masses used is different in each redshift bin (see
Figure~\ref{fig:mass_limits}) and the color cut does not account for
evolution, so we leave a detailed analysis for
future work. Because our galaxy sample is magnitude-limited,
the low mass bins at high redshift are incomplete and plotted as open
symbols. We make no corrections for incompleteness here, which likely
leaves the sample in these bins biased toward the detection of blue
galaxies that tend to have lower mass-to-light ratios than red galaxies,
so we consider the red fractions in incomplete bins to be lower limits.

We detect a clear dependence of galaxy color on environment, even at fixed
stellar mass and high redshift. Figure~\ref{fig:quenchz} uses the
member catalog determined with photometric redshifts only, but
including the available spectroscopic redshifts does not affect the
results. Our results are similarly insensitive to the choice of a
probability threshold for membership; changing the cut on $P_{\rm
  mem}>0.5$ to 0.3 or 0.7 or weighting objects by $P_{\rm mem}$ instead
of choosing a threshold moves the red fractions by no more than a
few percent in any bin. Using the color
cuts described in \citet{Bundy2010} to separate passive galaxies in
COSMOS from dusty star-forming galaxies also gives qualitatively similar
results. Though the absolute fraction of red galaxies measured depends
on the specific cuts used, the relative trends with stellar mass and
environment are similar; galaxy groups have a larger proportion of red
galaxies than the field, and groups dominated by blue galaxies are
rare, though some appear to exist. 

We note that while our results confirm a significant relationship
between color and environment out to the redshift limit of our sample,
the cause of this relationship remains unknown. More detailed analysis
distinguishing between physical processes happening before and after
galaxies join groups is necessary to determine the role that groups
play in this process.

We now compare Figure~\ref{fig:quenchz} to results from the literature.
Our findings are consistent with results at low-redshift that show a
decrease in star formation at high stellar masses, as well as a clear
connection between star-formation and environment even after
accounting for stellar mass differences 
\citep[e.g.,][]{Kauffmann2004, Baldry2006}. At $z\sim0.4$,
\citet{McGee2011} report an environmental trend in the GEEC
survey that is smaller than a comparable low-redshift sample, but
still significant. However, \citet{Poggianti2008} do not detect a significant separation in
the fraction of star-forming galaxies in cluster and field environments at
$z=0.4-0.8$ after matching stellar mass distributions. 

At the highest redshift range covered by our group sample, our
findings are consistent with those of \citet{Cooper2010} and 
\citet{Peng2010} who detect a significant color-density relation in
a similar stellar mass range covered by DEEP2 and
zCOSMOS, respectively. These results appear to be at odds 
with some claims from VVDS and zCOSMOS analyses that such a relation
could be attributed solely to the existence of more massive galaxies
in dense environments
\citep[e.g.,][]{Scodeggio2009,Cucciati2010,Iovino2010}. Each of these
studies consider the environmental effect on color at fixed stellar
mass, as we have done here. \citet{Cooper2010}
emphasize that systematics in the selection of dense environments tend
to wash out the measured signal of environmental dependence, so that if such
correlations are seen they are likely to be real.
They also suggest that the non-detection of environmental trends in other
surveys is likely due to lower spectroscopic sampling rates and
reliance on less confident redshifts which comprise a significant
fraction of the sample at high redshift. The photometric
redshifts used in the present work are certainly less precise than
spectroscopic redshifts, but have a much higher sampling density. 

An important distinction with this study is that we are using a unique sample of groups;
X-ray detections ensure a robust sample of structures that are
virialized, a trait which may not be true of optically-selected
groups. The X-ray groups are also more massive than the typical
spectroscopically-selected groups, and may have formed earlier giving
them a longer time to suppress star formation in member
galaxies. Other studies of this sample of X-ray groups also detect a
significant environmental effect on galaxy colors (Giodini et al., in
prep.; Tanaka et al., in prep.). \citet{Finoguenov2010} have shown
that the number density of this X-ray group sample is in reasonable
agreement with that expected for halos of corresponding masses within
our cosmological model. The sample is therefore unlikely to be an extreme
population of groups, though we cannot rule out subtle differences
between X-ray-selected groups and the full population of groups in
this mass range.

Because of differences in analysis methods, it is difficult to
determine whether the qualitatively distinct findings in these
environmental studies are also quantitatively inconsistent, or whether
different results are simply due to measuring different quantities.
We now consider aspects of the analysis methods and
definitions of environments that may contribute
to the differing results, focusing our attention to field studies at $z\sim
1$ where the results appear most discrepant. For instance, we have shown in
Figure~\ref{fig:memstat_surveys_centroid} that centroiding errors can
influence the purity and completeness of a 
group sample, and our lensing tests (see Paper II) ensure a reliable
determination of the center of mass. Additionally, we use only galaxies and groups in
mass-complete bins, avoiding the need for volumetric corrections made
in some of the previous analyses.

Our group-based definition of 
dense environments is most similar to that of \citet{Iovino2010}, who
studied optically selected groups which are less massive on average
than the X-ray groups studied here. We leave a detailed comparison
between optically and X-ray selected groups in COSMOS to a
future paper (Finoguenov et. al, in prep.), but Figure 22 of \citet{Kovac2010a}
shows that these X-ray selected groups tend to be in more dense regions
on average than the optically selected ones. The richest optical
groups (with 4 or more members) show a good correspondence with X-ray
groups in the density field and are more likely to have similar halo
masses. Figure 12 of \citet{Iovino2010} shows that the colors of
stellar mass-selected samples do not significantly depend on group
richness (except perhaps at low stellar mass and redshift), so the
difference in halo masses between X-ray and optical groups is not
obviously the cause of discrepancy between our results and those of
\citet{Iovino2010}. \citet{Knobel2009} tested the purity of the
optical group sample with mock catalogs and found a contamination
fraction of roughly $25\%$, improving to about $15\%$ for richer
groups. These values are similar to our estimate of $16\%$
contamination within $0.5\rvir$ for the sample used in our analysis,
so the correct identification of group members is not a clear cause for
the difference in results either.

\citet{Cucciati2010} quantify environment using the local galaxy
overdensity $\delta=(\rho-\bar{\rho})/\bar{\rho}$, where $\rho$ is the
local density computed from the distance to the fifth nearest neighbor
and $\bar{\rho}$ is the mean density at a given redshift. Their Figure
10 shows a red fraction that is roughly constant across quartiles
of the local overdensity distribution for $z>0.5$ and
$\log(M_{\star}/M_{\odot})>10.85$, with a weak overdensity dependence at lower
redshift. With the same data and density indicator, \citet{Peng2010}
study the density field $\rho$ 
instead of the overdensity field $\delta$ and show a
significant color-density relation at $z\sim0.5$ and state that it continues to at
least $z=1$. They attribute the
difference in results to the fact that a lower fraction of galaxies at
$z\sim1$ live in regions with high $\delta$, and so the highest overdensity
quartile used in the analysis of \citet{Cucciati2010} is presumably too
broad at $z\sim1$ to isolate the small population of galaxies in
regions dense enough to produce strong environmental
effects. \citet{Cooper2010} use a third nearest neighbor density
estimator on DEEP2 data and identify a difference in the
distribution of colors between galaxies in the upper $10\%$ and lower
$50\%$ of the density distribution, so perhaps a cut more stringent
than the upper quartile of the density distribution is needed to detect
environmental effects at $z\sim1$.  Referring again to Figure 22 of
\citet{Kovac2010a} however, we see that X-ray groups live in roughly
the same range of overdensities as the upper
quartile of $\delta$ used in the \citet{Cucciati2010} analysis
($\log(1+\delta)\gtrsim1$). Thus the explanation from \citet{Peng2010}
for the non-detection of environmental trends at $z\sim1$ by
\citet{Cucciati2010} is not obviously applicable since we see a clear
environmental signal in this overdensity range.

The scale on which environment is defined also differs between these
analyses. The fifth nearest neighbor estimator used by
\citet{Cucciati2010} and \citet{Peng2010} and third nearest neighbor
used by \citet{Cooper2010} measure environment on a
scale that varies with density and is typically of order
$1~\rm{Mpc}$. \citet{Scodeggio2009} measures a density field on a
significantly larger scale of $8~\rm{Mpc}$, so correlations on the
smaller scales of group halos may be washed out. The group sample in
Figure~\ref{fig:quenchz} is restricted to $0.5\rvir$ but the
environmental signal is not significantly different when all selected
members out to \rvir\ are included, despite the higher contamination
fraction. In either case, \rvir\ is typically about $500~\rm{kpc}$ for
these groups, so our sample is likely probing the environment on
smaller scales than studies using the galaxy density field.

Differences in the colors used to identify galaxies as red or blue
could also contribute to the contrasting results. The other studies
discussed here typically use rest-frame $U-B$ or $B-I$ colors,
sometimes with a mass or redshift dependent cut to account for varying
populations. We use an extinction-corrected rest-frame $NUV-R$ color
to account for unquenched galaxies that appear red due to dust. To
test the effect of this correction, we have tried redefining the
sample of red galaxies using the cuts $NUV-R>3.5$ or $U-B>1$ without
any extinction correction. The red fraction increases due to the influence of dust
redding, and the separation between group and field values at $z>0.5$
is reduced by up to a factor of $2$, but we still see a clear
difference between the red fraction in group and field environments.

We have not identified an obvious single factor to explain why previous
analyses did not detect an environmental effect on galaxy color, but
suspect the cause to be a combination of factors mentioned above. To
avoid confusion when discussing environmental effects and to properly
detected these trends, it is clearly important to specify what is
meant by ``environment'' and to measure it carefully. 

\subsection{Star Forming Galaxies and the SZ Power Spectrum}

Recent high-resolution, ground-based experiments have probed the power
spectrum of the CMB to unprecedented fine scales (e.g., $\ell \gtrsim 2000$;
\citealt{Lueker2010, Fowler2010, Shirokoff2010, Das2011}), which are
sensitive to a variety of secondary 
anisotropies, such as radio and submillimeter point sources, and the SZ
effect from groups and clusters.  It was found that the power
due to the SZ effect was 50\% or less than predictions of most of the models
\citep{Lueker2010, Dunkley2010}, which could be indicative of our incomplete
understanding of the properties of groups and clusters, especially the low mass
($<10^{14} M_\odot$) systems at $z>0.5$, as they are believed to contribute
half of the SZ power at $\ell\approx 3000$
\citep[e.g.,][]{Komatsu2002, Shaw2010, Trac2011}.  It is possible
that the hot gas pressure profile of distant groups behaves differently from
that of clusters or that cluster profiles deviate from expectations at
large radii, although it was recently shown that local groups obey the
``universal'' pressure profile \citep{Sun2011} and at least one nearby
cluster obeys the profile out to \rvir\ after accounting for gas clumping
\citep{Simionescu2011}.  Another possibility is that star 
formation (SF) activity is elevated in high-$z$ systems, and the contribution
of unresolved SF galaxies in the submillimeter regime could fill in the SZ decrement at $\sim
150$ GHz, thus reducing the SZ power \citep[e.g.,][]{Hall2010}.  Using
the $z=0.5-1.0$ COSMOS groups, we are in a good position to
investigate the contamination due to SF galaxies in groups.

For each group, we cross-matched all candidate member galaxies with the MIPS
$24~\mu$m source catalog \citep{Sanders2007,LeFloch2009}, using a matching radius of $2''$.  For the
matched objects, we assumed a starburst SED (spanning from $3600$\AA\ to $1~\rm{cm}$)
taken from \citet{Lagache2003}, and compared the $24~\mu$m to 148 GHz flux ratio.
More specifically, we approximated the MIPS $24~\mu$m band as a tophat spanning
20.8 to 26.1~$\mu$m, and we used 18 GHz as the bandwidth for the 148 GHz
channel of the Atacama Cosmology Telescope (ACT).  Given the mass of
our groups, we estimated the SZ flux for our 
groups, following \citet{Majumdar2004}. We find that the sum of the fluxes from SF
galaxies at 148 GHz is negligible (typically $\lesssim 0.3\%$) compared to the magnitude of the SZ
effect from these groups. 

We selected the SED from the set of templates of \citet{Lagache2003}
that maximizes the 148 GHz to $24~\mu$m flux ratio to set a
conservative upper limit, but uncertainties in the spectral model
could allow for a larger flux at millimeter wavelengths from these
sources. \citet{Lee2010} compare stacked MIPS measurements at 24, 
70, and 160~$\mu$m to empirical templates from \citet{Chary2001,
  Dale2002, Lagache2003} and theoretical models from
\citet{Siebenmorgen2007}. For the stacked MIPS fluxes of sources at
$z=0.5-1$, the best-fitting template from \citet{Lagache2003} tends to
predict a flux at wavelengths longer than $300~\mu$m that can be an order of
magnitude lower than other models. Even accounting for these uncertainties
in the spectral models, our upper limit to the contamination of
star-forming galaxies to the SZ signal from our sample of groups is no
more than a few percent.

Our group sample is limited to $z<1$.  Several studies have reported
elevated SF activity in centers of clusters at $z\gtrsim 1.4$ \citep[e.g.,
][]{Hilton2010,Tran2010,Tanaka2010}.  It remains to be seen if the SF galaxies could
be abundant enough at higher redshifts to make a significant impact on
the SZ signal in such systems.


\section{Summary and Conclusions}
\label{s:conclusion}

We have presented a catalog of member galaxies in X-ray selected
groups in the COSMOS field, carefully taking into account photo-$z$
errors and attempting to avoid biasing the sample with assumptions
about the properties of galaxies which have not previously been
well-constrained in this mass and redshift range. We have thoroughly
characterized the quality of the selection algorithm using tests with
mock catalogs and spectroscopic redshifts. In these tests we
discovered contamination from galaxies in projection in the outskirts
of groups, but selection in the central regions is relatively clean. 
We have also studied the prospects for applying this selection
algorithm to future multi-wavelength data sets, estimating the purity
and completeness of the member selection as a function of redshift and
centering uncertainties.

Analyzing this sample of group members, we have shown that both
stellar mass and environment play a role in determining galaxy colors
at $z\sim 1$. We emphasize that there are many ways to smear out
environmental correlations and that these factors must be properly
controlled in order to detect environmental trends. The X-ray
groups studied here provide a clean sample of dense environments for
which we can determine halo masses and centers.

Following our finding of suppressed star formation in group
environments at all redshifts sampled, we investigated the possibility
that clustered dusty star-forming galaxies could reduce the detected power in the
high-$\ell$ CMB power spectrum by filling in SZ decrements in
groups, and found that the effect must be quite small in this group
sample. In contrast with the results relating to the suppression of
star formation in groups, we have identified several blue
central galaxies, which warrant further study.

In Paper II of this series on our sample of galaxy groups and members, we
will describe weak lensing tests used to optimize the centering by
finding tracers that best locate the center of mass. Further work
will analyze galaxy properties in groups with respect to the distance
from these centers, providing constraints on models describing the
evolution of galaxies in dense environments. With a
carefully selected sample of member galaxies in groups with
well-constrained masses and centers, we can hope to map the course by
which galaxies transform.


\acknowledgments
We thank Joanne Cohn, Eliot Quataert, Eli Rykoff,
David Schlegel, Uros Seljak, Erik Shirokoff, Andrew Wetzel, and Martin
White for helpful conversations. We also thank Michael Cooper and Marc
Davis for providing software and template spectra used in our
spectroscopic analysis, as well as comments on the paper. MRG is supported by a Graduate Research
Fellowship from the National Science Foundation. 

This work is partly based on observations made with ESO Telescopes at
Paranal Observatory under program ID 084.B-0523. We gratefully
acknowledge the contributions of the entire COSMOS 
colaboration consisting of more than 70 scientists. More information
on the COSMOS survey is available at {\bf
  \url{http://www.astro.caltech.edu/~cosmos}}. This work is based on observations
with the NASA/ESA {\em Hubble Space Telescope}, obtained at the Space Telescope Science
Institute, which is operated by AURA Inc, under NASA contract NAS
5-26555; also based on data collected at: the Subaru Telescope,
which is operated by the National Astronomical Observatory of Japan;
the XMM-Newton, an ESA science mission with instruments and
contributions directly funded by ESA Member States and NASA; the
European Southern Observatory under Large Program 175.A-0839, Chile;
Kitt Peak National Observatory, Cerro Tololo Inter-American
Observatory, and the National Optical Astronomy Observatory, which
are operated by the Association of Universities for Research in
Astronomy, Inc. (AURA) under cooperative agreement with the National
Science Foundation; and and the Canada-France-Hawaii Telescope with
MegaPrime/MegaCam operated as a joint project by the CFHT Corporation,
CEA/DAPNIA, the National Research Council of Canada, the Canadian
Astronomy Data Centre, the Centre National de la Recherche
Scientifique de France, TERAPIX and the University of Hawaii.

\mbox{~} 



\bibliographystyle{apj}

\bibliography{catalog}

\begin{thebibliography}{136}
\expandafter\ifx\csname natexlab\endcsname\relax\def\natexlab#1{#1}\fi

\bibitem[{Adami {et~al.}(2010)Adami, Mazure, Pierre, Sprimont, Libbrecht,
  Pacaud, Clerc, Sadibekova, Surdej, Altieri, Duc, Galaz, Gueguen, Guennou,
  Hertling, Ilbert, LeF\`{e}vre, Quintana, Valtchanov, Willis, Akiyama, Aussel,
  Chiappetti, Detal, Garilli, LeBrun, LeF\`{e}vre, Maccagni, Melin, Ponman,
  Ricci, \& Tresse}]{Adami2010}
Adami, C., {et~al.} 2010, arxiv:1010.6195

\bibitem[{Baldry {et~al.}(2006)Baldry, Balogh, Bower, Glazebrook, Nichol,
  Bamford, \& Budavari}]{Baldry2006}
Baldry, I.~K., Balogh, M.~L., Bower, R.~G., Glazebrook, K., Nichol, R.~C.,
  Bamford, S.~P., \& Budavari, T. 2006, \mnras, 373, 469

\bibitem[{{Balogh} {et~al.}(2004){Balogh}, {Eke}, {Miller}, {Lewis}, {Bower},
  {Couch}, {Nichol}, {Bland-Hawthorn}, {Baldry}, {Baugh}, {Bridges}, {Cannon},
  {Cole}, {Colless}, {Collins}, {Cross}, {Dalton}, {de Propris}, {Driver},
  {Efstathiou}, {Ellis}, {Frenk}, {Glazebrook}, {Gomez}, {Gray}, {Hawkins},
  {Jackson}, {Lahav}, {Lumsden}, {Maddox}, {Madgwick}, {Norberg}, {Peacock},
  {Percival}, {Peterson}, {Sutherland}, \& {Taylor}}]{Balogh2004}
{Balogh}, M., {et~al.} 2004, \mnras, 348, 1355

\bibitem[{Bamford {et~al.}(2009)Bamford, Nichol, Baldry, Land, Lintott,
  Schawinski, Slosar, Szalay, Thomas, Torki, Andreescu, Edmondson, Miller,
  Murray, Raddick, \& Vandenberg}]{Bamford2009}
Bamford, S.~P., {et~al.} 2009, \mnras, 393, 1324

\bibitem[{Banerji {et~al.}(2008)Banerji, Abdalla, Lahav, \& Lin}]{Banerji2008}
Banerji, M., Abdalla, F.~B., Lahav, O., \& Lin, H. 2008, \mnras, 386, 1219

\bibitem[{{Bellagamba} {et~al.}(2011){Bellagamba}, {Maturi}, {Hamana},
  {Meneghetti}, {Miyazaki}, \& {Moscardini}}]{Bellagamba2011}
{Bellagamba}, F., {Maturi}, M., {Hamana}, T., {Meneghetti}, M., {Miyazaki}, S.,
  \& {Moscardini}, L. 2011, \mnras, 217

\bibitem[{{Berlind} {et~al.}(2006){Berlind}, {Frieman}, {Weinberg}, {Blanton},
  {Warren}, {Abazajian}, {Scranton}, {Hogg}, {Scoccimarro}, {Bahcall},
  {Brinkmann}, {Gott}, {Kleinman}, {Krzesinski}, {Lee}, {Miller}, {Nitta},
  {Schneider}, {Tucker}, \& {Zehavi}}]{Berlind2006}
{Berlind}, A.~A., {et~al.} 2006, \apjs, 167, 1

\bibitem[{Blanton \& Berlind(2007)}]{Blanton2007}
Blanton, M.~R., \& Berlind, A.~A. 2007, \apj, 664, 791

\bibitem[{Blanton {et~al.}(2005)Blanton, Eisenstein, Hogg, Schlegel, \&
  Brinkmann}]{Blanton2005}
Blanton, M.~R., Eisenstein, D., Hogg, D.~W., Schlegel, D.~J., \& Brinkmann, J.
  2005, \apj, 629, 143

\bibitem[{Blanton \& Moustakas(2009)}]{Blanton2009}
Blanton, M.~R., \& Moustakas, J. 2009, \araa, 47, 159

\bibitem[{{Brunner} \& {Lubin}(2000)}]{Brunner2000}
{Brunner}, R.~J., \& {Lubin}, L.~M. 2000, \aj, 120, 2851

\bibitem[{Bruzual \& Charlot(2003)}]{Bruzual2003}
Bruzual, G., \& Charlot, S. 2003, \mnras, 344, 1000

\bibitem[{Bundy {et~al.}(2006)Bundy, Ellis, Conselice, Taylor, Cooper, Willmer,
  Weiner, Coil, Noeske, \& Eisenhardt}]{Bundy2006a}
Bundy, K., {et~al.} 2006, \apj, 651, 120

\bibitem[{Bundy {et~al.}(2010)Bundy, Scarlata, Carollo, Ellis, Drory, Hopkins,
  Salvato, Leauthaud, Koekemoer, Murray, Ilbert, Oesch, Ma, Capak, Pozzetti, \&
  Scoville}]{Bundy2010}
---. 2010, \apj, 719, 1969

\bibitem[{Butcher \& Oemler(1984)}]{Butcher1984}
Butcher, H., \& Oemler, A. 1984, \apj, 285, 426

\bibitem[{Capak {et~al.}(2007{\natexlab{a}})Capak, Abraham, Ellis, Mobasher,
  Scoville, Sheth, \& Koekemoer}]{Capak2007a}
Capak, P., Abraham, R.~G., Ellis, R.~S., Mobasher, B., Scoville, N., Sheth, K.,
  \& Koekemoer, A. 2007{\natexlab{a}}, \apjs, 172, 284

\bibitem[{Capak {et~al.}(2007{\natexlab{b}})Capak, Aussel, Ajiki, McCracken,
  Mobasher, Scoville, Shopbell, Taniguchi, Thompson, Tribiano, Sasaki, Blain,
  Brusa, Carilli, Comastri, Carollo, Cassata, Colbert, Ellis, Elvis,
  Giavalisco, Green, Guzzo, Hasinger, Ilbert, Impey, Jahnke, Kartaltepe, Kneib,
  Koda, Koekemoer, Komiyama, Leauthaud, {Le Fevre}, Lilly, Liu, Massey,
  Miyazaki, Murayama, Nagao, Peacock, Pickles, Porciani, Renzini, Rhodes, Rich,
  Salvato, Sanders, Scarlata, Schiminovich, Schinnerer, Scodeggio, Sheth,
  Shioya, Tasca, Taylor, Yan, \& Zamorani}]{Capak2007b}
Capak, P., {et~al.} 2007{\natexlab{b}}, \apjs, 172, 99

\bibitem[{{Capak} {et~al.}(2010){Capak}, {Scoville}, {Sanders}, {Mobasher},
  {Masters}, \& {Salvato}}]{Capak2010}
{Capak}, P.~L., {Scoville}, N.~Z., {Sanders}, D.~B., {Mobasher}, B., {Masters},
  D., \& {Salvato}, M. 2010, in Bulletin of the American Astronomical Society,
  Vol.~42, American Astronomical Society Meeting Abstracts 215, 410.05

\bibitem[{{Cappelluti} {et~al.}(2009){Cappelluti}, {Brusa}, {Hasinger},
  {Comastri}, {Zamorani}, {Finoguenov}, {Gilli}, {Puccetti}, {Miyaji},
  {Salvato}, {Vignali}, {Aldcroft}, {B{\"o}hringer}, {Brunner}, {Civano},
  {Elvis}, {Fiore}, {Fruscione}, {Griffiths}, {Guzzo}, {Iovino}, {Koekemoer},
  {Mainieri}, {Scoville}, {Shopbell}, {Silverman}, \& {Urry}}]{Cappelluti2009}
{Cappelluti}, N., {et~al.} 2009, \aap, 497, 635

\bibitem[{{Chabrier}(2003)}]{Chabrier2003}
{Chabrier}, G. 2003, \pasp, 115, 763

\bibitem[{{Chary} \& {Elbaz}(2001)}]{Chary2001}
{Chary}, R., \& {Elbaz}, D. 2001, \apj, 556, 562

\bibitem[{{Cohn} \& {White}(2009)}]{Cohn2009}
{Cohn}, J.~D., \& {White}, M. 2009, \mnras, 393, 393

\bibitem[{{Coil} {et~al.}(2010){Coil}, {Blanton}, {Burles}, {Cool},
  {Eisenstein}, {Moustakas}, {Wong}, {Zhu}, {Aird}, {Bernstein}, {Bolton}, \&
  {Hogg}}]{Coil2010}
{Coil}, A.~L., {et~al.} 2010, ArXiv e-prints

\bibitem[{{Cooper} {et~al.}(2006){Cooper}, {Newman}, {Croton}, {Weiner},
  {Willmer}, {Gerke}, {Madgwick}, {Faber}, {Davis}, {Coil}, {Finkbeiner},
  {Guhathakurta}, \& {Koo}}]{Cooper2006}
{Cooper}, M.~C., {et~al.} 2006, \mnras, 370, 198

\bibitem[{{Cooper} {et~al.}(2007){Cooper}, {Newman}, {Coil}, {Croton}, {Gerke},
  {Yan}, {Davis}, {Faber}, {Guhathakurta}, {Koo}, {Weiner}, \&
  {Willmer}}]{Cooper2007}
---. 2007, \mnras, 376, 1445

\bibitem[{Cooper {et~al.}(2010)Cooper, Coil, Gerke, Newman, Bundy, Conselice,
  Croton, Davis, Faber, Guhathakurta, Koo, Lin, Weiner, Willmer, \&
  Yan}]{Cooper2010}
Cooper, M.~C., {et~al.} 2010, \mnras, 409, 337

\bibitem[{{Cortese} {et~al.}(2007){Cortese}, {Marcillac}, {Richard},
  {Bravo-Alfaro}, {Kneib}, {Rieke}, {Covone}, {Egami}, {Rigby}, {Czoske}, \&
  {Davies}}]{Cortese2007}
{Cortese}, L., {et~al.} 2007, \mnras, 376, 157

\bibitem[{Csabai {et~al.}(2003)Csabai, Budav\'{a}ri, Connolly, Szalay, Győry,
  Ben\'{\i}tez, Annis, Brinkmann, Eisenstein, Fukugita, Gunn, Kent, Lupton,
  Nichol, \& Stoughton}]{Csabai2003}
Csabai, I., {et~al.} 2003, \aj, 125, 580

\bibitem[{Cucciati {et~al.}(2006)Cucciati, Iovino, Marinoni, Ilbert, Bardelli,
  Franzetti, {Le F\`{e}vre}, Pollo, Zamorani, Cappi, Guzzo, McCracken, Meneux,
  Scaramella, Scodeggio, Tresse, Zucca, Bottini, Garilli, {Le Brun}, Maccagni,
  Picat, Vettolani, Zanichelli, Adami, Arnaboldi, Arnouts, Bolzonella, Charlot,
  Ciliegi, Contini, Foucaud, Gavignaud, Marano, Mazure, Merighi, Paltani,
  Pell\`{o}, Pozzetti, Radovich, Bondi, Bongiorno, Busarello, de~la Torre,
  Gregorini, Lamareille, Mathez, Mellier, Merluzzi, Ripepi, Rizzo, Temporin, \&
  Vergani}]{Cucciati2006}
Cucciati, O., {et~al.} 2006, \aap, 458, 39

\bibitem[{Cucciati {et~al.}(2010)Cucciati, Iovino, Kova\v{c}, Scodeggio, Lilly,
  Bolzonella, Bardelli, Vergani, Tasca, Zucca, Zamorani, Pozzetti, Knobel,
  Oesch, Lamareille, Caputi, Kampczyk, Tresse, Maier, Carollo, Contini, Kneib,
  {Le F\`{e}vre}, Mainieri, Renzini, Bongiorno, Coppa, de~la Torre, de~Ravel,
  Franzetti, Garilli, {Le Borgne}, {Le Brun}, Mignoli, Pell\`{o}, Peng,
  Perez-Montero, Ricciardelli, Silverman, Tanaka, Koekemoer, Scoville, Abbas,
  Bottini, Cappi, Cassata, Cimatti, Guzzo, Leauthaud, Maccagni, Marinoni,
  McCracken, Memeo, Meneux, Porciani, \& Scaramella}]{Cucciati2010}
---. 2010, \aap, 524, 2

\bibitem[{{Dale} \& {Helou}(2002)}]{Dale2002}
{Dale}, D.~A., \& {Helou}, G. 2002, \apj, 576, 159

\bibitem[{{Das} {et~al.}(2011){Das}, {Marriage}, {Ade}, {Aguirre}, {Amiri},
  {Appel}, {Barrientos}, {Battistelli}, {Bond}, {Brown}, {Burger}, {Chervenak},
  {Devlin}, {Dicker}, {Bertrand Doriese}, {Dunkley}, {D{\"u}nner},
  {Essinger-Hileman}, {Fisher}, {Fowler}, {Hajian}, {Halpern}, {Hasselfield},
  {Hern{\'a}ndez-Monteagudo}, {Hilton}, {Hilton}, {Hincks}, {Hlozek},
  {Huffenberger}, {Hughes}, {Hughes}, {Infante}, {Irwin}, {Baptiste Juin},
  {Kaul}, {Klein}, {Kosowsky}, {Lau}, {Limon}, {Lin}, {Lupton}, {Marsden},
  {Martocci}, {Mauskopf}, {Menanteau}, {Moodley}, {Moseley}, {Netterfield},
  {Niemack}, {Nolta}, {Page}, {Parker}, {Partridge}, {Reid}, {Sehgal},
  {Sherwin}, {Sievers}, {Spergel}, {Staggs}, {Swetz}, {Switzer}, {Thornton},
  {Trac}, {Tucker}, {Warne}, {Wollack}, \& {Zhao}}]{Das2011}
{Das}, S., {et~al.} 2011, \apj, 729, 62

\bibitem[{Davis {et~al.}(1985)Davis, Efstathiou, Frenk, \& White}]{Davis1985}
Davis, M., Efstathiou, G., Frenk, C.~S., \& White, S. D.~M. 1985, \apj, 292,
  371

\bibitem[{{De Propris} {et~al.}(2004){De Propris}, Colless, Peacock, Couch,
  Driver, Balogh, Baldry, Baugh, Bland-Hawthorn, Bridges, Cannon, Cole,
  Collins, Cross, Dalton, Efstathiou, Ellis, Frenk, Glazebrook, Hawkins,
  Jackson, Lahav, Lewis, Lumsden, Maddox, Madgwick, Norberg, Percival,
  Peterson, Sutherland, \& Taylor}]{DePropris2004}
{De Propris}, R., {et~al.} 2004, \mnras, 351, 125

\bibitem[{Dressler(1980)}]{Dressler1980}
Dressler, A. 1980, \apj, 236, 351

\bibitem[{Dressler {et~al.}(1997)Dressler, Oemler, Couch, Smail, Ellis, Barger,
  Butcher, Poggianti, \& Sharples}]{Dressler1997}
Dressler, A., {et~al.} 1997, \apj, 490, 577

\bibitem[{Driver {et~al.}(2009)Driver, Norberg, Baldry, Bamford, Hopkins,
  Liske, Loveday, Peacock, Hill, Kelvin, Robotham, Cross, Parkinson, Prescott,
  Conselice, Dunne, Brough, Jones, Sharp, van Kampen, Oliver, Roseboom,
  Bland-Hawthorn, Croom, Ellis, Cameron, Cole, Frenk, Couch, Graham, Proctor,
  {De Propris}, Doyle, Edmondson, Nichol, Thomas, Eales, Jarvis, Kuijken,
  Lahav, Madore, Seibert, Meyer, Staveley-Smith, Phillipps, Popescu, Sansom,
  Sutherland, Tuffs, \& Warren}]{Driver2009}
Driver, S.~P., {et~al.} 2009, Astronomy and Geophysics, 50, 12

\bibitem[{Drory {et~al.}(2009)Drory, Bundy, Leauthaud, Scoville, Capak, Ilbert,
  Kartaltepe, Kneib, McCracken, Salvato, Sanders, Thompson, \&
  Willott}]{Drory2009}
Drory, N., {et~al.} 2009, \apj, 707, 1595

\bibitem[{{Dunkley} {et~al.}(2009){Dunkley}, {Komatsu}, {Nolta}, {Spergel},
  {Larson}, {Hinshaw}, {Page}, {Bennett}, {Gold}, {Jarosik}, {Weiland},
  {Halpern}, {Hill}, {Kogut}, {Limon}, {Meyer}, {Tucker}, {Wollack}, \&
  {Wright}}]{Dunkley2009}
{Dunkley}, J., {et~al.} 2009, \apjs, 180, 306

\bibitem[{{Dunkley} {et~al.}(2010){Dunkley}, {Hlozek}, {Sievers}, {Acquaviva},
  {Ade}, {Aguirre}, {Amiri}, {Appel}, {Barrientos}, {Battistelli}, {Bond},
  {Brown}, {Burger}, {Chervenak}, {Das}, {Devlin}, {Dicker}, {Bertrand
  Doriese}, {Dunner}, {Essinger-Hileman}, {Fisher}, {Fowler}, {Hajian},
  {Halpern}, {Hasselfield}, {Hernandez-Monteagudo}, {Hilton}, {Hilton},
  {Hincks}, {Huffenberger}, {Hughes}, {Hughes}, {Infante}, {Irwin}, {Juin},
  {Kaul}, {Klein}, {Kosowsky}, {Lau}, {Limon}, {Lin}, {Lupton}, {Marriage},
  {Marsden}, {Mauskopf}, {Menanteau}, {Moodley}, {Moseley}, {Netterfield},
  {Niemack}, {Nolta}, {Page}, {Parker}, {Partridge}, {Reid}, {Sehgal},
  {Sherwin}, {Spergel}, {Staggs}, {Swetz}, {Switzer}, {Thornton}, {Trac},
  {Tucker}, {Warne}, {Wollack}, \& {Zhao}}]{Dunkley2010}
---. 2010, ArXiv e-prints

\bibitem[{{Eke} {et~al.}(2004){Eke}, {Baugh}, {Cole}, {Frenk}, {Norberg},
  {Peacock}, {Baldry}, {Bland-Hawthorn}, {Bridges}, {Cannon}, {Colless},
  {Collins}, {Couch}, {Dalton}, {de Propris}, {Driver}, {Efstathiou}, {Ellis},
  {Glazebrook}, {Jackson}, {Lahav}, {Lewis}, {Lumsden}, {Maddox}, {Madgwick},
  {Peterson}, {Sutherland}, \& {Taylor}}]{Eke2004}
{Eke}, V.~R., {et~al.} 2004, \mnras, 348, 866

\bibitem[{Elvis {et~al.}(2009)Elvis, Civano, Vignali, Puccetti, Fiore,
  Cappelluti, Aldcroft, Fruscione, Zamorani, Comastri, Brusa, Gilli, Miyaji,
  Damiani, Koekemoer, Finoguenov, Brunner, Urry, Silverman, Mainieri, Hasinger,
  Griffiths, Carollo, Hao, Guzzo, Blain, Calzetti, Carilli, Capak, Ettori,
  Fabbiano, Impey, Lilly, Mobasher, Rich, Salvato, Sanders, Schinnerer,
  Scoville, Shopbell, Taylor, Taniguchi, \& Volonteri}]{Elvis2009}
Elvis, M., {et~al.} 2009, \apjs, 184, 158

\bibitem[{Evrard {et~al.}(2008)Evrard, Bialek, Busha, White, Habib, Heitmann,
  Warren, Rasia, Tormen, Moscardini, Power, Jenkins, Gao, Frenk, Springel,
  White, \& Diemand}]{Evrard2008}
Evrard, A.~E., {et~al.} 2008, \apj, 672, 122

\bibitem[{{Feruglio} {et~al.}(2010){Feruglio}, {Aussel}, {Le Floc'h}, {Ilbert},
  {Salvato}, {Capak}, {Fiore}, {Kartaltepe}, {Sanders}, {Scoville},
  {Koekemoer}, \& {Ideue}}]{Feruglio2010}
{Feruglio}, C., {et~al.} 2010, \apj, 721, 607

\bibitem[{Finoguenov {et~al.}(2007)Finoguenov, Guzzo, Hasinger, Scoville,
  Aussel, Bohringer, Brusa, Capak, Cappelluti, Comastri, Giodini, Griffiths,
  Impey, Koekemoer, Kneib, Leauthaud, {Le Fevre}, Lilly, Mainieri, Massey,
  McCracken, Mobasher, Murayama, Peacock, Sakelliou, Schinnerer, Silverman,
  Smol\v{c}i\'{c}, Taniguchi, Tasca, Taylor, Trump, \&
  Zamorani}]{Finoguenov2007}
Finoguenov, a., {et~al.} 2007, \apjs, 172, 182

\bibitem[{Finoguenov {et~al.}(2009)Finoguenov, Connelly, Parker, Wilman,
  Mulchaey, Saglia, Balogh, Bower, \& McGee}]{Finoguenov2009}
---. 2009, \apj, 704, 564

\bibitem[{{Finoguenov} {et~al.}(2010){Finoguenov}, {Watson}, {Tanaka},
  {Simpson}, {Cirasuolo}, {Dunlop}, {Peacock}, {Farrah}, {Akiyama}, {Ueda},
  {Smol{\v c}i{\'c}}, {Stewart}, {Rawlings}, {van Breukelen}, {Almaini},
  {Clewley}, {Bonfield}, {Jarvis}, {Barr}, {Foucaud}, {McLure}, {Sekiguchi}, \&
  {Egami}}]{Finoguenov2010}
{Finoguenov}, A., {et~al.} 2010, \mnras, 403, 2063

\bibitem[{Fowler {et~al.}(2010)Fowler, Acquaviva, Ade, Aguirre, Amiri, Appel,
  Barrientos, Battistelli, Bond, Brown, Burger, Chervenak, Das, Devlin, Dicker,
  Doriese, Dunkley, D\"{u}nner, Essinger-Hileman, Fisher, Hajian, Halpern,
  Hasselfield, Hern\'{a}ndez-Monteagudo, Hilton, Hilton, Hincks, Hlozek,
  Huffenberger, Hughes, Hughes, Infante, Irwin, Jimenez, Juin, Kaul, Klein,
  Kosowsky, Lau, Limon, Lin, Lupton, Marriage, Marsden, Martocci, Mauskopf,
  Menanteau, Moodley, Moseley, Netterfield, Niemack, Nolta, Page, Parker,
  Partridge, Quintana, Reid, Sehgal, Sievers, Spergel, Staggs, Swetz, Switzer,
  Thornton, Trac, Tucker, Verde, Warne, Wilson, Wollack, \& Zhao}]{Fowler2010}
Fowler, J.~W., {et~al.} 2010, \apj, 722, 1148

\bibitem[{{Gavazzi} {et~al.}(2001){Gavazzi}, {Boselli}, {Mayer},
  {Iglesias-Paramo}, {V{\'{\i}}lchez}, \& {Carrasco}}]{Gavazzi2001}
{Gavazzi}, G., {Boselli}, A., {Mayer}, L., {Iglesias-Paramo}, J.,
  {V{\'{\i}}lchez}, J.~M., \& {Carrasco}, L. 2001, \apjl, 563, L23

\bibitem[{Gerke {et~al.}(2005)Gerke, Newman, Davis, Marinoni, Yan, Coil,
  Conroy, Cooper, Faber, Finkbeiner, Guhathakurta, Kaiser, Koo, Phillips,
  Weiner, \& Willmer}]{Gerke2005}
Gerke, B.~F., {et~al.} 2005, \apj, 625, 6

\bibitem[{Gerke {et~al.}(2007)Gerke, Newman, Faber, Cooper, Croton, Davis,
  Willmer, Yan, Coil, Guhathakurta, Koo, \& Weiner}]{Gerke2007}
---. 2007, \mnras, 376, 1425

\bibitem[{{Gillis} \& {Hudson}(2011)}]{Gillis2011}
{Gillis}, B.~R., \& {Hudson}, M.~J. 2011, \mnras, 410, 13

\bibitem[{Giodini {et~al.}(2009)Giodini, Pierini, Finoguenov, Pratt,
  Boehringer, Leauthaud, Guzzo, Aussel, Bolzonella, Capak, Elvis, Hasinger,
  Ilbert, Kartaltepe, Koekemoer, Lilly, Massey, McCracken, Rhodes, Salvato,
  Sanders, Scoville, Sasaki, Smolcic, Taniguchi, Thompson, \& {the COSMOS
  Collaboration}}]{Giodini2009}
Giodini, S., {et~al.} 2009, \apj, 703, 982

\bibitem[{Gladders \& Yee(2005)}]{Gladders2005}
Gladders, M.~D., \& Yee, H. K.~C. 2005, \apjs, 157, 1

\bibitem[{Goto {et~al.}(2003)Goto, Okamura, Sekiguchi, Bernardi, Brinkmann,
  G\'{o}mez, Harvanek, Kleinman, Krzesinski, Long, Loveday, Miller, Neilsen,
  Newman, Nitta, Sheth, Snedden, \& Yamauchi}]{Goto2003}
Goto, T., {et~al.} 2003, \pasj, 55, 757

\bibitem[{Grove {et~al.}(2009)Grove, Benoist, \& Martel}]{Grove2009}
Grove, L.~F., Benoist, C., \& Martel, F. 2009, \aap, 494, 845

\bibitem[{{Haas} {et~al.}(2011){Haas}, {Schaye}, \& {Jeeson-Daniel}}]{Haas2011}
{Haas}, M.~R., {Schaye}, J., \& {Jeeson-Daniel}, A. 2011, ArXiv e-prints

\bibitem[{{Hall} {et~al.}(2010){Hall}, {Keisler}, {Knox}, {Reichardt}, {Ade},
  {Aird}, {Benson}, {Bleem}, {Carlstrom}, {Chang}, {Cho}, {Crawford}, {Crites},
  {de Haan}, {Dobbs}, {George}, {Halverson}, {Holder}, {Holzapfel}, {Hrubes},
  {Joy}, {Lee}, {Leitch}, {Lueker}, {McMahon}, {Mehl}, {Meyer}, {Mohr},
  {Montroy}, {Padin}, {Plagge}, {Pryke}, {Ruhl}, {Schaffer}, {Shaw},
  {Shirokoff}, {Spieler}, {Stalder}, {Staniszewski}, {Stark}, {Switzer},
  {Vanderlinde}, {Vieira}, {Williamson}, \& {Zahn}}]{Hall2010}
{Hall}, N.~R., {et~al.} 2010, \apj, 718, 632

\bibitem[{{Hansen} {et~al.}(2005){Hansen}, {McKay}, {Wechsler}, {Annis},
  {Sheldon}, \& {Kimball}}]{Hansen2005}
{Hansen}, S.~M., {McKay}, T.~A., {Wechsler}, R.~H., {Annis}, J., {Sheldon},
  E.~S., \& {Kimball}, A. 2005, \apj, 633, 122

\bibitem[{{Hansen} {et~al.}(2009){Hansen}, {Sheldon}, {Wechsler}, \&
  {Koester}}]{Hansen2009}
{Hansen}, S.~M., {Sheldon}, E.~S., {Wechsler}, R.~H., \& {Koester}, B.~P. 2009,
  \apj, 699, 1333

\bibitem[{Hasinger {et~al.}(2007)Hasinger, Cappelluti, Brunner, Brusa,
  Comastri, Elvis, Finoguenov, Fiore, Franceschini, Gilli, Griffiths, Lehmann,
  Mainieri, Matt, Matute, Miyaji, Molendi, Paltani, Sanders, Scoville, Tresse,
  Urry, Vettolani, \& Zamorani}]{Hasinger2007}
Hasinger, G., {et~al.} 2007, \apjs, 172, 29

\bibitem[{Hilton {et~al.}(2010)Hilton, Lloyd-Davies, Stanford, Stott, Collins,
  Romer, Hosmer, Hoyle, Kay, Liddle, Mehrtens, Miller, Sahl\'{e}n, \&
  Viana}]{Hilton2010}
Hilton, M., {et~al.} 2010, \apj, 718, 133

\bibitem[{{Hoaglin} {et~al.}(1983){Hoaglin}, {Mosteller}, \&
  {Tukey}}]{Hoaglin1983}
{Hoaglin}, D.~C., {Mosteller}, F., \& {Tukey}, J.~W. 1983, {Understanding
  robust and exploratory data anlysis}, ed. {Hoaglin, D.~C., Mosteller, F., \&
  Tukey, J.~W.}

\bibitem[{Ilbert {et~al.}(2009)Ilbert, Capak, Salvato, Aussel, McCracken,
  Sanders, Scoville, Kartaltepe, Arnouts, Floc'h, Mobasher, Taniguchi,
  Lamareille, Leauthaud, Sasaki, Thompson, Zamojski, Zamorani, Bardelli,
  Bolzonella, Bongiorno, Brusa, Caputi, Carollo, Contini, Cook, Coppa,
  Cucciati, de~la Torre, de~Ravel, Franzetti, Garilli, Hasinger, Iovino,
  Kampczyk, Kneib, Knobel, Kovac, {Le Borgne}, {Le Brun}, F\`{e}vre, Lilly,
  Looper, Maier, Mainieri, Mellier, Mignoli, Murayama, Pell\`{o}, Peng,
  P\'{e}rez-Montero, Renzini, Ricciardelli, Schiminovich, Scodeggio, Shioya,
  Silverman, Surace, Tanaka, Tasca, Tresse, Vergani, \& Zucca}]{Ilbert2009}
Ilbert, O., {et~al.} 2009, \apj, 690, 1236

\bibitem[{Ilbert {et~al.}(2010)Ilbert, Salvato, {Le Floc'h}, Aussel, Capak,
  McCracken, Mobasher, Kartaltepe, Scoville, Sanders, Arnouts, Bundy, Cassata,
  Kneib, Koekemoer, {Le F\`{e}vre}, Lilly, Surace, Taniguchi, Tasca, Thompson,
  Tresse, Zamojski, Zamorani, \& Zucca}]{Ilbert2010}
---. 2010, \apj, 709, 644

\bibitem[{Iovino {et~al.}(2010)Iovino, Cucciati, Scodeggio, Knobel, Kova\v{c},
  Lilly, Bolzonella, Tasca, Zamorani, Zucca, Caputi, Pozzetti, Oesch,
  Lamareille, Halliday, Bardelli, Finoguenov, Guzzo, Kampczyk, Maier, Tanaka,
  Vergani, Carollo, Contini, Kneib, {Le F\`{e}vre}, Mainieri, Renzini,
  Bongiorno, Coppa, de~la Torre, de~Ravel, Franzetti, Garilli, {Le Borgne}, {Le
  Brun}, Mignoli, Pell\`{o}, Peng, Perez-Montero, Ricciardelli, Silverman,
  Tresse, Abbas, Bottini, Cappi, Cassata, Cimatti, Koekemoer, Leauthaud,
  Maccagni, Marinoni, McCracken, Memeo, Meneux, Porciani, Scaramella,
  Schiminovich, \& Scoville}]{Iovino2010}
Iovino, A., {et~al.} 2010, \aap, 509, 40

\bibitem[{{Johnston} {et~al.}(2007){Johnston}, {Sheldon}, {Wechsler}, {Rozo},
  {Koester}, {Frieman}, {McKay}, {Evrard}, {Becker}, \& {Annis}}]{Johnston2007}
{Johnston}, D.~E., {et~al.} 2007, ArXiv e-prints

\bibitem[{Kauffmann {et~al.}(2004)Kauffmann, White, Heckman, M\'{e}nard,
  Brinchmann, Charlot, Tremonti, \& Brinkmann}]{Kauffmann2004}
Kauffmann, G., White, S. D.~M., Heckman, T.~M., M\'{e}nard, B., Brinchmann, J.,
  Charlot, S., Tremonti, C., \& Brinkmann, J. 2004, \mnras, 353, 713

\bibitem[{{Kenney} {et~al.}(1995){Kenney}, {Rubin}, {Planesas}, \&
  {Young}}]{Kenney1995}
{Kenney}, J.~D.~P., {Rubin}, V.~C., {Planesas}, P., \& {Young}, J.~S. 1995,
  \apj, 438, 135

\bibitem[{Knobel {et~al.}(2009)Knobel, Lilly, Iovino, Porciani, Kova\v{c},
  Cucciati, Finoguenov, Kitzbichler, Carollo, Contini, Kneib, {Le F\`{e}vre},
  Mainieri, Renzini, Scodeggio, Zamorani, Bardelli, Bolzonella, Bongiorno,
  Caputi, Coppa, de~la Torre, de~Ravel, Franzetti, Garilli, Kampczyk,
  Lamareille, {Le Borgne}, {Le Brun}, Maier, Mignoli, Pello, Peng, Montero,
  Ricciardelli, Silverman, Tanaka, Tasca, Tresse, Vergani, Zucca, Abbas,
  Bottini, Cappi, Cassata, Cimatti, Fumana, Guzzo, Koekemoer, Leauthaud,
  Maccagni, Marinoni, McCracken, Memeo, Meneux, Oesch, Pozzetti, \&
  Scaramella}]{Knobel2009}
Knobel, C., {et~al.} 2009, \apj, 697, 1842

\bibitem[{Koekemoer {et~al.}(2007)Koekemoer, Aussel, Calzetti, Capak,
  Giavalisco, Kneib, Leauthaud, {Le F\`{e}vre}, McCracken, Massey, Mobasher,
  Rhodes, Scoville, \& Shopbell}]{Koekemoer2007}
Koekemoer, A.~M., {et~al.} 2007, \apjs, 172, 196

\bibitem[{{Koester} {et~al.}(2007){Koester}, {McKay}, {Annis}, {Wechsler},
  {Evrard}, {Rozo}, {Bleem}, {Sheldon}, \& {Johnston}}]{Koester2007}
{Koester}, B.~P., {et~al.} 2007, \apj, 660, 221

\bibitem[{{Komatsu} \& {Seljak}(2002)}]{Komatsu2002}
{Komatsu}, E., \& {Seljak}, U. 2002, \mnras, 336, 1256

\bibitem[{Kova\v{c} {et~al.}(2010{\natexlab{a}})Kova\v{c}, Lilly, Knobel,
  Bolzonella, Iovino, Carollo, Scarlata, Sargent, Cucciati, Zamorani, Pozzetti,
  Tasca, Scodeggio, Kampczyk, Peng, Oesch, Zucca, Finoguenov, Contini, Kneib,
  {Le F\`{e}vre}, Mainieri, Renzini, Bardelli, Bongiorno, Caputi, Coppa, de~la
  Torre, de~Ravel, Franzetti, Garilli, Lamareille, {Le Borgne}, {Le Brun},
  Maier, Mignoli, Pello, {Perez Montero}, Ricciardelli, Silverman, Tanaka,
  Tresse, Vergani, Abbas, Bottini, Cappi, Cassata, Cimatti, Fumana, Guzzo,
  Koekemoer, Leauthaud, Maccagni, Marinoni, McCracken, Memeo, Meneux, Porciani,
  Scaramella, \& Scoville}]{Kovac2010}
Kova\v{c}, K., {et~al.} 2010{\natexlab{a}}, \apj, 718, 86

\bibitem[{Kova\v{c} {et~al.}(2010{\natexlab{b}})Kova\v{c}, Lilly, Cucciati,
  Porciani, Iovino, Zamorani, Oesch, Bolzonella, Knobel, Finoguenov, Peng,
  Carollo, Pozzetti, Caputi, Silverman, Tasca, Scodeggio, Vergani, Scoville,
  Capak, Contini, Kneib, {Le F\`{e}vre}, Mainieri, Renzini, Bardelli,
  Bongiorno, Coppa, de~la Torre, de~Ravel, Franzetti, Garilli, Guzzo, Kampczyk,
  Lamareille, {Le Borgne}, {Le Brun}, Maier, Mignoli, Pello, {Perez Montero},
  Ricciardelli, Tanaka, Tresse, Zucca, Abbas, Bottini, Cappi, Cassata, Cimatti,
  Fumana, Koekemoer, Maccagni, Marinoni, McCracken, Memeo, Meneux, \&
  Scaramella}]{Kovac2010a}
---. 2010{\natexlab{b}}, \apj, 708, 505

\bibitem[{Lagache {et~al.}(2003)Lagache, Dole, \& Puget}]{Lagache2003}
Lagache, G., Dole, H., \& Puget, J.-L. 2003, \mnras, 338, 555

\bibitem[{{Le Floc'h} {et~al.}(2009){Le Floc'h}, {Aussel}, {Ilbert},
  {Riguccini}, {Frayer}, {Salvato}, {Arnouts}, {Surace}, {Feruglio},
  {Rodighiero}, {Capak}, {Kartaltepe}, {Heinis}, {Sheth}, {Yan}, {McCracken},
  {Thompson}, {Sanders}, {Scoville}, \& {Koekemoer}}]{LeFloch2009}
{Le Floc'h}, E., {et~al.} 2009, \apj, 703, 222

\bibitem[{Leauthaud {et~al.}(2011{\natexlab{a}})Leauthaud, Tinker, Behroozi,
  Busha, \& Wechsler}]{Leauthaud2011a}
Leauthaud, A., Tinker, J., Behroozi, P.~S., Busha, M.~T., \& Wechsler, R.
  2011{\natexlab{a}}, arxiv:1103.2077

\bibitem[{Leauthaud {et~al.}(2007)Leauthaud, Massey, Kneib, Rhodes, Johnston,
  Capak, Heymans, Ellis, Koekemoer, {Le Fevre}, Mellier, Refregier, Robin,
  Scoville, Tasca, Taylor, \& {Van Waerbeke}}]{Leauthaud2007}
Leauthaud, A., {et~al.} 2007, \apjs, 172, 219

\bibitem[{Leauthaud {et~al.}(2010)Leauthaud, Finoguenov, Kneib, Taylor, Massey,
  Rhodes, Ilbert, Bundy, Tinker, George, Capak, Koekemoer, Johnston, Zhang,
  Cappelluti, Ellis, Elvis, Giodini, Heymans, {Le F\`{e}vre}, Lilly, McCracken,
  Mellier, R\'{e}fr\'{e}gier, Salvato, Scoville, Smoot, Tanaka, {Van Waerbeke},
  \& Wolk}]{Leauthaud2010}
---. 2010, \apj, 709, 97

\bibitem[{Leauthaud {et~al.}(2011{\natexlab{b}})Leauthaud, Tinker, Bundy,
  Behroozi, Massey, Rhodes, George, Kneib, Benson, Wechsler, Busha, Capak,
  Cortes, Ilbert, Koekemoer, Fevre, Lilly, McCracken, Salvato, Schrabback,
  Scoville, Smith, \& Taylor}]{Leauthaud2011b}
---. 2011{\natexlab{b}}, arxiv:1104.0928

\bibitem[{{Lee} {et~al.}(2010){Lee}, {Le Floc'h}, {Sanders}, {Frayer},
  {Arnouts}, {Ilbert}, {Aussel}, {Salvato}, {Scoville}, \&
  {Kartaltepe}}]{Lee2010}
{Lee}, N., {et~al.} 2010, \apj, 717, 175

\bibitem[{Lewis {et~al.}(2002)Lewis, Balogh, {De Propris}, Couch, Bower, Offer,
  Bland-Hawthorn, Baldry, Baugh, Bridges, Cannon, Cole, Colless, Collins,
  Cross, Dalton, Driver, Efstathiou, Ellis, Frenk, Glazebrook, Hawkins,
  Jackson, Lahav, Lumsden, Maddox, Madgwick, Norberg, Peacock, Percival,
  Peterson, Sutherland, \& Taylor}]{Lewis2002}
Lewis, I., {et~al.} 2002, \mnras, 334, 673

\bibitem[{Lilly {et~al.}(2007)Lilly, {Le F\`{e}vre}, Renzini, Zamorani,
  Scodeggio, Contini, Carollo, Hasinger, Kneib, Iovino, {Le Brun}, Maier,
  Mainieri, Mignoli, Silverman, Tasca, Bolzonella, Bongiorno, Bottini, Capak,
  Caputi, Cimatti, Cucciati, Daddi, Feldmann, Franzetti, Garilli, Guzzo,
  Ilbert, Kampczyk, Kovac, Lamareille, Leauthaud, Borgne, McCracken, Marinoni,
  Pello, Ricciardelli, Scarlata, Vergani, Sanders, Schinnerer, Scoville,
  Taniguchi, Arnouts, Aussel, Bardelli, Brusa, Cappi, Ciliegi, Finoguenov,
  Foucaud, Franceschini, Halliday, Impey, Knobel, Koekemoer, Kurk, Maccagni,
  Maddox, Marano, Marconi, Meneux, Mobasher, Moreau, Peacock, Porciani,
  Pozzetti, Scaramella, Schiminovich, Shopbell, Smail, Thompson, Tresse,
  Vettolani, Zanichelli, \& Zucca}]{Lilly2007}
Lilly, S.~J., {et~al.} 2007, \apjs, 172, 70

\bibitem[{{Lin} {et~al.}(2004){Lin}, {Mohr}, \& {Stanford}}]{Lin2004}
{Lin}, Y.-T., {Mohr}, J.~J., \& {Stanford}, S.~A. 2004, \apj, 610, 745

\bibitem[{Lueker {et~al.}(2010)Lueker, Reichardt, Schaffer, Zahn, Ade, Aird,
  Benson, Bleem, Carlstrom, Chang, Cho, Crawford, Crites, de~Haan, Dobbs,
  George, Hall, Halverson, Holder, Holzapfel, Hrubes, Joy, Keisler, Knox, Lee,
  Leitch, McMahon, Mehl, Meyer, Mohr, Montroy, Padin, Plagge, Pryke, Ruhl,
  Shaw, Shirokoff, Spieler, Stalder, Staniszewski, Stark, Vanderlinde, Vieira,
  \& Williamson}]{Lueker2010}
Lueker, M., {et~al.} 2010, \apj, 719, 1045

\bibitem[{Majumdar \& Mohr(2004)}]{Majumdar2004}
Majumdar, S., \& Mohr, J.~J. 2004, \apj, 613, 41

\bibitem[{Marinoni {et~al.}(2002)Marinoni, Davis, Newman, \&
  Coil}]{Marinoni2002}
Marinoni, C., Davis, M., Newman, J.~A., \& Coil, A.~L. 2002, \apj, 580, 122

\bibitem[{McCracken {et~al.}(2010)McCracken, Capak, Salvato, Aussel, Thompson,
  Daddi, Sanders, Kneib, Willott, Mancini, Renzini, Cook, {Le F\`{e}vre},
  Ilbert, Kartaltepe, Koekemoer, Mellier, Murayama, Scoville, Shioya, \&
  Tanaguchi}]{McCracken2010}
McCracken, H.~J., {et~al.} 2010, \apj, 708, 202

\bibitem[{{McGee} {et~al.}(2011){McGee}, {Balogh}, {Wilman}, {Bower},
  {Mulchaey}, {Parker}, \& {Oemler}}]{McGee2011}
{McGee}, S.~L., {Balogh}, M.~L., {Wilman}, D.~J., {Bower}, R.~G., {Mulchaey},
  J.~S., {Parker}, L.~C., \& {Oemler}, A. 2011, \mnras, 413, 996

\bibitem[{{Mei} {et~al.}(2009){Mei}, {Holden}, {Blakeslee}, {Ford}, {Franx},
  {Homeier}, {Illingworth}, {Jee}, {Overzier}, {Postman}, {Rosati}, {Van der
  Wel}, \& {Bartlett}}]{Mei2009}
{Mei}, S., {et~al.} 2009, \apj, 690, 42

\bibitem[{{Milkeraitis} {et~al.}(2010){Milkeraitis}, {van Waerbeke}, {Heymans},
  {Hildebrandt}, {Dietrich}, \& {Erben}}]{Milkeraitis2010}
{Milkeraitis}, M., {van Waerbeke}, L., {Heymans}, C., {Hildebrandt}, H.,
  {Dietrich}, J.~P., \& {Erben}, T. 2010, \mnras, 406, 673

\bibitem[{{Miller} {et~al.}(2005){Miller}, {Nichol}, {Reichart}, {Wechsler},
  {Evrard}, {Annis}, {McKay}, {Bahcall}, {Bernardi}, {Boehringer}, {Connolly},
  {Goto}, {Kniazev}, {Lamb}, {Postman}, {Schneider}, {Sheth}, \&
  {Voges}}]{Miller2005}
{Miller}, C.~J., {et~al.} 2005, \aj, 130, 968

\bibitem[{Mulchaey {et~al.}(2003)Mulchaey, Davis, Mushotzky, \&
  Burstein}]{Mulchaey2003}
Mulchaey, J.~S., Davis, D.~S., Mushotzky, R.~F., \& Burstein, D. 2003, \apjs,
  145, 39

\bibitem[{Mulchaey \& Zabludoff(1998)}]{Mulchaey1998}
Mulchaey, J.~S., \& Zabludoff, A.~I. 1998, \apj, 496, 73

\bibitem[{Navarro {et~al.}(1996)Navarro, Frenk, \& White}]{Navarro1996}
Navarro, J.~F., Frenk, C.~S., \& White, S. D.~M. 1996, \apj, 462, 563

\bibitem[{Oemler(1974)}]{Oemler1974}
Oemler, A. 1974, \apj, 194, 1

\bibitem[{Olsen {et~al.}(2007)Olsen, Benoist, Cappi, Maurogordato, Mazure,
  Slezak, Adami, Ferrari, \& Martel}]{Olsen2007}
Olsen, L.~F., {et~al.} 2007, \aap, 461, 81

\bibitem[{Osmond \& Ponman(2004)}]{Osmond2004}
Osmond, J. P.~F., \& Ponman, T.~J. 2004, \mnras, 350, 1511

\bibitem[{{Peng} {et~al.}(2010){Peng}, {Lilly}, {Kova{\v c}}, {Bolzonella},
  {Pozzetti}, {Renzini}, {Zamorani}, {Ilbert}, {Knobel}, {Iovino}, {Maier},
  {Cucciati}, {Tasca}, {Carollo}, {Silverman}, {Kampczyk}, {de Ravel},
  {Sanders}, {Scoville}, {Contini}, {Mainieri}, {Scodeggio}, {Kneib}, {Le
  F{\`e}vre}, {Bardelli}, {Bongiorno}, {Caputi}, {Coppa}, {de la Torre},
  {Franzetti}, {Garilli}, {Lamareille}, {Le Borgne}, {Le Brun}, {Mignoli},
  {Perez Montero}, {Pello}, {Ricciardelli}, {Tanaka}, {Tresse}, {Vergani},
  {Welikala}, {Zucca}, {Oesch}, {Abbas}, {Barnes}, {Bordoloi}, {Bottini},
  {Cappi}, {Cassata}, {Cimatti}, {Fumana}, {Hasinger}, {Koekemoer},
  {Leauthaud}, {Maccagni}, {Marinoni}, {McCracken}, {Memeo}, {Meneux}, {Nair},
  {Porciani}, {Presotto}, \& {Scaramella}}]{Peng2010}
{Peng}, Y.-j., {et~al.} 2010, \apj, 721, 193

\bibitem[{Poggianti {et~al.}(1999)Poggianti, Smail, Dressler, Couch, Barger,
  Butcher, Ellis, \& Oemler}]{Poggianti1999}
Poggianti, B.~M., Smail, I., Dressler, A., Couch, W.~J., Barger, A.~J.,
  Butcher, H., Ellis, R.~S., \& Oemler, A. 1999, \apj, 518, 576

\bibitem[{{Poggianti} {et~al.}(2008){Poggianti}, {Desai}, {Finn}, {Bamford},
  {De Lucia}, {Varela}, {Arag{\'o}n-Salamanca}, {Halliday}, {Noll}, {Saglia},
  {Zaritsky}, {Best}, {Clowe}, {Milvang-Jensen}, {Jablonka}, {Pell{\'o}},
  {Rudnick}, {Simard}, {von der Linden}, \& {White}}]{Poggianti2008}
{Poggianti}, B.~M., {et~al.} 2008, \apj, 684, 888

\bibitem[{Postman {et~al.}(1996)Postman, Lubin, Gunn, Oke, Hoessel, Schneider,
  \& Christensen}]{Postman1996}
Postman, M., Lubin, L., Gunn, J., Oke, J., Hoessel, J., Schneider, D., \&
  Christensen, J. 1996, \aj, 111, 615

\bibitem[{{Postman} {et~al.}(2005){Postman}, {Franx}, {Cross}, {Holden},
  {Ford}, {Illingworth}, {Goto}, {Demarco}, {Rosati}, {Blakeslee}, {Tran},
  {Ben{\'{\i}}tez}, {Clampin}, {Hartig}, {Homeier}, {Ardila}, {Bartko},
  {Bouwens}, {Bradley}, {Broadhurst}, {Brown}, {Burrows}, {Cheng}, {Feldman},
  {Golimowski}, {Gronwall}, {Infante}, {Kimble}, {Krist}, {Lesser}, {Martel},
  {Mei}, {Menanteau}, {Meurer}, {Miley}, {Motta}, {Sirianni}, {Sparks}, {Tran},
  {Tsvetanov}, {White}, \& {Zheng}}]{Postman2005}
{Postman}, M., {et~al.} 2005, \apj, 623, 721

\bibitem[{{Prescott} {et~al.}(2006){Prescott}, {Impey}, {Cool}, \&
  {Scoville}}]{Prescott2006}
{Prescott}, M.~K.~M., {Impey}, C.~D., {Cool}, R.~J., \& {Scoville}, N.~Z. 2006,
  \apj, 644, 100

\bibitem[{{Rozo} {et~al.}(2011){Rozo}, {Rykoff}, {Koester}, {Nord}, {Wu},
  {Evrard}, \& {Wechsler}}]{Rozo2011}
{Rozo}, E., {Rykoff}, E., {Koester}, B., {Nord}, B., {Wu}, H.-Y., {Evrard}, A.,
  \& {Wechsler}, R. 2011, ArXiv e-prints

\bibitem[{{Rykoff} {et~al.}(2011){Rykoff}, {Koester}, {Rozo}, {Annis},
  {Evrard}, {Hansen}, {Hao}, {Johnston}, {McKay}, \& {Wechsler}}]{Rykoff2011}
{Rykoff}, E.~S., {et~al.} 2011, ArXiv e-prints

\bibitem[{Salvato {et~al.}(2009)Salvato, Hasinger, Ilbert, Zamorani, Brusa,
  Scoville, Rau, Capak, Arnouts, Aussel, Bolzonella, Buongiorno, Cappelluti,
  Caputi, Civano, Cook, Elvis, Gilli, Jahnke, Kartaltepe, Impey, Lamareille,
  {Le Floc'h}, Lilly, Mainieri, McCarthy, McCracken, Mignoli, Mobasher,
  Murayama, Sasaki, Sanders, Schiminovich, Shioya, Shopbell, Silverman,
  Smol\v{c}i\'{c}, Surace, Taniguchi, Thompson, Trump, Urry, \&
  Zamojski}]{Salvato2009}
Salvato, M., {et~al.} 2009, \apj, 690, 1250

\bibitem[{{Sanders} {et~al.}(2007){Sanders}, {Salvato}, {Aussel}, {Ilbert},
  {Scoville}, {Surace}, {Frayer}, {Sheth}, {Helou}, {Brooke}, {Bhattacharya},
  {Yan}, {Kartaltepe}, {Barnes}, {Blain}, {Calzetti}, {Capak}, {Carilli},
  {Carollo}, {Comastri}, {Daddi}, {Ellis}, {Elvis}, {Fall}, {Franceschini},
  {Giavalisco}, {Hasinger}, {Impey}, {Koekemoer}, {Le F{\`e}vre}, {Lilly},
  {Liu}, {McCracken}, {Mobasher}, {Renzini}, {Rich}, {Schinnerer}, {Shopbell},
  {Taniguchi}, {Thompson}, {Urry}, \& {Williams}}]{Sanders2007}
{Sanders}, D.~B., {et~al.} 2007, \apjs, 172, 86

\bibitem[{Scarlata {et~al.}(2007)Scarlata, Carollo, Lilly, Sargent, Feldmann,
  Kampczyk, Porciani, Koekemoer, Scoville, Kneib, Leauthaud, Massey, Rhodes,
  Tasca, Capak, Maier, McCracken, Mobasher, Renzini, Taniguchi, Thompson,
  Sheth, Ajiki, Aussel, Murayama, Sanders, Sasaki, Shioya, \&
  Takahashi}]{Scarlata2007}
Scarlata, C., {et~al.} 2007, \apjs, 172, 406

\bibitem[{Scodeggio {et~al.}(2009)Scodeggio, Vergani, Cucciati, Iovino,
  Franzetti, Garilli, Lamareille, Bolzonella, Pozzetti, Abbas, Marinoni,
  Contini, Bottini, {Le Brun}, {Le F\`{e}vre}, Maccagni, Scaramella, Tresse,
  Vettolani, Zanichelli, Adami, Arnouts, Bardelli, Cappi, Charlot, Ciliegi,
  Foucaud, Gavignaud, Guzzo, Ilbert, McCracken, Marano, Mazure, Meneux,
  Merighi, Paltani, Pell\`{o}, Pollo, Radovich, Zamorani, Zucca, Bondi,
  Bongiorno, Brinchmann, {de La Torre}, de~Ravel, Gregorini, Memeo,
  Perez-Montero, Mellier, Temporin, \& Walcher}]{Scodeggio2009}
Scodeggio, M., {et~al.} 2009, \aap, 501, 21

\bibitem[{Scoville {et~al.}(2007)Scoville, Aussel, Benson, Blain, Calzetti,
  Capak, Ellis, El‐Zant, Finoguenov, Giavalisco, Guzzo, Hasinger, Koda, {Le
  Fevre}, Massey, McCracken, Mobasher, Renzini, Rhodes, Salvato, Sanders,
  Sasaki, Schinnerer, Sheth, Shopbell, Taniguchi, Taylor, \&
  Thompson}]{Scoville2007b}
Scoville, N., {et~al.} 2007, \apjs, 172, 150

\bibitem[{{Scoville} {et~al.}(2007){Scoville}, {Aussel}, {Brusa}, {Capak},
  {Carollo}, {Elvis}, {Giavalisco}, {Guzzo}, {Hasinger}, {Impey}, {Kneib},
  {LeFevre}, {Lilly}, {Mobasher}, {Renzini}, {Rich}, {Sanders}, {Schinnerer},
  {Schminovich}, {Shopbell}, {Taniguchi}, \& {Tyson}}]{Scoville2007a}
{Scoville}, N., {et~al.} 2007, \apjs, 172, 1

\bibitem[{{Shaw} {et~al.}(2010){Shaw}, {Nagai}, {Bhattacharya}, \&
  {Lau}}]{Shaw2010}
{Shaw}, L.~D., {Nagai}, D., {Bhattacharya}, S., \& {Lau}, E.~T. 2010, \apj,
  725, 1452

\bibitem[{{Shirokoff} {et~al.}(2010){Shirokoff}, {Reichardt}, {Shaw}, {Millea},
  {Ade}, {Aird}, {Benson}, {Bleem}, {Carlstrom}, {Chang}, {Cho}, {Crawford},
  {Crites}, {de Haan}, {Dobbs}, {Dudley}, {George}, {Halverson}, {Holder},
  {Holzapfel}, {Hrubes}, {Joy}, {Keisler}, {Knox}, {Lee}, {Leitch}, {Lueker},
  {Luong-Van}, {McMahon}, {Mehl}, {Meyer}, {Mohr}, {Montroy}, {Padin},
  {Plagge}, {Pryke}, {Ruhl}, {Schaffer}, {Spieler}, {Staniszewski}, {Stark},
  {Story}, {Vanderlinde}, {Vieira}, {Williamson}, \& {Zahn}}]{Shirokoff2010}
{Shirokoff}, E., {et~al.} 2010, ArXiv e-prints

\bibitem[{{Siebenmorgen} \& {Kr{\"u}gel}(2007)}]{Siebenmorgen2007}
{Siebenmorgen}, R., \& {Kr{\"u}gel}, E. 2007, \aap, 461, 445

\bibitem[{Simionescu {et~al.}(2011)Simionescu, Allen, Mantz, Werner, Takei,
  Morris, Fabian, Sanders, Nulsen, George, \& Taylor}]{Simionescu2011}
Simionescu, A., {et~al.} 2011, Science, 331, 1576

\bibitem[{{Skibba} {et~al.}(2011){Skibba}, {van den Bosch}, {Yang}, {More},
  {Mo}, \& {Fontanot}}]{Skibba2011}
{Skibba}, R.~A., {van den Bosch}, F.~C., {Yang}, X., {More}, S., {Mo}, H., \&
  {Fontanot}, F. 2011, \mnras, 410, 417

\bibitem[{{Smith} {et~al.}(2005){Smith}, {Treu}, {Ellis}, {Moran}, \&
  {Dressler}}]{Smith2005}
{Smith}, G.~P., {Treu}, T., {Ellis}, R.~S., {Moran}, S.~M., \& {Dressler}, A.
  2005, \apj, 620, 78

\bibitem[{Sun {et~al.}(2011)Sun, Sehgal, Voit, Donahue, Jones, Forman,
  Vikhlinin, \& Sarazin}]{Sun2011}
Sun, M., Sehgal, N., Voit, G.~M., Donahue, M., Jones, C., Forman, W.,
  Vikhlinin, A., \& Sarazin, C. 2011, \apj, 727, L49

\bibitem[{Sun {et~al.}(2009)Sun, Voit, Donahue, Jones, Forman, \&
  Vikhlinin}]{Sun2009}
Sun, M., Voit, G.~M., Donahue, M., Jones, C., Forman, W., \& Vikhlinin, A.
  2009, \apj, 693, 1142

\bibitem[{Sunyaev \& Zeldovich(1972)}]{Sunyaev1972}
Sunyaev, R.~A., \& Zeldovich, Y.~B. 1972, Comments on Astrophysics and Space
  Physics, 4, 173

\bibitem[{Tanaka {et~al.}(2010)Tanaka, Finoguenov, \& Ueda}]{Tanaka2010}
Tanaka, M., Finoguenov, A., \& Ueda, Y. 2010, \apj, 716, L152

\bibitem[{{Tanaka} {et~al.}(2005){Tanaka}, {Kodama}, {Arimoto}, {Okamura},
  {Umetsu}, {Shimasaku}, {Tanaka}, \& {Yamada}}]{Tanaka2005}
{Tanaka}, M., {Kodama}, T., {Arimoto}, N., {Okamura}, S., {Umetsu}, K.,
  {Shimasaku}, K., {Tanaka}, I., \& {Yamada}, T. 2005, \mnras, 362, 268

\bibitem[{Tasca {et~al.}(2009)Tasca, Kneib, Iovino, {Le F\`{e}vre}, Kova\v{c},
  Bolzonella, Lilly, Abraham, Cassata, Cucciati, Guzzo, Tresse, Zamorani,
  Capak, Garilli, Scodeggio, Sheth, Zucca, Carollo, Contini, Mainieri, Renzini,
  Bardelli, Bongiorno, Caputi, Coppa, {de La Torre}, de~Ravel, Franzetti,
  Kampczyk, Knobel, Koekemoer, Lamareille, {Le Borgne}, {Le Brun}, Maier,
  Mignoli, Pello, Peng, {Perez Montero}, Ricciardelli, Silverman, Vergani,
  Tanaka, Abbas, Bottini, Cappi, Cimatti, Ilbert, Leauthaud, Maccagni,
  Marinoni, McCracken, Memeo, Meneux, Oesch, Porciani, Pozzetti, Scaramella, \&
  Scarlata}]{Tasca2009}
Tasca, L. A.~M., {et~al.} 2009, \aap, 503, 379

\bibitem[{{Tinker} {et~al.}(2008){Tinker}, {Kravtsov}, {Klypin}, {Abazajian},
  {Warren}, {Yepes}, {Gottl{\"o}ber}, \& {Holz}}]{Tinker2008}
{Tinker}, J., {Kravtsov}, A.~V., {Klypin}, A., {Abazajian}, K., {Warren}, M.,
  {Yepes}, G., {Gottl{\"o}ber}, S., \& {Holz}, D.~E. 2008, \apj, 688, 709

\bibitem[{Trac {et~al.}(2011)Trac, Bode, \& Ostriker}]{Trac2011}
Trac, H., Bode, P., \& Ostriker, J.~P. 2011, \apj, 727, 94

\bibitem[{Tran {et~al.}(2001)Tran, Simard, Zabludoff, \& Mulchaey}]{Tran2001}
Tran, K.-V.~H., Simard, L., Zabludoff, A.~I., \& Mulchaey, J.~S. 2001, \apj,
  549, 172

\bibitem[{Tran {et~al.}(2010)Tran, Papovich, Saintonge, Brodwin, Dunlop,
  Farrah, Finkelstein, Finkelstein, Lotz, McLure, Momcheva, \&
  Willmer}]{Tran2010}
Tran, K.-V.~H., {et~al.} 2010, \apj, 719, L126

\bibitem[{{van der Wel} {et~al.}(2007){van der Wel}, {Holden}, {Franx},
  {Illingworth}, {Postman}, {Kelson}, {Labb{\'e}}, {Wuyts}, {Blakeslee}, \&
  {Ford}}]{VanDerWel2007}
{van der Wel}, A., {et~al.} 2007, \apj, 670, 206

\bibitem[{Vikhlinin {et~al.}(1998)Vikhlinin, McNamara, Forman, Jones, Quintana,
  \& Hornstrup}]{Vikhlinin1998}
Vikhlinin, A., McNamara, B.~R., Forman, W., Jones, C., Quintana, H., \&
  Hornstrup, A. 1998, \apj, 502, 558

\bibitem[{Weinmann {et~al.}(2006)Weinmann, van~den Bosch, Yang, \&
  Mo}]{Weinmann2006}
Weinmann, S.~M., van~den Bosch, F.~C., Yang, X., \& Mo, H.~J. 2006, \mnras,
  366, 2

\bibitem[{Yang {et~al.}(2005)Yang, Mo, {van Den Bosch}, \& Jing}]{Yang2005}
Yang, X., Mo, H.~J., {van Den Bosch}, F.~C., \& Jing, Y.~P. 2005, \mnras, 356,
  1293

\bibitem[{{Yip} {et~al.}(2011){Yip}, {Szalay}, {Carliles}, \&
  {Budav{\'a}ri}}]{Yip2011}
{Yip}, C., {Szalay}, A.~S., {Carliles}, S., \& {Budav{\'a}ri}, T. 2011, \apj,
  730, 54

\bibitem[{Zabludoff \& Mulchaey(1998)}]{Zabludoff1998}
Zabludoff, A.~I., \& Mulchaey, J.~S. 1998, \apj, 496, 39

\bibitem[{Zhao {et~al.}(2009)Zhao, Jing, Mo, \& B\"{o}rner}]{Zhao2009}
Zhao, D.~H., Jing, Y.~P., Mo, H.~J., \& B\"{o}rner, G. 2009, \apj, 707, 354

\end{thebibliography}


\end{document}